\documentclass{article}

\usepackage[affil-it]{authblk}

\title{Interpreting Radial Correlation Doppler Reflectometry using Gyrokinetic Simulations}

\author[1]{J. Ruiz Ruiz}
\author[1]{F. I. Parra}
\author[1]{V. H. Hall-Chen}
\author[1]{N. Christen}
\author[1]{M. Barnes}
\author[2]{J. Candy}
\author[3]{J. Garcia}
\author[4]{C. Giroud}
\author[5]{W. Guttenfelder}
\author[4]{J. C. Hillesheim}
\author[6]{C. Holland}
\author[7]{N. T. Howard}
\author[5]{Y. Ren}
\author[7]{A. E. White}
\author[8]{JET contributors} 

\affil[1]{Rudolf Peierls Centre for Theoretical Physics, University of Oxford, OX1 3NP, UK}
\affil[2]{General Atomics, P.O. Box 85608, San Diego, CA, USA}
\affil[3]{CEA, IRFM, F-13108 Saint-Paul-lez-Durance, France}
\affil[4]{CCFE, Culham Science Centre, Abingdon, Oxon OX14 3DB, UK}
\affil[5]{Princeton Plasma Physics Laboratory, Princeton, New Jersey 08543, USA}
\affil[6]{Center for Energy Research, University of California, San Diego, La Jolla, California 92093-0417, USA}
\affil[7]{MIT-Plasma Science and Fusion Center, Cambridge, Massachusetts 02139, USA}
\affil[8]{See the author list of  'Overview of JET results for optimising ITER operation' by J. Mailloux et al. to be published in Nuclear Fusion Special issue: Overview and Summary Papers from the 28th Fusion Energy Conference (Nice, France, 10-15 May 2021).}

\date{}

\usepackage{amssymb,amsthm,amsfonts,amscd}
\usepackage{amsmath, amssymb}
\usepackage{mathtools}   
\usepackage{vmargin}
\usepackage{times}
\usepackage{textcomp}
\usepackage{esvect}      
\usepackage{stmaryrd}
\usepackage{dsfont}      
\usepackage{fancyhdr}    
\usepackage{lastpage}
\usepackage{eurosym}    
\usepackage{indentfirst}
\usepackage{graphicx}
\usepackage{cite}
\usepackage{esvect}
\usepackage{hhline}
\usepackage{bm}
\def\-{\raisebox{.75pt}{-}}

\usepackage{array}
\usepackage{booktabs}
\usepackage{xcolor}
\DeclareMathOperator{\f}{f}

\usepackage[font=small,labelfont=bf]{caption} 

\begin{document}

\maketitle


\setlength{\parskip}{\baselineskip}%
\setlength{\parindent}{0pt}%

\begin{abstract}

A linear response, local model for the DBS amplitude applied to gyrokinetic simulations shows that radial correlation Doppler reflectometry measurements (RCDR, Schirmer \emph{et al}., Plasma Phys. Control. Fusion \textbf{49} 1019 (2007) \cite{schirmer_ppcf_2007}) are not sensitive to the average turbulence radial correlation length, but to a correlation length that depends on the binormal wavenumber $k_\perp$ selected by the Doppler backscattering (DBS) signal. Nonlinear gyrokinetic simulations show that the turbulence naturally exhibits a non-separable power law spectrum in wavenumber space, leading to a power law dependence of the radial correlation length with binormal wavenumber $l_r \sim C k_\perp^{-\alpha} (\alpha \approx 1)$ which agrees with the inverse proportionality relationship between the measured $l_r$ and $k_\perp $ in experiments (Fernández-Marina \emph{et al}., Nucl. Fusion \textbf{54} 072001 (2014) \cite{fernandezmarina_nf_2014}). This offers the possibility of characterizing the eddy aspect ratio in the perpendicular plane to the magnetic field and motivates future use of a non-separable turbulent spectrum to quantitatively interpret RCDR and potentially other turbulence diagnostics. The radial correlation length is only measurable when the radial resolution at the cutoff location $W_n$ satisfies $W_n \ll l_r$, while the measurement becomes dominated by $W_n$ for $W_n \gg l_r$. This suggests that $l_r$ is likely inaccessible for electron-scale DBS measurements ($k_\perp\rho_s > 1$). The effect of $W_n$ on ion-scale radial correlation lengths could be non-negligible.

\end{abstract}

\section{Introduction}
\label{intro}


Decades of development in fusion energy research have provided strong experimental evidence \cite{liewer_nf_1985, tynan_ppcf_2009} and theoretical support \cite{horton_revmodphys_1999, garbet_ppcf_2001} that micro-scale turbulent fluctuations are ultimately responsible for driving anomalous transport in the core of magnetic confinement devices. The turbulence is driven by the large density and temperature gradients existing in the tokamak core, which provide free energy to drive micro-scale instabilities that ultimately develop into turbulence. Turbulence can manifest itself in the levels of anomalous transport of particles and heat that it drives out of the core confining region, degrading confinement. First-principles nonlinear gyrokinetic simulations of the turbulence have routinely been successful at predicting the experimental levels of anomalous transport in some plasma scenarios \cite{candy_prl_2003}. These numerical and theoretical studies yield understanding on the turbulent processes, but can only provide indirect evidence of the turbulence. The most definitive experimental evidence can be provided by fluctuation diagnostics, which can measure fluctuations in density, temperature, magnetic field and flows that are produced by the turbulence. Our understanding relies on making quantitative interpretations of the fluctuation measurements, which has proven to be a great challenge for the validation of the existing turbulence and transport models \cite{terry_pop_2008, greenwald_pop_2010, holland_pop_2016}. The highly sophisticated nature of the measurement process has motivated the development of synthetic diagnostics to understand not only the turbulent fluctuations, but also the measurement process taking place in fluctuation measurements. 

The radial correlation length is an important intrinsic characteristic of the turbulence with a direct relation to transport. Several diagnostics have been successful at measuring turbulence radial correlation lengths in the core plasma, such as beam emission spectroscopy (BES \cite{fonck_rsi_1990, fonck_prl_1993, mckee_nf_2001}), phase-contrast imaging/interferometry \cite{weisen_rsi_1990, coda_rsi_1992, coda_rsi_1995, coda_prl_2001, porkolab_ieee_2006}, correlation electron cyclotron emission \cite{sattler_prl_1994, cima_pop_1995, white_pop_2008} and radial correlation reflectometry (RCR \cite{cripwell_eps_1989, costley_rsi_1990, nazikian_rsi_1995, nazikian_pop_2001, rhodes_pop_2002}). Measurements using radial correlation reflectometry have proven particularly challenging to interpret, often yielding unphysically large or small values \cite{mazzucato_prl_1993, rhodes_rsi_1992, sanchez_branas_rsi_1993, hanson_rsi_1990, hutchinson_ppcf_1992, conway_ppcf_1997, nazikian_pop_2001}. In the literature, the unusually short correlation lengths are often attributed to the nonlinear response of the reflectometer to the density fluctuation amplitude \cite{gusakov_ppcf_2004_nlrcr}, while the unusually large values have mostly been attributed to the presence of a small angle forward-scattering contribution due to long radial scale turbulent fluctuations \cite{gusakov_ppcf_2002_linrcr}. 

A more recent evolution of radial correlation reflectometry is Radial Correlation Doppler Reflectometry (RCDR \cite{schirmer_ppcf_2007}), implemented via radial correlation of Doppler backscattering signals. Doppler backscattering \cite{holzhauer_ppcf_1998, hirsch_rsi_2001} is a standard and versatile diagnostic able to measure the turbulent wavenumber spectrum \cite{hennequin_nf_2006, hillesheim_nf_2015b}, zonal and equilibrium flows \cite{hirsch_ppcf_2004, hennequin_rsi_2004, hillesheim_prl_2016} as well as the turbulent correlation length \cite{schirmer_ppcf_2007}. As in standard reflectometry, the Doppler backscattering technique launches a microwave beam into the core plasma, but with a finite incidence angle with respect to the horizontal. The beam propagates in the plasma until it encounters a cutoff surface, where the forward beam is refracted away from the detector and only the backscattering contribution is collected. This scheme is believed to remove, or greatly reduce, the small angle forward scattering contribution to the radial correlation in the linear regime \cite{gusakov_ppcf_2014, gusakov_pop_2017} that hinders standard reflectometry. The finite incidence beam launch angle makes the backscattered signal sensitive to a specific turbulence wavenumber $\mathbf{k}_{\perp}$ that is selected by the scattering process, which is related to the central ray wave-vector value $\mathbf{k}_i$ (taken at the cutoff location) by the Bragg condition for backscattering at the cutoff $\mathbf{k}_\perp = -2 \mathbf{k}_i$ \cite{holzhauer_ppcf_1998, hirsch_rsi_2001} (note the absence of the incidence beam launch angle). 

Experiments measuring the radial correlation length via the RCDR technique have provided radial correlation lengths characteristic of ion-scale eddies $l_r \approx 3 \textendash 10 \ \rho_s$ \cite{schirmer_ppcf_2007, gusakov_ppcf_2013, fernandezmarina_nf_2014, altukhov_ppcf_2016, altukhov_pop_2018a, altukhov_pop_2018b, prisiazhniuk_ppcf_2018, krutkin_nf_2019}, which are more consistent with theory than its reflectometer counterpart ($\rho_s = c_s/\omega_{ci}$ is the ion-sound gyro-radius computed using the local toroidal field $B = |\mathbf{B_\varphi}|$ and electron temperature $T_e$, $c_s = \sqrt{T_e/m_i}$ is the sound speed, $\omega_{ci} = {e} B / m_i c$ is the gyro-frequency, $m_i$ is the main ion mass, ${e}$ is the proton charge and $c$ the speed of light). However, open questions still remain in the interpretation of the measured values. The 'common-wisdom' is that experimentally relevant values of the selected $k_\perp$ by DBS lie on the intermediate to electron-scale range, that is $k_\perp\rho_s \gtrsim 1$. This would imply that the measured fluctuations by DBS are in fact intermediate-to-electron scales in the binormal $k_\perp$ direction, but exhibit radial correlation lengths that are ion-scales. Part of the present work seeks to clarify the origin of this result of DBS measurements. 

\begin{figure}
	\begin{center}
		\includegraphics[height=8cm]{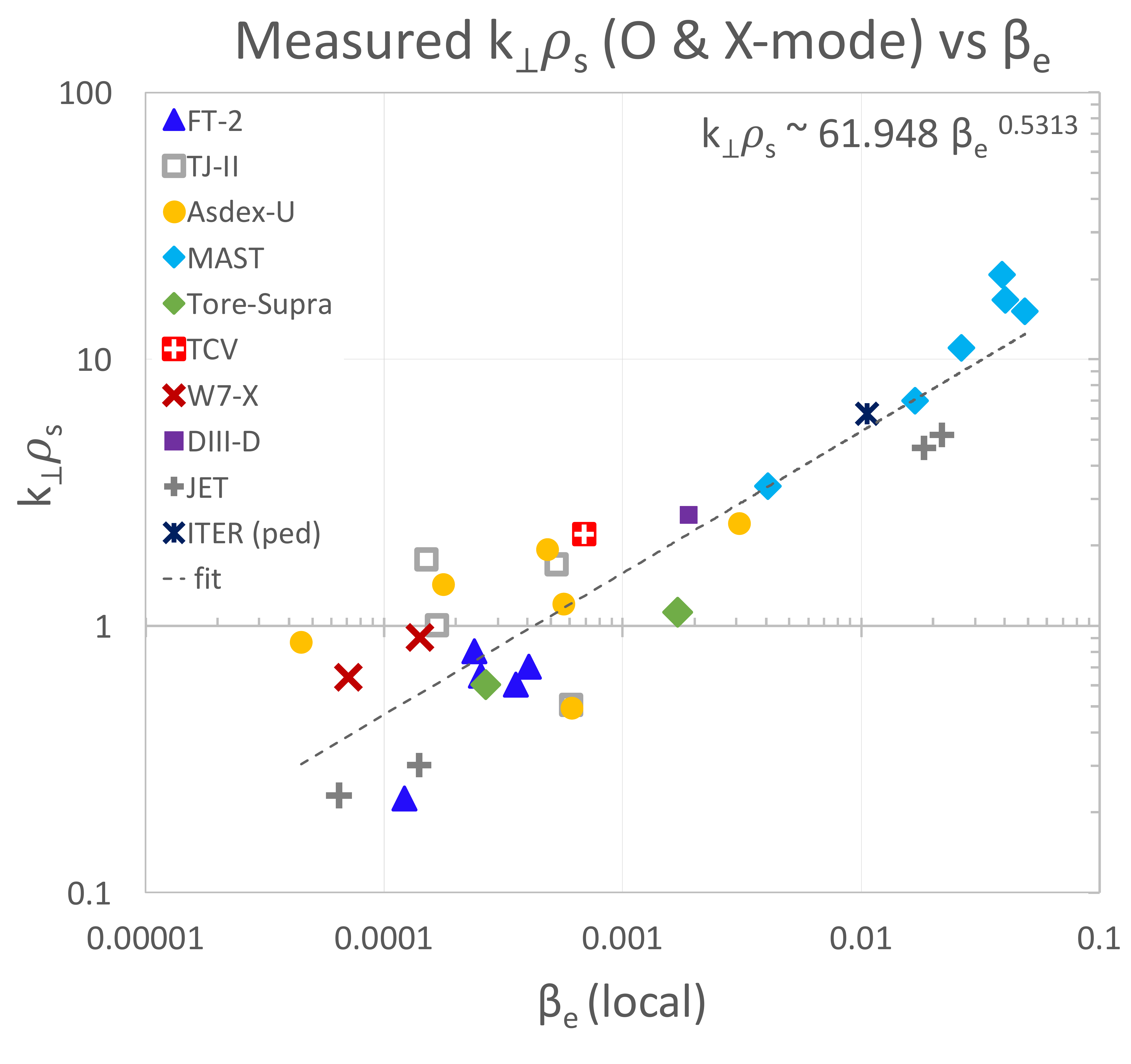}
	\end{center}
	\caption{Measured $k_\perp\rho_s$ from a series of magnetic confinement devices $vs.$ the local values of $\beta_e$ \cite{gusakov_ppcf_2013, altukhov_ppcf_2016, altukhov_pop_2018a, altukhov_pop_2018b, fernandezmarina_nf_2014, happel_pop_2011, conway_ppcf_2004, schirmer_ppcf_2007, hillesheim_nf_2015a, hillesheim_nf_2015b, bourdelle_nf_2011, casati_prl_2009, meijere_ppcf_2014, windisch_rsi_2018, deboo_pop_2010, hillesheim_prl_2016, silva_ppcf_2018, silva_ppcf_2019, conway_irw_2011}. The scaling of $k_\perp\rho_s$ with $\beta_e$ can be explained by the Bragg condition for scattering and the cutoff condition, yielding $k_\perp\rho_s \sim (m_D/m_e)^{0.5} \beta_e^{0.5}$ (O-mode). Plasma regimes characterized by low $\beta_e$ ($\beta_e \approx 10^{-5} \textendash 10^{-4}$ at the pedestal, high-field machines or low performance discharges) will predominantly measure ion-scale wavenumbers $k_\perp\rho_s < 1$, while high $\beta_e$ scenarios ($\beta_e \approx 10^{-2} \textendash 10^{-1}$ in high performance regimes or spherical tokamaks) will likely be sensitive to intermediate to electron-scales $k_\perp\rho_s \gtrsim 1$. ITER values correspond to pedestal projections from \cite{conway_irw_2011}. The figure is not an accurate representation of the measurement capabilities of specific devices, but indicative of the measured $k_\perp\rho_s$ for a specific value of $\beta_e$. }   
	\label{tokamaks_kperprho_betae}
\end{figure}

Careful analysis of DBS systems in different magnetic confinement devices shows that disparate values of $k_\perp$ are accessible via DBS, ranging from electron to ion scales, as shown in figure \ref{tokamaks_kperprho_betae}. The measured $k_\perp\rho_s$ is shown to scale closely with the local value of electron beta to the $1/2$ power, $k_\perp\rho_s \propto \beta_e^{0.5}$, which can be explained as follows. Due to the backscattering Bragg condition, the measured $k_\perp $ scales with the incident $k_i$ of the microwave beam $k_\perp\rho_s \sim k_i \rho_s$. In turn, the incident $k_i$ is related to the frequency of the cutoff surface $\omega$ due to the cutoff condition, that is $k_i \sim \omega/c$. To simplify the argument, we assume $\omega \approx \omega_{pe}$ the electron plasma frequency (strictly valid for normal incidence, O-mode polarization). Thus, $k_\perp\rho_s \sim \omega_{pe}\rho_s/c$, which is rewritten as $k_\perp\rho_s \sim ( m_D/m_e)^{0.5} \beta_e^{0.5} $ for a deuterium plasma. Although seemingly simplistic, this relationship between the selected $k_\perp\rho_s$ and the local value of $\beta_e$ is shown to perform fairly satisfactorily when contrasted with DBS measurements from different machines in figure \ref{tokamaks_kperprho_betae}. Figure \ref{tokamaks_kperprho_betae} is constructed using values of $k_\perp\rho_s$ and $\beta_e$ that are either provided, or computed using alternate quantities provided in references \cite{gusakov_ppcf_2013, altukhov_ppcf_2016, altukhov_pop_2018a, altukhov_pop_2018b, fernandezmarina_nf_2014, happel_pop_2011, conway_ppcf_2004, schirmer_ppcf_2007, hillesheim_nf_2015a, hillesheim_nf_2015b, bourdelle_nf_2011, casati_prl_2009, meijere_ppcf_2014, windisch_rsi_2018, deboo_pop_2010, hillesheim_prl_2016, silva_ppcf_2018, silva_ppcf_2019, conway_irw_2011}. There is natural scatter in the values. There are likely errors in the measurement and in the computation of $k_\perp\rho_s$ and $\beta_e$. Perhaps more importantly, this scaling does not take into account the dependence on the incident launch angle of the microwave beam in a specific scattering experiment, which can admittedly yield differences on the order of $\sim 1 \textendash 10$ in the measured $k_\perp$. However, when spanning four orders of magnitude in $\beta_e$, these important factors appear not to have a dramatic effect on this scaling, as suggested by a least-squares power law fit $k_\perp\rho_s \sim 61.95 \ \beta_e^{0.53}$, which remains close to the naive scaling $k_\perp\rho_s \sim ( m_D/m_e)^{0.5} \beta_e^{0.5}$ presented. 

This dimensionless scaling is \emph{not} an equality relating $k_\perp\rho_s$, $( m_D/m_e)^{0.5}$ and $\beta_e$. Figure \ref{tokamaks_kperprho_betae} does \emph{not} represent the measured $k_\perp$ range achievable by a specific machine (except possibly the range reported for JET \footnote{Low $\beta_e$ values for JET are from low performance edge measurements \cite{hillesheim_prl_2016, silva_ppcf_2018, silva_ppcf_2019}, while the high $\beta_e$ values are taken from the inner core of reference discharges 94042 and 95274 ($\Psi_N \approx 0.1 \textendash 0.2$, $\beta_N \approx 1.8$).}). However, when looking at the global range of DBS measurements and not at specific plasma conditions or machines, figure \ref{tokamaks_kperprho_betae} shows that DBS measurements taking place in low $\beta_e$ plasma regimes ($\beta_e \approx 10^{-5} \textendash 10^{-4}$) will likely be sensitive to ion-scale wavenumbers $k_\perp\rho_s < 1$ (at the pedestal, high-field machines or low performance discharges), while high $\beta_e$ scenarios ($\beta_e \approx 10^{-2} \textendash 10^{-1}$ as in high performance regimes or spherical tokamaks) will likely be sensitive to intermediate to electron scales $k_\perp\rho_s \gtrsim 1$. This confirms that DBS measurements can in fact be sensitive to ion and electron-scale wavenumbers $k_\perp$, and motivates separately studying ion vs. electron scale turbulence, and their differing implications to radial correlation length measurements via RCDR. 


In this work we seek to answer two questions: the dependence of the radial correlation length on the measured $k_\perp$, and the potential diagnostic resolution effects on radial correlation length measurements via RCDR in the linear response regime \cite{gusakov_ppcf_2004, gusakov_eps_2011, gusakov_ppcf_2014}. The rest of this manuscript proceeds as follows: section \ref{modelling} describes the modelling approach employed. We use nonlinear gyrokinetic simulations and a synthetic model DBS developed for gyrokinetic simulations. In section \ref{escale_gyro} we study the synthetic and average radial correlation lengths $l_r$ from electron-scale turbulent fluctuations computed from a strongly-driven ETG regime in the NSTX core \cite{ruizruiz_pop_2015, ruizruiz_ppcf_2019, ruizruiz_ppcf_2020, ruizruiz_pop_2020, ren_nf_2020} using the GYRO code \cite{gyro, gyro_guide}. In section \ref{iscale_gs2} we carry out a parallel study of the synthetic and average radial correlation lengths $l_r$ from ion-scale turbulent fluctuations computed from a fully developed ITG-driven turbulence regime in the JET core \cite{christen_jpp_2021} using the GS2 code \cite{kotschenreuther_cpc_1995}. In section \ref{sec5} we discuss the implication of employing separable turbulence spectra to model the DBS response. We conclude in section \ref{conclusions}.

\section{Modelling approach}
\label{modelling}
 
In this section we describe the modelling approach developed to calculate synthetic RCDR correlation lengths from simulated turbulence output from nonlinear gyrokinetic simulations. We also present the description of a model DBS suitable for filtering the density fluctuation field from gyrokinetic simulations.  

\subsection{Gyrokinetic simulations and wavenumber definitions}

Owing to the disparate wavenumber range accessible by DBS measurements (figure \ref{tokamaks_kperprho_betae}), we analyze two different plasma turbulence conditions: a highly unstable ETG turbulence regime ($k_\perp\rho_s > 1$) in the outer core of a modest-beta NSTX NBI heated H-mode (section \ref{escale_gyro}, analyzed in \cite{ruizruiz_pop_2015, ruizruiz_ppcf_2019, ruizruiz_ppcf_2020, ren_nf_2020}), and a fully developed ITG turbulence regime ($k_\perp\rho_s < 1$) from the mid-core of a JET NBI heated L-mode plasma (section \ref{iscale_gs2}, \cite{christen_jpp_2021}). Seeking generality in the approach, we have implemented and deployed a model DBS valid for both nonlinear gyrokinetic simulations from the GYRO code \cite{gyro, gyro_guide} corresponding to the NSTX ETG regime, and GS2 \cite{kotschenreuther_cpc_1995} simulations for the JET ITG regime. 

\begin{figure}
	\begin{center}
		\includegraphics[height=8cm]{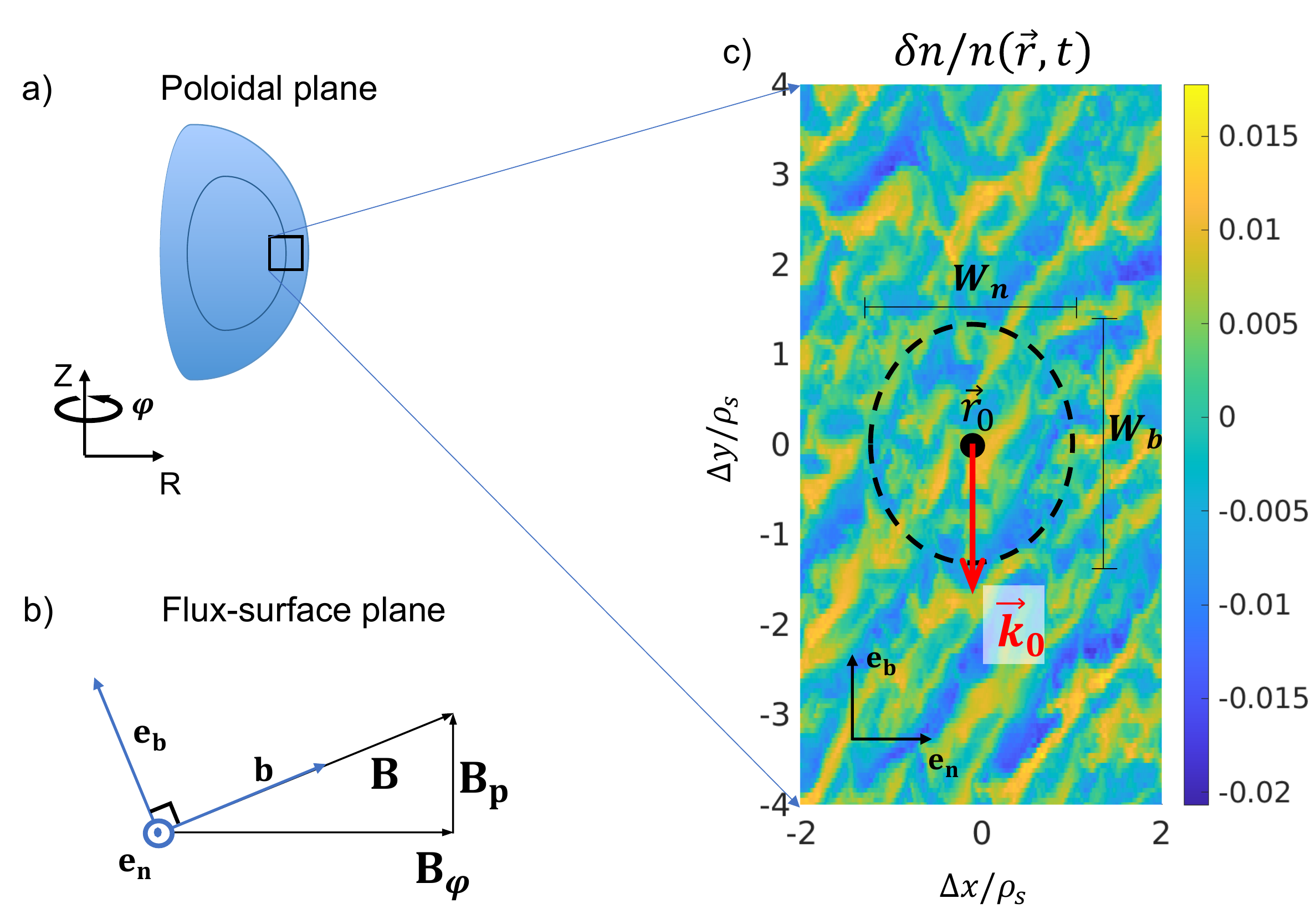}
	\end{center}
	\caption{Schematic representation of the model DBS. \textbf{a)} Poloidal plane cross-section. \textbf{b)} Flux-surface plane showing the orthonormal vectors $\mathbf{b}$ along $\mathbf{B}$, normal direction $\mathbf{e_n}$ and binormal $\mathbf{e_b}$. $\mathbf{B_\varphi}$ and $\mathbf{B_p}$ are the toroidal and poloidal magnetic field directions, schematically shown for reference. \textbf{c)} Density fluctuation field $\delta n$ plotted in terms of real-space perpendicular directions $\mathbf{e_n}$ and binormal $\mathbf{e_b}$. $W_n $ and $W_b$ are the filter characteristic length scales along $\mathbf{e_n}$ and $\mathbf{e_b}$ ($W_b \gg 2\pi/k_{b0}$). The lengths in figure \textbf{c)} are lengths as would be viewed by an external observer. Accordingly, $\rho_s$ is the ion-sound gyro-radius evaluated at the local toroidal field $B = |\mathbf{B_\varphi}|$.}   
	\label{dbs_turb_omp_wxwy}
\end{figure}

Local, nonlinear gyrokinetic simulations from GYRO and GS2 solve the nonlinear, gyrokinetic equation \cite{sugama_pop_1998} in the local limit ($\rho_s/a \rightarrow 0$), where $a$ is the plasma minor radius. While GYRO and GS2 use different definitions for the reference magnetic field ($B_\text{unit}$ \cite{gyro, gyro_guide} for GYRO and $B_\text{ref}$ for GS2, appendix \ref{app2}), in this manuscript the quoted $\rho_s$ is computed using local toroidal magnetic field magnitude $B = |\mathbf{B_\varphi}|$ (figure \ref{dbs_turb_omp_wxwy}.\textbf{b)}) and electron temperature $T_e$ values at the outboard midplane (unless otherwise specified). The electron density fluctuation field $\delta n$ calculated from nonlinear gyrokinetic simulations is used in this analysis to calculate the scattered amplitude and radial correlation lengths (the quantity $\delta n/n$ is normalized to $[\rho_s/a]$ in GS2, but not in GYRO). 

Both GYRO and GS2 use a field-aligned coordinate system in which the fluctuating field $\delta n$ exhibits a slowly varying dependence along the field-line, described by the coordinate $\theta$, and rapidly varying components perpendicular to the field-line, described by $(x,y)$. Here $x$ is the physical distance in the radial, normal direction to the flux surface, and $y$ is in the binormal direction (on the flux-surface and perpendicular to the magnetic field $\bold{B}$). In this work, the coordinates $(x, y)$ represent real lengths in the laboratory frame, and are transformed from the internal GYRO and GS2 coordinates perpendicular to $\mathbf{B}$ (resp. $(k_r, k_\theta)$ in GYRO, $(k_\text{x}, k_\text{y})$ in GS2, details in appendix \ref{app2}). In this manuscript we restrict ourselves to fluctuations taken at the outboard midplane ($\theta=0$), which is the location where traditional electrostatic drift-wave ballooning type instabilities tend to exhibit highest amplitude and where the vast majority of fluctuation measurements take place. Off-midplane locations might exhibit lower fluctuation amplitude for ballooning instabilities, and tilted turbulent eddies in the perpendicular plane for finite magnetic shear. These effects are not considered here for simplicity. For the purpose of analyzing the electron density field $\delta n$ that predominantly contributes to scattering, we decompose the perpendicular wave-vector into its normal and binormal components $\bold{k}_\perp = k_n \bold{e_n} + k_b \bold{e_b}$. Here we note the normal direction $\mathbf{e_n}$ as the perpendicular direction to the flux surface ($x$-direction), and the binormal direction $\mathbf{e_b}$ as the perpendicular direction to the magnetic field $\mathbf{B}$ and to $\mathbf{e_n}$ ($y$-direction), which lies in the flux surface (additional details in appendix \ref{app2}). 

Given the wavenumber definitions for $(k_n, k_b)$ [m$^{-1}$] and the physical space $(x,y)$ [m], we use the characteristic expansion of the density fluctuation field $\delta n(x,y,\theta,t)$ as treated in a local flux-tube (applicable to GYRO and GS2 in particular), given by

\begin{equation}
	\delta n(x,y,\theta=0,t) = \sum_{k_n, k_b} \delta \hat{n}(k_n, k_b, \theta=0, t) \text{e}^{ i k_n x + i k_b y}.
	\label{dn_fluxtube}
\end{equation}

Equation (\ref{dn_fluxtube}) states that $(k_n, k_b)$ and $(x,y)$ are simply Fourier conjugate variables (note the distinction between Euler number $\text{e}$ and the electric charge $e$ made throughout the manuscript). An example density fluctuation amplitude is given in figure \ref{dbs_turb_omp_wxwy}.\textbf{c)}.

The use of the $(x, y)$ coordinates and the normal $k_n$ and binormal $k_b$ wavenumber components of $\bold{k_\perp}$ has two main motivations in this work. First, $k_n$ and $k_b$ [m$^{-1}$] are routinely provided in experimental measurements via ray-tracing or beam-tracing methods (at the cutoff location), and correspond to real inverse lengths in physical space (the difference between the binormal $k_b$ and poloidal component of $\mathbf{k}$ might be small in conventional tokamak scenarios, but is substantial at the outboard midplane in spherical tokamak plasmas where the field-line pitch angle can reach $\sim 45^o$). Correspondingly, $(x, y)$ [m] correspond to real lengths in physical space. This should make the content of the present manuscript accessible for interpretation to the experimental plasma physicist and diagnostic expert. Secondly, the components $k_n$ and $k_b$ are independent of the internal wavenumber definitions in a specific gyrokinetic code. This makes $k_n$ and $k_b$ a general point of reference for synthetic diagnostics deployed in different gyrokinetic codes. In this manuscript, we assume a measurement at the outboard midplane, and we normalize $k_n$ and $k_b$ by the local value of the ion-sound gyro-radius, noted $\rho_s$ from now on, which is also independent of the reference magnetic field definitions used in gyrokinetic codes. For reference, $\rho_s$ takes the value $\rho_s \approx 0.7 $ [cm] in the NSTX ETG case, while $\rho_s \approx 0.33 $ [cm] in the JET ITG case analyzed.

\subsection{Model DBS}

The model DBS implemented for gyrokinetic simulations can be derived from a first principles, linear response, beam-tracing DBS model developed by Hall-Chen \emph{et al}. \cite{valerian_ppcf_2021}, and is in line with previous synthetic DBS developments implemented for GYRO \cite{holland_pop_2009, holland_nf_2012, hillesheim_rsi_2012}. In this work, our model assumes fixed wave polarization of the scattering process (O-mode or X-mode), beam localization modelled via a Gaussian filter (response localized to the cutoff), no scattering along the beam path ($k_n = 0$) and no mismatch angle between the binormal direction and the microwave beam wave-vector $\bold{k_i}$ \cite{rhodes_rsi_2006, hillesheim_nf_2015b, valerian_ppcf_2021}. The model is only applicable in the \emph{linear} response regime \cite{gusakov_ppcf_2004, gusakov_eps_2011, gusakov_ppcf_2014} of the Doppler reflectometer. The scattering amplitude is modelled by

\begin{equation}
	A_s(\bold{r_0}, \bold{k}_0, t) = \int{d^3\bold{r} \ \delta n(\bold{r}, t) U(\bold{r-r_0}) \text{e}^{-i\bold{k}_0 \cdot \bold{r}}} = \text{e}^{-i \mathbf{k_{0} \cdot r_0}} \sum_{\bold{k}_\perp} \delta \hat{n}(\bold{k}_\perp, \theta, t) W(\bold{k}_\perp - \bold{k}_{0}). 	
	\label{scatt_amp}
\end{equation}

The function $U$ is the filter in real space, while $W$ is the filter in $\bold{k}$-space. These are related to each other via a Fourier transform. The sum $\sum_{\bold{k_\perp}}$ is made over the perpendicular wavenumbers $k_n$ and $k_b$. The dot product $\bold{k_0} \cdot \bold{r} = k_{b0} y$ only exhibits a component in the binormal $y$-direction, emphasizing no finite $k_n$ corrections to the DBS signal are retained in the model. Here $k_{b0}$ is the \emph{measured} $k_\perp$ routinely quoted in experiments, modelling and simulation of Doppler Backscattering. We take the filter $U$ to have a Gaussian shape in the perpendicular directions $(x, y)$, while no structure is given along the field line (expressions are evaluated at the outboard midplane $\theta=0$). As a result the filter $W$ is also Gaussian in \textbf{k}-space, namely 

\begin{equation}
	\begin{aligned}
		& U(\bold{r-r_0}) = \frac{1}{\pi W_n W_b} \text{e}^{-(x-x_0)^2/W_n^2} \text{e}^{-(y-y_0)^2/W_b^2}, \qquad \text{and} \\
		& W(\bold{k}_\perp - \bold{k}_{0}) = \text{e}^{i(k_b-k_{b0})y_0 + ik_n x_0} \text{e}^{-(k_b-k_{b0})^2/\Delta k_b^2} \text{e}^{-k_n^2/\Delta k_n^2}, \\
	\end{aligned}
	\label{uw_filters}
\end{equation}

where $(x_0, y_0)$ is the location of scattering. The radial spot size $W_n$ in the normal direction $\bold{e_n}$ is generally known as the radial resolution of the diagnostic. Its Gaussian shape is the result of assuming that the microwave field pattern is accurately represented by a Gaussian beam \cite{valerian_ppcf_2021}. The Gaussian shape in $y$ can be interpreted as the localization of the DBS signal at the cutoff location. As we will see, the specific shape of the filter $U$ in the $y$-direction is unimportant in this work as long as $W_b \gg 2\pi/k_{b0}$, which is largely satisfied in all DBS experiments. The model also assumes a localized signal at the cutoff, that is $W_b$ is much smaller than a typical equilibrium length scale, which neglects scattering contributions along the beam path and is consistent with the selected wavenumber component $k_{n} = 0$. $\Delta k_b$ is known as the wavenumber resolution of the diagnostic (commonly noted $\Delta k_\perp$ in the literature), and is related to the filter spot-size $W_b$ at the cutoff via $\Delta k_b = 2/W_b$. The radial filter size $W_n$, or radial resolution, is related to the normal wavenumber resolution via $\Delta k_n = 2/W_n$. The radial resolution $W_n$ can be considered to include a combination of the physical beam width and curvature.


For the purpose of studying the radial correlation properties of the turbulence via RCDR, we define three different cross-correlation functions (CCF):

\begin{equation}
	\begin{aligned}
		& \text{CCF}^{\text{avg}}(\Delta x) = \frac{\langle \delta n(x+\Delta x, y) \delta n(x, y) \rangle_{x_0, y_0, T}}{ \langle |\delta n(x, y)|^2 \rangle_{x_0, y_0, T} } &= \qquad &\text{Average turbulence CCF,} \\
		& \text{CCF}^{k_b}(\Delta x) = \frac{\langle \delta \tilde{n}(x+\Delta x, k_b) \delta \tilde{n}(x, k_b)^* \rangle_{x_0, T}}{ \langle |\delta \hat{n}(x, k_b)|^2 \rangle_{x_0, T} } &= \qquad &\text{Scale-dependent CCF for $k_b$,} \\
		& \text{CCF}^{\text{syn}}(\Delta x) = \frac{\langle A_s(\bold{r_0} + \Delta x \ \bold{e_n}, k_{b0} \bold{e_b} ) A_s(\bold{r_0}, k_{b0} \bold{e_b})^* \rangle_{x_0, y_0, T}}{ \langle |A_s(\bold{r_0}, k_{b0} \bold{e_b} )|^2 \rangle_{x_0, y_0, T} } &= \qquad &\text{CCF of synthetic DBS,} \\
		\end{aligned}
	\label{ccf_defs}
\end{equation}

where the $\theta$ and $t$ coordinates are omitted from $\delta n$ for clarity (recall $\theta=0$). The field $\delta \tilde{n}$ is Fourier transformed in $y $ but not in $x$ (appendix \ref{app4}). The subscripts $\langle . \rangle_{x_0, ...}$ denote the ensemble averaging performed in $x_0, y_0$ within the perpendicular simulation domain, and in time $t$ during an interval $T$. Specifically, $\langle . \rangle_{x_0} = 1/L_{x}\int{dx_0(.)} $, and equivalently for $y_0$ and $t$ ($L_x$ is the simulation radial box-size). In what follows we define the radial correlation length $l_r$ by the $1/\text{e}$ value of the corresponding $\text{CCF}$. 

We call the quantity $\text{CCF}^{\text{avg}}$ the \emph{average} turbulence cross-correlation function, computed via radial correlation of the full density field $\delta n$ from the gyrokinetic code for different radial separations $\Delta x$. As a result, $\text{CCF}^{\text{avg}}$ knows about the full turbulence spectrum. It is independent of any experimental factor and is intrinsic to the specific turbulence under consideration. The quantity $\text{CCF}^{k_b} $ in equation (\ref{ccf_defs}) is the scale-dependent correlation function corresponding to a specific turbulence wavenumber $k_b$, and informs about the characteristic radial correlation of turbulent eddies with binormal wavenumber $k_{b}$. This quantity does not anymore know about the full turbulence spectrum, but is intrinsic to $k_b$. This means that each binormal wavenumber $k_b$ in the turbulence, or equivalently each binormal wavelength, has a specific radial structure associated with it, and it is given by $\text{CCF}^{k_b}$. The varying $\text{CCF}^{k_b}$ for different $k_b$ is an intrinsic characteristic of the turbulence, and it is \emph{not} related to a diagnostic effect. In this manuscript we argue that this is the quantity that can be measured via RCDR under the right conditions (when the resolution $W_n \ll l_r(k_{b0})$, where $l_r(k_{b0})$ is the radial correlation length of wavenumber $k_{b0}$). The last quantity in equation (\ref{ccf_defs}) is $\text{CCF}^{\text{syn}}$, the synthetic cross-correlation function, computed via radial correlation of the scattered amplitude $A_s$ (equation (\ref{scatt_amp})). Notably this cross-correlation function is directly dependent on specific experimental parameters, which in this model are $k_{b0}, W_n$ and $W_b$ (recall here $W_b \gg 2\pi/k_{b0}$). Part of the work that follows seeks to explain how the measured $k_{b0}$ and $W_n$ affect the synthetic cross-correlation function $\text{CCF}^{\text{syn}}$. This should prove highly useful for the interpretation of the radial correlation length measured via RCDR. 

 

\section{Interpreting $l_r$ from electron-scale turbulence}
\label{escale_gyro}

In this section we analyse the radial correlation properties of electron-scale turbulent fluctuations ($k_\perp\rho_s > 1$). A strongly-driven ETG regime is chosen to perform local, electron-scale nonlinear gyrokinetic simulations using the GYRO code. The simulations represent the outer-core ($r/a \approx 0.7$) of an NSTX NBI-heated H-mode plasma (NSTX discharge 141767, $P_{NBI} \approx 2$ MW). This radial location has been the object of thorough analysis of turbulent transport and high-k density fluctuations in previous works \cite{ruizruiz_pop_2015, ruizruiz_ppcf_2019, ruizruiz_ppcf_2020, ren_nf_2020, ruizruiz_pop_2020}, where it was referred to as the 'strong ETG' regime. Ion thermal transport was close to neoclassical levels as computed by NEO \cite{belli_ppcf_2008} and subdominant to electron thermal transport, consistent with ion-scale turbulence being stabilized by strong $E \times B$ shear. Importantly, electron-scale simulations were shown to reproduce the experimental electron heat flux estimate computed via TRANSP \cite{transp} as well as the wavenumber spectrum shape and fluctuation level ratio \cite{ruizruiz_ppcf_2019}, demonstrating the experimental relevance of the current ETG-driven turbulence simulations. 

Simulations were performed at $r/a \approx 0.7$ ($r$ is the minor radius coordinate normalized to the last closed flux surface value $a$), resolving three gyrokinetic species ($e^{-}$, D, C), including electron collisions ($\nu_{ei} \approx 1\ a/c_s$, but not ion collisions), background flow and flow shear (Mach number $M \approx 0.2$, $E\times B$ shearing rate $\gamma_E \approx 0.1 \textendash 0.2$ a/c$_s$, and parallel velocity gradient $\gamma_p \approx 1 $ a/c$_s$), and fully electromagnetic fluctuations including electrostatic potential $\delta \phi$, shear $\delta A_{||}$ and compressional $\delta B_{||}$ magnetic field perturbations. Linear background profiles were simulated employing nonperiodic boundary conditions in the radial direction with typical buffer widths $\Delta_b \approx 1\ \rho_{s,\text{unit}}$. The simulation domain is characterized by radial and poloidal box sizes of dimensions $(L_r, L_\theta) \approx (7, 7) \rho_{s,\text{unit}}$, equating to radial and poloidal wavenumber resolutions $k_r\rho_{s,\text{unit}} \ \epsilon \ [0.9, 62]$, $k_\theta\rho_{s,\text{unit}} \ \epsilon \ [0.9, 65]$. Here $k_r$ and $k_\theta$ are the radial and poloidal wavenumbers internally defined in GYRO, which are related to $k_n$ and $k_b$, and $\rho_{s,\text{unit}}$ uses the GYRO internal $B_\text{unit}$ definition (\cite{gyro, gyro_guide}, appendix \ref{app2}). Parallel resolution employed 14 poloidal grid points ($\times$ 2 signs of parallel velocity), 12 energies and 12 pitch-angles (6 passing + 6 trapped). This choice of numerical grids was made according to previous convergence and accuracy tests for the GYRO code simulating micro-instabilities in the core of NSTX \cite{guttenfelder_nf_2013}, and were also tested for convergence in the present conditions \cite{ruizruiz_ppcf_2019}. The flux surface geometry is described by a Miller equilibrium \cite{miller_pop_1998}. 


\subsection{Scale-dependent, physical correlation length.}
\label{escale_gyro1}

\begin{figure}
	\begin{center}
		\includegraphics[height=7cm]{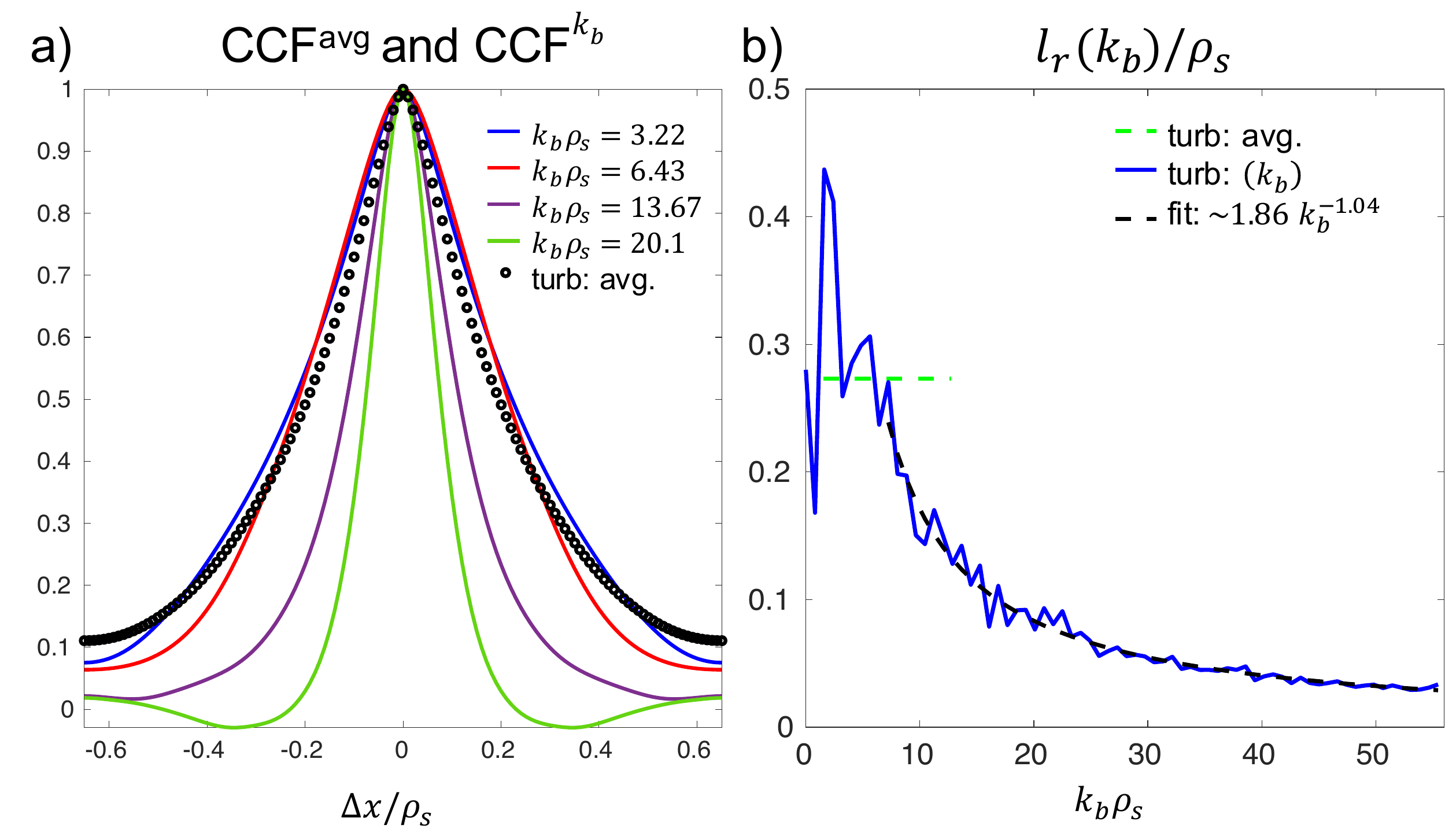}
	\end{center}
	\caption{\textbf{a)} Average turbulence cross-correlation function $\text{CCF}^{\text{avg}}$ from electron-scale turbulence (open black circles) and scale-dependent cross-correlation function for different $k_b$ $\text{CCF}^{k_b}$ in the electron-scale gyrokinetic simulation (definitions in eq. (\ref{ccf_defs}) and appendix \ref{app4}). \textbf{b)} Scale-dependent radial correlation length $l_r(k_b)$ corresponding to each binormal wavenumber in the simulation (blue curve) along with the average correlation length of the turbulence (green dashed line). This curve describes the scale by scale dependence of $l_r$ on $k_b$ (no synthetic DBS is applied). The ion sound gyro-radius $\rho_s \approx 0.7 $ [cm] is calculated with the local $B$ and $T_e$, and $\gamma_E=0$. }   
	\label{ccfx_lrky_escale_exb}
\end{figure}

Before applying the model DBS to gyrokinetic simulations, we discuss some important correlation properties in the electron-scale turbulence condition by analyzing the 'raw' electron density fluctuating field $\delta n$ output from the gyrokinetic simulation. We start by considering the case where the background $E\times B$ shear is 0 ($\gamma_E=0$, shown in figures \ref{ccfx_lrky_escale_exb}.\textbf{a)} and \textbf{b)}). Figure \ref{ccfx_lrky_escale_exb}.\textbf{a)} shows the average turbulence radial correlation function $\text{CCF}^\text{avg}$ (black empty circles) overlaid to a series of scale-dependent correlation functions $\text{CCF}^{k_b}$ corresponding to different wavenumbers $k_b\rho_s = 3.22, 6.43, 13.67, 20.1$, computed using the definitions from equation (\ref{ccf_defs}). The radial correlation function corresponding to each individual $k_b$ is different from the average turbulence correlation function, and it becomes wider for smaller $k_b$. This indicates that smaller turbulent $k_b$ (larger eddies in the $y$ direction) exhibit wider radial correlation function, that is larger radial correlation lengths. This can be more quantitatively assessed in figure \ref{ccfx_lrky_escale_exb}.\textbf{b)}. 

Figure \ref{ccfx_lrky_escale_exb}.\textbf{b)} shows the scale-dependent radial correlation length $l_r$ computed from each individual $k_b$ in the gyrokinetic simulation (blue curve) along with the average correlation length of the turbulence (green dashed line, both computed as the $1/\text{e}$ value of the corresponding $\text{CCF}$). The radial correlation length is largest for a finite $k_b\rho_s \approx 1.6$, and it becomes a decreasing function of $k_b$ for larger wavenumbers. A power law $\sim C  k_b^{-\alpha}$ is fitted to the $l_r$ curve for larger wavenumbers ($k_b\rho_s \gtrsim 7$), giving an exponent $\alpha \approx 1.04 $. This fit characterizes the scale-by-scale dependence of the radial correlation length corresponding to each binormal wavenumber $k_b$ (or binormal wavelength). Together, figures \ref{ccfx_lrky_escale_exb}.\textbf{a)} and \ref{ccfx_lrky_escale_exb}.\textbf{b)} show that the average turbulent correlation length is different from the correlation length corresponding to each $k_b$. For the purpose of understanding radial correlation lengths measured via RCDR, this discussion shows that it is of crucial importance to take into account the scale by scale dependence of the radial correlation length as only one wavenumber $k_b$ is selected by a Doppler backscattering system. 

The relationship $l_r \sim Ck_b^{-\alpha}$ with $\alpha \approx 1$ implies that the radial characteristic length $l_r$ and binormal  wavenumber $k_b$ preserve the same proportionality relation scale by scale, that is turbulent eddies in the power law fitting region (inertial range of the turbulent cascade) preserve their aspect ratio in $x$ and $y$ scale by scale (for $k_b\rho_s \gtrsim 7$). Larger $k_b$ indicates smaller eddies in the binormal direction, which are also proportionally smaller in the radial direction. The constant of proportionality $C$ can be interpreted as a rough measure of the eddy aspect ratio in the perpendicular $(x, y)$ plane. Heuristically one can define a 'binormal correlation length' $l_b$ as the $1/\text{e}$ value of the sinusoidal oscillation corresponding to $k_b$, that is $l_b/\rho_s = \arccos{(1/\text{e})}/k_b\rho_s \approx 1.19/k_b\rho_s$. A power law relationship of the form $l_r \sim Ck_b^{-\alpha}$ amounts to $l_r \sim C_r l_b^\alpha$ ($l_r/l_b \approx C_r = C\arccos(1/\text{e})^{-\alpha} \approx l_rk_b/1.19$). For $\alpha \approx 1$, the coefficient $C_r \approx l_r/l_b$ can be interpreted as the aspect ratio of the turbulent eddy between the radial and binormal directions. In this case we find $C \approx 1.86$, yielding $l_r/l_b \approx 1.54 \approx 3/2$. This suggests that ETG-driven eddies \emph{hypothetically} measured by DBS would exhibit an aspect ratio $l_r/l_b \approx 3/2$ in the $(x, y)$ plane (at the outboard midplane of this specific NSTX condition). In RCDR experiments, a good approximation to the eddy aspect ratio is $l_r/l_b \approx l_r k_b/\arccos{(1/ \text{e} )} \approx l_r k_b/1.19$, where $l_r$ and $k_b$ can be easily obtained via the standard experimental analysis and tools. 

 
The strong ETG drive of the present condition seems to contrast with the rather tight aspect ratio of $l_r/l_b \approx 3/2$. At first sight, this appears not entirely consistent with the radially elongated and poloidally thin nature of ETG 'streamers' \cite{drake_prl_1988, jenko_pop_2000, dorland_prl_2000}, and corresponds to lower aspect ratios than those inferred from ETG-driven turbulence in realistic gyrokinetic simulations of MAST $l_r/l_b \gtrsim 5$ \cite{roach_ppcf_2005, roach_ppcf_2009} and reported in NSTX $l_r/l_b \approx 4$ \cite{guttenfelder_pop_2011}, as well as for the cyclone base case $l_r/l_b \approx 2$ \cite{nevins_pop_2006} and early simulations for the W7-AS stellarator \cite{jenko_pop_2002_w7as} $l_r/l_b \approx 2$. In fact, the eddy aspect ratio in real, physical space $(x, y)$ can be strongly deformed when using the internally defined field-aligned coordinates of local flux-tube simulations. The difference between the physical space $(x,y)$ and the flux-tube coordinates is accentuated in strongly shaped flux surfaces characteristic of the spherical tokamak. Using the GYRO internally defined field-aligned coordinate system one finds an aspect ratio $\approx 9$ for the present condition, which would be characteristic of a radially elongated ETG 'streamer' (a factor of $\times 6$ difference). However, we emphasize that internally defined field-aligned coordinates do \emph{not} yield a real aspect-ratio in physical space, due to geometric effects such as flux-surface elongation, Shafranov's shift, etc. (details in table \ref{geo_coeffs_table}, appendix \ref{app5}). This realization further motivates the use of the normal and binormal $(x, y)$ coordinates and the $(k_n, k_b)$ wavenumber components for the purpose of quantitatively interpreting and projecting experimental measurements using gyrokinetic simulations.

Using the $l_r(k_b)$ relationship for calculating the aspect ratio of the turbulent eddies is accurate in the case where eddies are aligned with the radial direction $x$. For up-down symmetric flux-surface geometries, this can be the case at the outboard midplane $\theta=0$ when the background $E\times B$ flow shearing rate $\gamma_E $ is zero. It is well known that finite background $E \times B$ shear flow can stabilize (ion-scale) turbulence by tilting eddies in the perpendicular direction $(x,y)$ and inducing radial decorrelation \cite{biglari_physflub_1990, burrell_pop_1997}. Theoretically, a finite background $E \times B$ shear can break an underlying symmetry in the gyrokinetic equation, manifesting itself in an asymmetric turbulent wavenumber spectrum \cite{parra_pop_2011}, yielding an eddy tilt. Accordingly, BES measurements in the core of DIII-D \cite{shafer_pop_2012} have yielded tilt angles $\theta_\text{tilt} \approx 10^o$, while in the MAST core $\theta_\text{tilt} \gtrsim 45^o$ have been reported \cite{fox_ppcf_2017_symbreak, fox_ppcf_2017_2dbes} in the presence of finite $E\times B$ shear. Background $E\times B$ flow shear can also affect electron-scale turbulence in a similar manner, although the effect is expected to be reduced due to the smaller correlation times. Even so, eddy tilt can be non-negligible in ETG-dominated regimes where ion-scale turbulence is suppressed by strong $E\times B$ flow shear, such as in low-to-modest $\beta$ spherical tokamak scenarios. This has been suggested by nonlinear gyrokinetic simulations of ETG turbulence in the MAST and NSTX spherical tokamak core \cite{roach_ppcf_2009, guttenfelder_pop_2011}. An example of the ETG tilted eddies can be seen in figure \ref{dbs_turb_omp_wxwy}.

\begin{figure}
	\begin{center}
		\includegraphics[height=6.2cm]{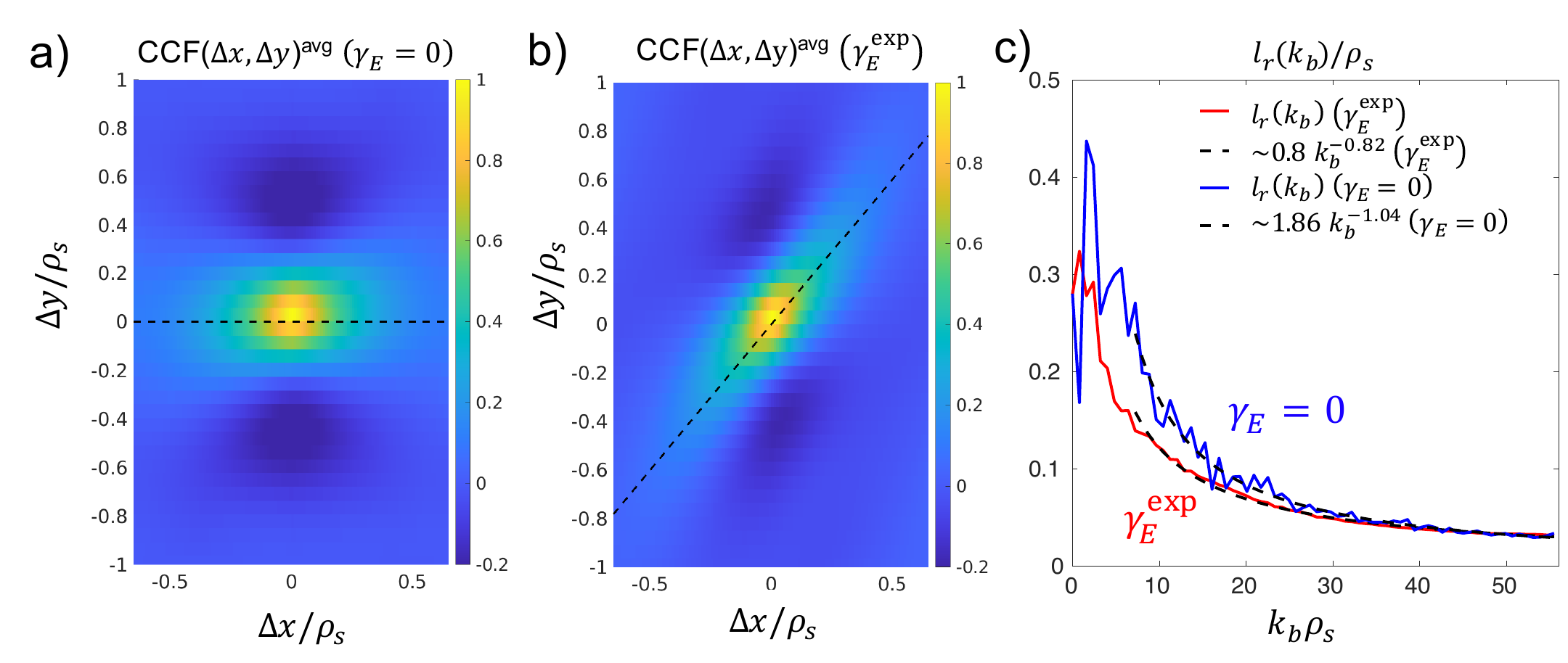}
	\end{center}
	\caption{\textbf{a)} 2D average turbulence correlation length corresponding to an ETG simulation with $\gamma_E=0$ (corresponding to figures \ref{ccfx_lrky_escale_exb}). $(x,y)$ are perpendicular to $\mathbf{B}$ and represent real lengths in the laboratory frame (they are not flux-tube coordinates; details in appendices \ref{app2}, \ref{app5}). \textbf{b)} 2D average turbulence correlation length for an ETG simulation run with the experimental value of $\gamma_E \approx 0.13 \ c_s/a$. Note the different tilt-angle in $a)$ vs. $b)$. \textbf{c)} Scale-dependent radial correlation length computed for each binormal wavenumber $k_b$ corresponding to simulations in $a)$ and $b)$. }   
	\label{ccfxy2d_exbexb0_lrky_escale}
\end{figure}

To understand the effect of the background $E \times B$ shearing rate $\gamma_E$ on the correlation length, an additional electron-scale simulation was performed, having identical input parameters and only differing by the finite, experimental value of the $E\times B$ shearing rate $\gamma_E$. Figure \ref{ccfxy2d_exbexb0_lrky_escale}.$\textbf{a)}$ shows the two-dimensional (2D) correlation function $\text{CCF}_{2D}^\text{avg}$ in $(x, y)$ computed from the density field $\delta n$ corresponding to the $\gamma_E=0$ case of figure \ref{ccfx_lrky_escale_exb}. $\text{CCF}_{2D}^\text{avg}(\Delta x, \Delta y)$ is the generalization of the one-dimensional $\text{CCF}^\text{avg}$ from equation (\ref{ccf_defs}) (details in appendix \ref{app4}). As expected, the 2D correlation function exhibits no tilt in the $(x,y)$ direction (noted by the black dashed line), showing eddies tend to be aligned along $x$ in the absence of $E\times B$ shear. The one-dimensional radial cross-correlation function $\text{CCF}^\text{avg}$ from figure \ref{ccfx_lrky_escale_exb}.\textbf{a)} (black empty circles) can be read in figure \ref{ccfxy2d_exbexb0_lrky_escale}.\textbf{a)} along the black dashed line ($\Delta y = 0$). Note that the 2D $\text{CCF}^\text{avg}_{2D}$ becomes negative in the binormal $y$ direction, which is indicative of the oscillatory nature of turbulence with finite $k_\text{b}$.

Figure \ref{ccfxy2d_exbexb0_lrky_escale}.$\textbf{b)}$ shows the 2D correlation function computed from the simulation using the experimental value of the $E \times B$ shearing rate ($\gamma_E \approx 0.13 \ c_s/a$). Notably, the 2D correlation function exhibits a finite tilt on the $(x,y)$ plane, attributed to the finite $\gamma_E$. The eddy tilt can be further quantified by calculating the preferred direction of inclination on the $(x, y)$ plane (details in appendix \ref{app5}). This yields a value of $\theta_{\text{tilt}} \approx 50^o$ in the present ETG condition. As in the discussion of the eddy aspect ratio, this tilt angle value is the physical angle, in the real space laboratory frame, made by turbulent eddies at the outboard midplane, which is again enabled by the use of the $(x, y)$ coordinates in this work. Using the GYRO field-aligned coordinate system at the outboard midplane would erroneously yield a tilt of $\theta_\text{tilt} \approx 11 ^o$. This emphasizes once more the need to map field-aligned, flux tube coordinates to real, physical space to directly compare to experimental measurements. Previously reported eddy tilt angle values in the core of MAST using a BES diagnostic \cite{fox_ppcf_2017_symbreak, fox_ppcf_2017_2dbes} seem to be in agreement with the value of $\theta_\text{tilt} \approx 50^o$ reported here. 

The $l_r(k_b)$ curves corresponding to \ref{ccfxy2d_exbexb0_lrky_escale}.$\textbf{a)}$-$\textbf{b)}$ are shown in figure \ref{ccfxy2d_exbexb0_lrky_escale}.$\textbf{c)}$. The blue curve corresponds to the $\gamma_E=0$ case, which exhibits a larger correlation length than the finite $\gamma_E$ case (red curve). A power law fit to the $l_r(k_b)$ curve with finite $\gamma_E$ shows that $l_r \sim 0.8 \ k_b^{-0.82}$, yielding an inferred eddy aspect ratio $C_r \approx 0.67$. As shown in appendix \ref{app5}, this value is wrong by factor $\times 2 \textendash 3$ with respect to the actual aspect ratio calculated along the tilted axes (black dashed lines in figures \ref{ccfxy2d_exbexb0_lrky_escale}.\textbf{a)}-\textbf{b)} and the corresponding perpendicular direction). In fact, the average aspect ratio of the turbulence in $(x,y)$, computed as the ratio between the $1/\text{e}$ length-scales along the tilted axes increases by $\sim 20\%$ from $\gamma_E=0$ to the finite $\gamma_E$ case. The disagreement with the $l_rk_b$ estimate for finite $\gamma_E$ (finite tilt) can be explained by the tilt of the eddies. A horizontal cut ($\Delta y=0$) of the $\text{CCF}^\text{avg}_{2D} (\Delta x, \Delta y)$ will yield a one dimensional correlation function that is affected by the oscillatory nature of the correlation function in the $y$-direction (with characteristic positive and negative values of the $\text{CCF}$). This artificially decreases the value of the turbulence average correlation length $l_r^\text{avg}$ as well as the scale-dependent correlation length $l_r(k_b)$, translating to an erroneous value of the aspect ratio in the estimate given by $\approx l_rk_b/1.19$. 

The main lesson to retain from this discussion is that the eddy aspect ratio could be measured in RCDR experiments from the quantity $l_rk_b/\arccos(1/\text{e}) \approx l_rk_b/1.19$, which will be accurate in conditions when the eddy tilt $\theta_\text{tilt} \approx 0$, but not in conditions of finite eddy tilt. In appendix \ref{app5} we provide additional details supporting and confirming these conclusions. Note that up-down asymmetric flux-surface geometries could introduce tilts in the $(x, y)$ plane similar to those shown here, as well as off-midplane locations due to the effect of magnetic shear $\hat{s}$, but these are not the object of the present study. In conditions of finite eddy tilt, a combination of the estimate proposed here to characterize the eddy aspect ratio $(l_r/l_b \sim l_rk_b/1.19)$ and that proposed by Pinzón \emph{et al}. to measure the eddy tilt angle \cite{pinzon_ppcf_2019, pinzon_nf_2019} could be used in future studies to simultaneously characterize the eddy aspect ratio and eddy tilt angle.

\subsection{Influence of the diagnostic resolution $W_n$.}

The previous section has highlighted the importance of taking into account the scale-by-scale variation of the radial correlation length of the turbulent eddies for different wavenumbers $k_b$, encapsulated in the $l_r(k_b)$ relation from figure \ref{ccfx_lrky_escale_exb}.\textbf{b)}. Importantly, the analysis was independent of the synthetic DBS diagnostic, and was merely a discussion of the correlation properties of the turbulence. In this subsection, we turn to understanding how a DBS measurement can affect the interpretation of the measured radial correlation length.

\begin{figure}
	\begin{center}
		\includegraphics[height=8cm]{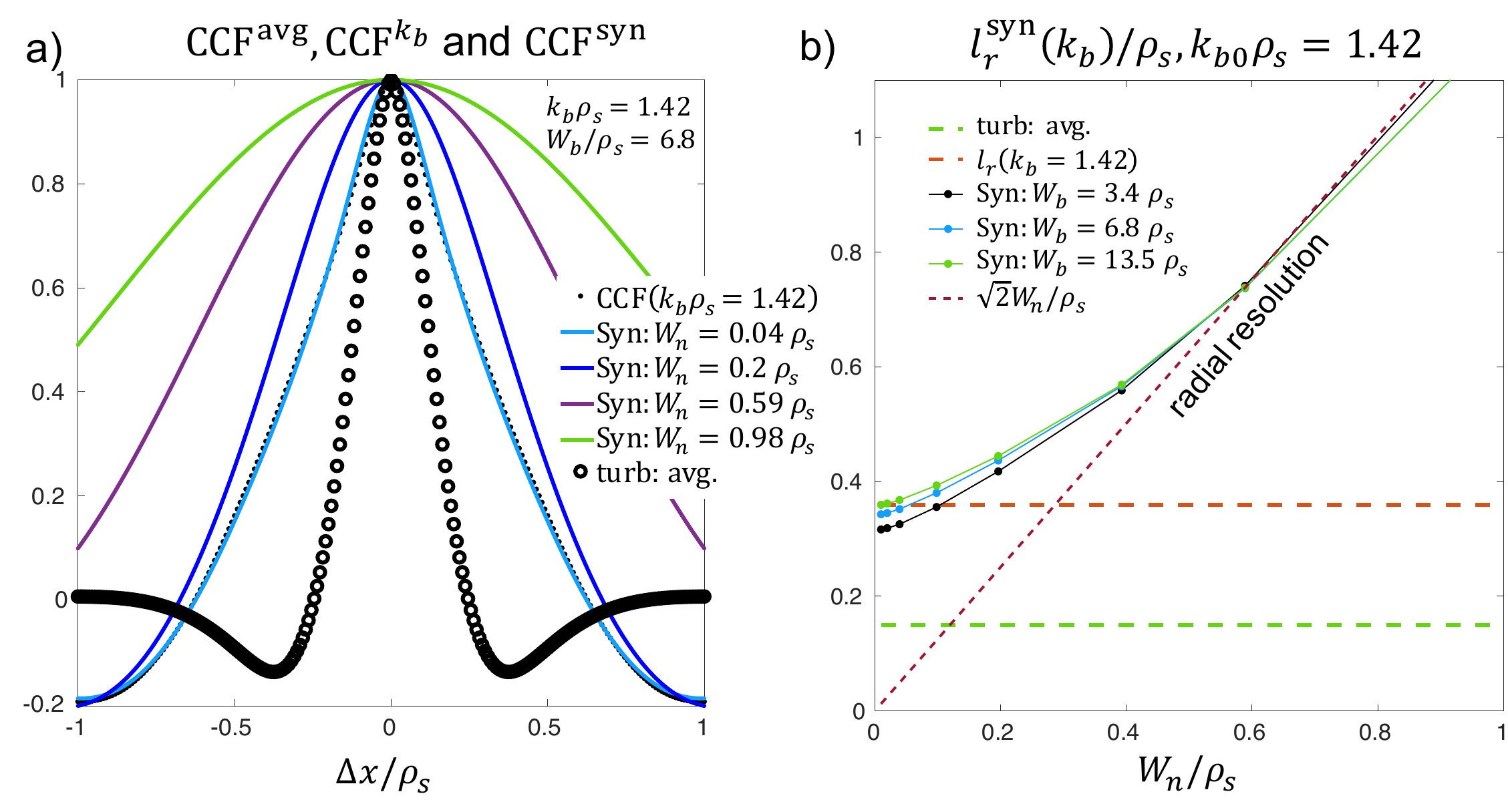}
	\end{center}
	\caption{\textbf{a)} Scale-dependent correlation function corresponding to electron-scale $k_b\rho_s = 1.42$ ($\text{CCF}^{k_b}$, black dots), synthetic correlation function for varying values of the radial resolution $W_n$ ($\text{CCF}^\text{syn}$, colored curves), and average turbulence correlation function ($\text{CCF}^\text{avg}$, black circles). \textbf{b)} Dependence of the synthetic radial correlation length computed as the $1/\text{e}$ value from the synthetic radial correlation function $\text{CCF}^\text{syn}$, for different values of $W_b$ ($W_b \gg 2\pi/k_b$). The radial correlation length from the selected wavenumber $k_{b}\rho_s = 1.42$ is only measurable when $l_r(k_b) \ll W_n$. $k_b, W_n$ and $W_b$ are all normalized by $\rho_s$.}   
	\label{ccfx_lrky_wrscan_ky2}
\end{figure}

We deploy the model DBS (equation (\ref{scatt_amp})) to compute synthetic radial correlation lengths corresponding to the strongly driven ETG condition. We seek to understand the effect of the filter's radial dimension $W_n$, and fix the binormal dimension $W_b = 6.8\ \rho_s$ (the specific value is unimportant as long as $W_b \gg 2\pi/k_{b0}$ but is in accordance with existing DBS systems \cite{hillesheim_nf_2015b}). For this study, we use a simulation with a larger spatial domain $(L_r, L_\theta) \approx (20, 20) \rho_{s,\text{unit}}$ to be able to scan realistic values of $W_n$. Figure \ref{ccfx_lrky_wrscan_ky2}.\textbf{a)} shows the cross-correlation functions $\text{CCF}^\text{avg}$, $\text{CCF}^{k_b}$ and $\text{CCF}^\text{syn}$ computed using the formulas from equation (\ref{ccf_defs}), corresponding to the specific binormal wavenumber $k_b\rho_s = 1.42$ which exhibited the largest correlation length in the simulation. The empty, black circles show the average turbulence correlation function. The black dots show the scale-dependent correlation function $\text{CCF}^{k_b}$ for $k_b\rho_s=1.42$. The colored curves show the synthetic cross-correlation functions $\text{CCF}^\text{syn}$ computed for the same selected wavenumber, for varying values of the radial resolution $W_n=0.04, 0.2, 0.59, 0.98\ \rho_s$ (and fixed binormal spot size $W_b = 6.8\ \rho_s$). Once more, the average turbulence correlation function is different from all the other correlation functions. In the case of small radial resolution $(W_n=0.04 \ \rho_s)$, the synthetic correlation function converges to the scale-dependent correlation function $\text{CCF}^{k_b}$ for $k_{b}\rho_s = 1.42$. This suggests that RCDR measurements are sensitive to the correlation function (and correlation length) of the specific, measured $k_{b0}$ by DBS (routinely denoted $k_\perp$ in the literature), but not to the average turbulence correlation function $\text{CCF}^\text{avg}$ as previously discussed. This is relevant for interpreting correlation lengths measured via RCDR. Additionally, the synthetic correlation functions for increasing values of $W_n$ are shown to become wider. Figure \ref{ccfx_lrky_wrscan_ky2}.\textbf{b)} shows the correlation lengths corresponding to the correlation functions in $\textbf{a)}$ as a function of $W_n$. Here, synthetic correlation lengths are also computed for different experimentally relevant values of the binormal spot size $W_b = 3.4, 6.8, 13.5 \ \rho_s$. Importantly, figure \ref{ccfx_lrky_wrscan_ky2}.\textbf{b)} shows that the correlation length for large $W_b \gg 2\pi/k_{b0}$ and small $W_n$ converges to the correlation length for $k_{b}\rho_s = 1.42$ (red dashed line). However, for increasing $W_n$, the synthetic correlation length is shown to asymptote linearly to the purple dashed line, which is the radial resolution $W_n$. This suggests that the diagnostic radial resolution $W_n$ can itself strongly affect the value of the 'measured' correlation length in DBS experiments if the width $W_n$ is large enough. The synthetic correlation length becomes dominated by the radial resolution when $W_n > l_r(k_{b0})$. 


This discussion has emphasized the significant impact that the diagnostic resolution $W_n$ can have on the measured DBS correlation length, where $W_n$ has been artificially varied. The value of $W_n$ remains an open question for Doppler backscattering. Previous work seems to suggest values on the order of the width of the Airy lobe $W_\text{Ai}$ or the beam width, whichever is larger. It is well known that standard normal incidence reflectometry experiments are characterized by a radial spot size $W_\text{Ai} \approx 0.5 L_\epsilon^{1/3} \lambda^{2/3}$ \cite{hutchinson_ppcf_1992, nazikian_pop_2001}, corresponding to the width of the Airy lobe in the vicinity of the cutoff. Here $L_\epsilon$ is the permittivity scale length, equal to the density gradient scale length $L_n$ in linear profile, slab geometry and O-mode polarization, while $\lambda $ is the vacuum wavelength of the incident microwave beam. In the case of Doppler reflectometry, it is possible that the Gaussian beam dimension near the cutoff might affect this reflectometry estimate $W_\text{Ai}$. The issue of the width of the beam near the reflection point is discussed analytically in \cite{maj_pop_2009, maj_ppcf_2010}, which showed that the exact solution of the Helmholtz equation for linear profile, slab geometry and O-mode polarization exhibits a $W_n$ at the cutoff location that becomes independent of the launching beam width for small incident launch angles, suggesting wave-effects can dominate, whereas beam width effects are more important for larger incidence angles. Those studies also showed that the cutoff width $W_n$ from beam tracing calculations becomes smaller than the $W_\text{Ai}$ value for small angles of incidence, but for larger incidence angles the beam tracing value depends on the initial beam width. Ray tracing, beam tracing and full-wave simulations carried out in \cite{conway_irw_2015, conway_irw_2019} for realistic toroidal geometries of AUG seemed to agree with \cite{maj_pop_2009, maj_ppcf_2010}, additionally suggesting that the Airy width $W_\text{Ai}$ should be considered as a minimum value of $W_n$ in experimental measurements. 

Motivated by these works, we calculate the value of the width of the Airy lobe $W_\text{Ai}$ for the present experimental condition in NSTX, yielding $W_\text{Ai} \approx 2 \textendash 4 \ \rho_s$. This value is notably larger than the largest correlation length in this strongly-driven ETG simulation condition (recall $l_r(k_{b0}\rho_s = 1.4) \approx 0.4\ \rho_s$). Importantly, this suggests that radial correlation length measurements carried out for electron-scale $k_\perp$ DBS measurements are likely to be dominated by the radial resolution $W_n$, that is $l_r^\text{syn} \rightarrow W_\text{Ai}$, since the condition $W_n \gg l_r(k_{b0})$ is easily fulfilled. 

These conclusions are reached using a highly simplified model for the DBS scattered amplitude (eqs. (\ref{scatt_amp}), (\ref{uw_filters})), based on a Gaussian filter in the normal $x$ and binormal $y$ directions. However, the specific shape of the filter envelope function (Gaussian, $\text{Ai}$ or other) should not affect the conclusion presented here, namely that measuring $l_r(k_{b0})$ using RCDR requires $l_r(k_{b0}) \ll W_n$, and that condition is likely not satisfied for electron-scale DBS measurements. In other words, we still expect the condition $W_n \gg l_r(k_{b0})$ to hold for non-Gaussian radial spot sizes, such as an Airy function. A resolution limit to the radial correlation measurement via RCDR was already pointed out in \cite{schirmer_ppcf_2007}, and recent full-wave simulations using a Gaussian turbulence spectrum \cite{prisiazhniuk_ppcf_2018} have also shown that the synthetic radial correlation length can be increased by diagnostic effects. A more detailed explanation is given in appendix \ref{app4}. Having studied the radial correlation from an electron-scale turbulence simulation, the following section addresses whether $W_n$ is also larger than $l_r(k_{b0})$ in conditions when a DBS reflectometer is sensitive to ion-scale wavenumbers.

\section{Interpreting $l_r$ for ion-scale turbulence}
\label{iscale_gs2}

In this section we apply the previous analysis to a fully-developped, ion-scale ITG-driven turbulence regime in the mid-core of a JET NBI-heated ($P_{NBI} = 17$ MW) L-mode plasma, which is currently the object of momentum transport studies \cite{christen_jpp_2021}. Local, flux-tube ion-scale nonlinear gyrokinetic simulations using the GS2 code \cite{kotschenreuther_cpc_1995} can simultaneously reproduce the ion thermal and momentum transport from the experimental estimates computed via TRANSP \cite{transp}, while the electron heat flux was under-predicted but subdominant to the ion heat flux. 

The gyrokinetic simulations were performed at $r/a \approx 0.5$, resolving two gyrokinetic species ($e^{-}$, D), including collisions ($\nu_{ee} \approx 0.02 \ a/v_{tr}, \nu_{ii} \approx 0.00026 \ a/v_{tr}$), background flow and flow shear ($M \approx -0.08, \gamma_E \approx -0.05 \ a/v_{tr}$, $\gamma_p \approx 0.2 \ a/v_{tr}$) and electrostatic fluctuations. The $r$ subscript in $v_{tr}$ refers to the reference species values, which was chosen to be deuterium in these simulations. Periodic boundary conditions were employed in the perpendicular dimensions ${x}$ and ${y}$, while the twist-and-shift boundary condition \cite{dimits_pre_1993, beer_pop_1995} was employed using a newly developed wavenumber-shifting method \cite{christen_jpp_2021}. A typical simulation domain for ion-scale turbulence was employed, with radial and poloidal box sizes on the order of $(L_\text{x}, L_\text{y}) \approx (77, 70) \rho_r$, equating to radial and poloidal wavenumber resolutions $k_\text{x}\rho_r \ \epsilon \ [0.08, 4.79]$, $k_\text{y}\rho_r \ \epsilon \ [0.09, 2.07]$. Here $k_\text{x}$ and $k_\text{y}$ are the radial and poloidal wavenumbers internally defined in GS2, which are related to $k_n$ and $k_b$, and $\rho_r$ is the GS2 internally defined gyroradius (appendix \ref{app2}). 

Parallel resolution employed 33 poloidal grid points in a $2\pi$ domain, 16 energies and 37 pitch-angles (20 passing, 17 trapped). This choice of numerical grids was made according to convergence and accuracy tests performed for the present condition \cite{christen_jpp_2021}. The flux surface geometry is described by a Miller equilibrium \cite{miller_pop_1998}. 

\subsection{Scale-dependent, physical correlation length.}
\label{iscale_gs2_2}

In the case of ITG-driven, ion-scale turbulent fluctuations ($k_\perp\rho_s \lesssim 1$), one naturally expects larger radial correlation lengths than for electron-turbulence. This begs the question whether the condition $l_r(k_{b0}) > W_n$ could be satisfied in these conditions. 

\begin{figure}
	\begin{center}
		\includegraphics[height=7cm]{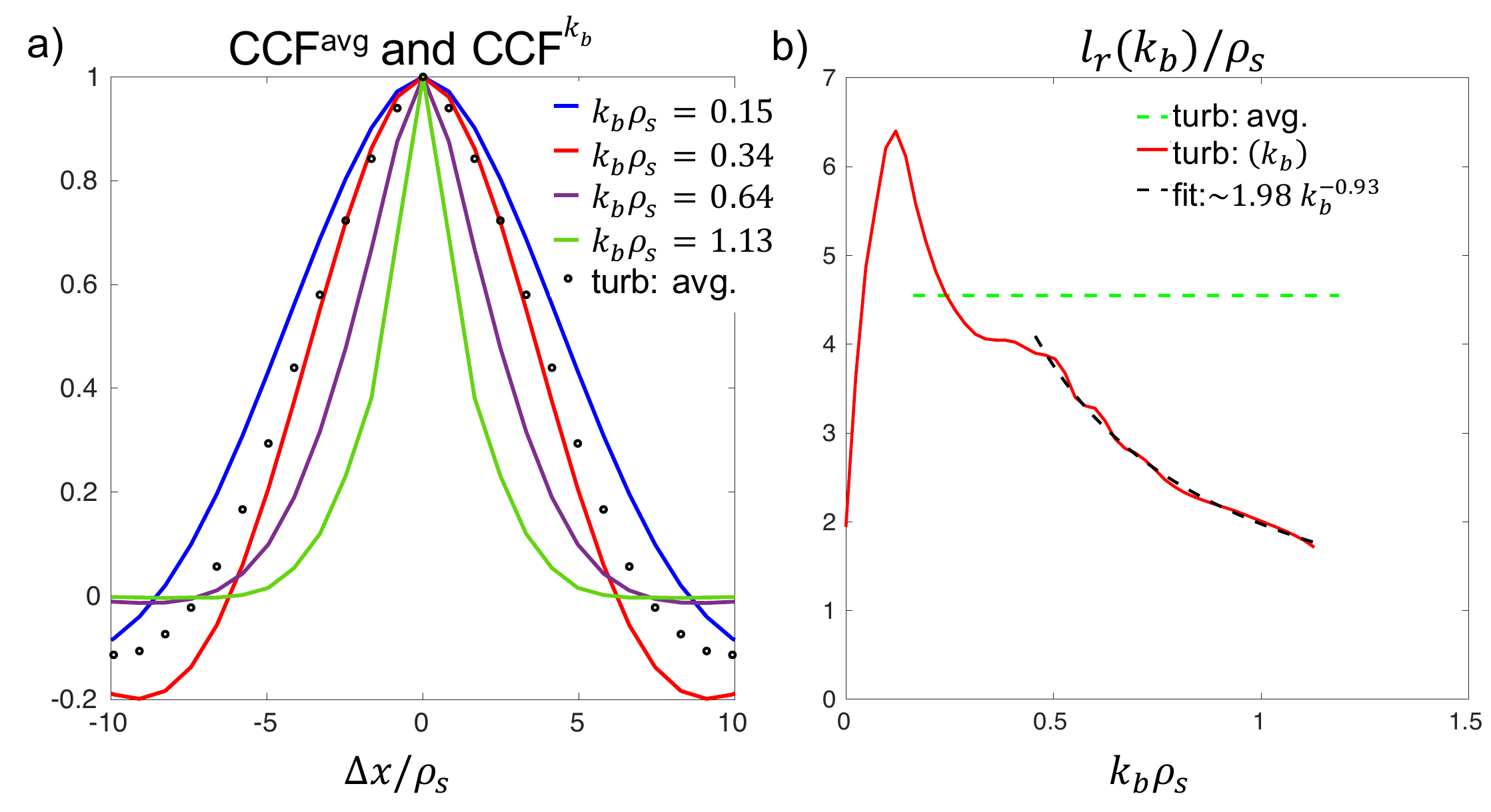}
	\end{center}
	\caption{Similar figure to \ref{ccfx_lrky_escale_exb} for ion-scale turbulence. \textbf{a)} Average turbulence cross-correlation function $\text{CCF}^{\text{avg}}$ (open black circles) and cross-correlation function for different $k_b$ $\text{CCF}^{k_b}$ in the ion-scale gyrokinetic simulation. \textbf{b)} Radial correlation length $l_r$ corresponding to each binormal wavenumber in the simulation. Note the radial correlation length corresponding to ion-scale $k_b$ (ITG-driven turbulence) exhibits a very similar dependence with $k_b$ as ETG-driven turbulence (fig. \ref{ccfx_lrky_escale_exb}). In this condition, $\rho_s \approx 0.33 $ [cm] and $\gamma_E=\gamma_E^\text{exp}$.}   
	\label{ccfx_lrky_iscale_exb}
\end{figure}

Similar features to the electron-scale turbulence can be observed in this ion-scale turbulence case. We consider the more realistic $\gamma_E = \gamma_E^\text{exp}$ case first. Figure \ref{ccfx_lrky_iscale_exb}.$\textbf{a)}$ shows the average turbulence correlation function $\text{CCF}^\text{avg}$ (black, empty circles) vs. the correlation function for a subset of binormal wavenumbers $k_b\rho_s = 0.15, 0.34, 0.64, 1.13$ (colored curves) corresponding to this ITG-driven turbulence condition. The average turbulence correlation function differs from the scale-dependent correlation function of each $k_b$, which becomes narrower for larger $k_b$, similar to what was observed analyzing electron-scale turbulence in figure \ref{ccfx_lrky_escale_exb}.$\textbf{a)}$. The $1/\text{e}$ correlation length for each $k_b$ is plotted in figure \ref{ccfx_lrky_iscale_exb}.\textbf{b)}, showing a peak correlation length for $k_b\rho_s \approx 0.14$, and decreasing for larger $k_b$. Fitting a power law to the correlation length for $k_b\rho_s \gtrsim 0.5$ yields $l_r \sim 1.98\ k_b^{-0.93}$. Inferring an eddy aspect ratio from such a measurement would yield $C_r \approx 1.68$, although we now know that the $l_rk_{b0}$ estimate is not accurate in the finite $\gamma_E$, finite $\theta_\text{tilt}$ case. A more detailed analysis of the 2D correlation function yields an average 'true' aspect ratio of $l_r/l_b \approx 3.1$ (along the tilted axes, appendix \ref{app5}). The scale-dependent correlation length in figure \ref{ccfx_lrky_iscale_exb}.\textbf{b)} exhibits a 'plateau' for $k_b\rho_s \approx 0.3 \textendash 0.5$. This could be a physical signature of the specific turbulent state, or it could be an artefact of the eddy tilt induced by the background $E\times B$ shear. For this reason we chose to fit the function for $k_b\rho_s \gtrsim 0.5$. 

\begin{figure}
	\begin{center}
		\includegraphics[height=6cm]{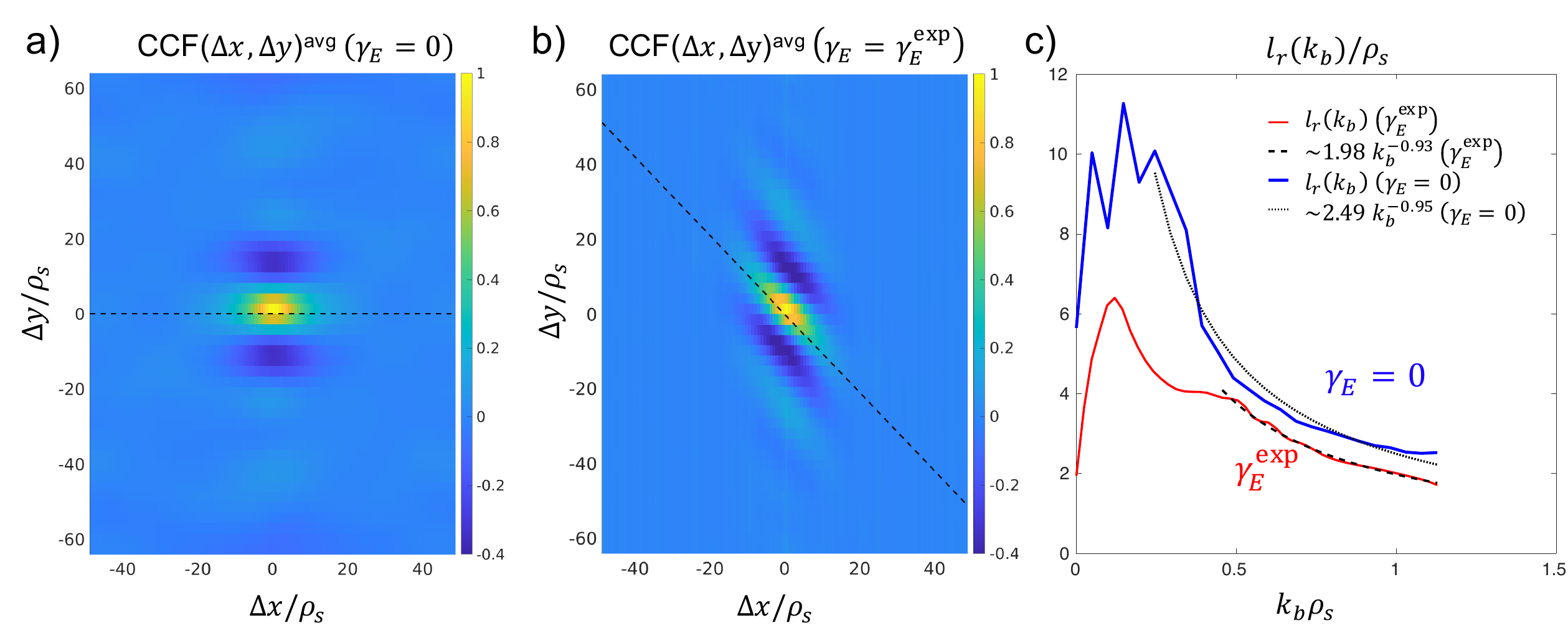}
	\end{center}
	\caption{Similar figure to \ref{ccfxy2d_exbexb0_lrky_escale} for ion-scale turbulence simulation in JET. \textbf{a)} 2D average turbulence correlation length run with the experimental value of $\gamma_E$. \textbf{b)} 2D average turbulence correlation length run with value of $\gamma_E = 0$. \textbf{c)} Radial correlation length computed for each binormal wavenumber $k_b$ corresponding to simulations in $a)$ and $b)$.}   
	\label{ccfxy2d_exbexb0_lrky_iscale}
\end{figure}

As in the electron-scale case, an additional nonlinear gyrokinetic simulation was performed with GS2, only differing from the first one by the zero $E\times B$ shearing rate value ($\gamma_E = 0$). The 2D correlation functions in the $(x, y)$ plane for both conditions are shown in figures \ref{ccfxy2d_exbexb0_lrky_iscale}.\textbf{a)} and \textbf{b)}, showing once more the tilting induced by the background $E\times B$ shear on the turbulence, while the simulation with zero $\gamma_E$ exhibits no appreciable tilt. Figure \ref{ccfxy2d_exbexb0_lrky_iscale}.\textbf{c)} shows the corresponding radial correlation length $l_r(k_b)$ from each binormal wavenumber $k_b$ in each simulation. The simulation without $E\times B$ shear (blue) exhibits a larger correlation length. The power law fit $l_r \sim 2.49\ k_b^{-0.95}$ suggests an eddy aspect ratio $l_r/l_b \approx 2.11$ when $\gamma_E = 0$. This aspect ratio is consistent with the average turbulence aspect ratio computed from the 2D correlation function, which yields $l_r/l_b \approx 1.88$ in this $\gamma_E=0$ case (appendix \ref{app5}). This value is in a similar range to previously measured eddy aspect ratios of ion-scale turbulent fluctuations using BES, showing $l_r/l_b \approx 2 \textendash 3$ in conventional tokamaks such as DIII-D \cite{mckee_rsi_2003, shafer_pop_2012} and TFTR \cite{fonck_prl_1993}. Interestingly, BES measurements in the MAST spherical tokamak \cite{ghim_prl_2013, field_ppcf_2014, fox_ppcf_2017_symbreak, fox_ppcf_2017_2dbes} have yielded inverted aspect ratios $l_r / l_b \approx 1/4 \textendash 1/2$, which could be related to the suppression of ion-scale turbulence in MAST, and are also likely affected by the strongly shaped nature of spherical tokamak flux surfaces (inducing radial compression and poloidal elongation at the outboard midplane). As in the previous section, the quoted aspect ratio $l_r/l_b \approx 1.88$ corresponds to the real, physical aspect ratio perpendicular to $\mathbf{B}$. Using the internal, field-aligned coordinates in GS2 for the present case would have yielded a value $l_r/l_b \approx 2.99$, a $\sim 60 \%$ difference with respect to the physical aspect ratio (for $\gamma_E=0$). The difference is lower than the factor of $\times 6$ error in the previous NSTX case due to the lower flux-surface shaping in this JET L-mode case. However, this suggests that quantitative comparisons of the eddy aspect ratio between experiments and simulation of conventional tokamaks require carefully considering physical $(x,y)$ coordinates (or equivalent), and not the internally defined field-aligned coordinates characteristic of gyrokinetic codes.

It is worth noting that the radial correlation length (along the physical $x$-direction) is reduced by the effect of flow shear (Figures \ref{ccfxy2d_exbexb0_lrky_escale}.\textbf{c)} and \ref{ccfxy2d_exbexb0_lrky_iscale}.\textbf{c)}), however the eddy aspect ratio $l_r/l_b$ (along the tilted axes) is enhanced by the effect of flow shear: $l_r/l_b \approx 1.88 $ for $\gamma_E=0$, while $l_r/l_b \approx 3.1$ for $\gamma_E^\text{exp}$. This is consistent with the induced eddy tilt and the effect of the binormal structure of the turbulence: the oscillatory nature of the correlation function in the $y$-direction has the effect of reducing the effective correlation length along the physical $x$-direction (Figures \ref{ccfxy2d_exbexb0_lrky_escale}.\textbf{c)} and \ref{ccfxy2d_exbexb0_lrky_iscale}.\textbf{c)}), however the actual aspect ratio along the tilted axes is increased for finite $\gamma_E$. This effect is qualitatively observable in Figures \ref{ccfxy2d_exbexb0_lrky_iscale}.\textbf{a)}-\textbf{b)}, as well as for the electron-scale case in Figures \ref{ccfxy2d_exbexb0_lrky_escale}.\textbf{a)}-\textbf{b)} from section \ref{escale_gyro}, where the aspect ratio along the tilted axes takes the values $l_r/l_b \approx 1.58$ for $\gamma_E=0$, while $l_r/l_b \approx 1.98$ for $\gamma_E^\text{exp}$.

Analysis shows that the eddy tilt for the experimental value of $E\times B$ shear in this JET mid-core plasma is $\theta_\text{tilt} \approx - (40^o \textendash 45^o)$. In the previous section we saw that the eddy tilt in physical space can be strongly affected by the strong shaping characteristic of a spherical tokamak (by a factor $\times 5 \textendash 6$ in the previous NSTX condition), which is not the case for the present JET condition. The tilt is also affected by the characteristic length and time scales of the particular eddy under consideration, which differs between electron and ion scale turbulence.

  
\subsection{Influence of the diagnostic resolution $W_n$.}

\begin{figure}
	\begin{center}
		\includegraphics[height=7cm]{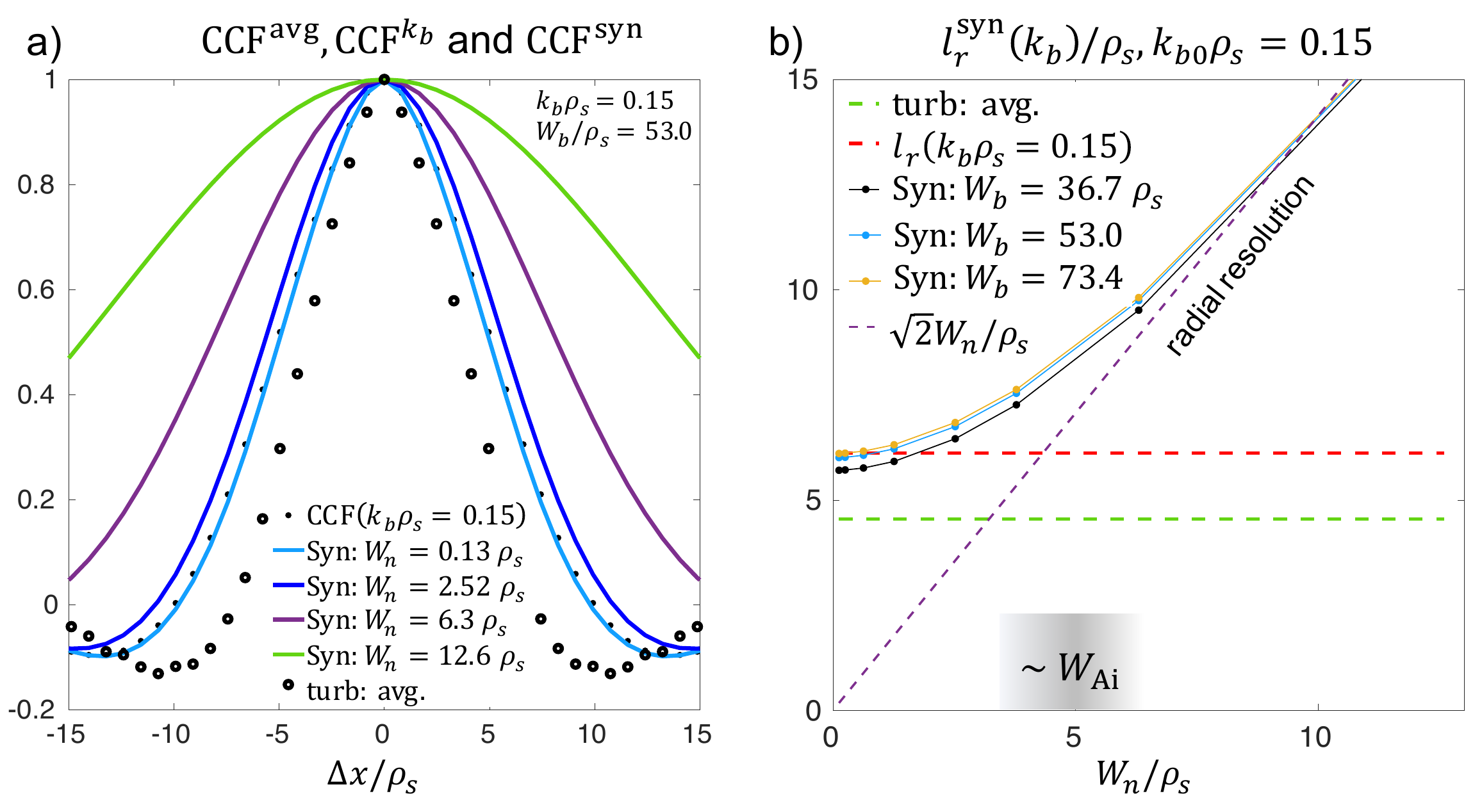}
	\end{center}
	\caption{Similar to figure \ref{ccfx_lrky_wrscan_ky2} for ion-scale turbulence simulation in the JET core. \textbf{a)} Scale-dependent correlation function corresponding to electron-scale $k_b\rho_s = 0.15$ ($\text{CCF}^{k_b}$, black dots), synthetic correlation function for varying values of the radial resolution $W_n$ ($\text{CCF}^\text{syn}$, colored curves), and average turbulence correlation function ($\text{CCF}^\text{avg}$, black circles). \textbf{b)} Dependence of the synthetic radial correlation length computed as the $1/\text{e}$ value from the synthetic radial correlation function $\text{CCF}^\text{syn}$, for different values of $W_b$ ($W_b \gg 2\pi/k_b$). The condition $l_r(k_b) \gtrsim W_n$ is more easily satisfied for ion scales than for electron scales, but suggests that the diagnostic radial resolution still likely affects $l_r^\text{syn}$. The estimated width of the first Airy lobe $W_\text{Ai}$ is indicated for reference.}   
	\label{ccfx_lrky_wrscan_ky02}
\end{figure}

In the case of ion-scale turbulence, the effect of finite diagnostic radial resolution $W_n$ is similar to electron-scale turbulence, meaning $l_r^\text{syn} \rightarrow W_n$ for $W_n \gg l_r(k_{b0})$. However, given ion-scale fluctuations exhibit larger correlation lengths, the condition $W_n < l_r(k_{b0})$ for a realistic $W_n \sim W_\text{Ai}$ is now more easily satisfied. Figure \ref{ccfx_lrky_wrscan_ky02}.\textbf{a)} shows the average turbulence correlation function (black circles), the scale-dependent correlation function for $k_b\rho_s = 0.15$ (black dots, which exhibits the largest correlation length in the simulation) and the synthetic correlation function for varying values of the radial resolution $W_n$, at fixed binormal spot size $W_b = 53.0 \ \rho_s$. As in figure \ref{ccfx_lrky_wrscan_ky2}.\textbf{a)}, the synthetic correlation function for small $W_n = 0.13 \ \rho_s$ converges to the scale-dependent correlation function for $k_b\rho_s = 0.15$, while for higher $W_n$, the synthetic correlation function gets wider and wider. Figure \ref{ccfx_lrky_wrscan_ky02}.\textbf{b)} shows the variation of the $1/\text{e}$ synthetic correlation length with $W_n$. Once again, the correlation length converges to the correlation length of $k_{b0}\rho_s = 0.15$ for all values of $W_b$ when $W_n \ll l_r(k_{b0})$ ($W_b \gg 2\pi/k_b$), and it asymptotes to the radial resolution $ W_n$ for $W_n \gg l_r(k_{b0})$. Note however that now the condition $W_n \approx l_r(k_{b0})$ happens for a larger correlation length $l_r(k_{b0}\rho_s = 0.15) \approx 5 \ \rho_s$. Evaluating the reflectometry estimate for the radial spot size $W_n \sim W_\text{Ai} \approx 0.5 L_n^{2/3} \lambda^{1/3}$ for the current conditions gives $W_\text{Ai} \approx 4 \textendash 6 \ \rho_s$. This suggests that experimentally relevant $W_n$ are likely to affect a measured RCDR correlation length, but in a less dramatic way than for electron-scale fluctuations. The effect is not completely negligible, as demonstrated by figure \ref{ccfx_lrky_wrscan_ky02}.\textbf{b)}. 

\section{Discussion of the use of separable turbulence spectra}
\label{sec5} 

In the previous sections, we have used a simple model DBS (equations (\ref{scatt_amp}) and (\ref{uw_filters})) and realistic turbulence spectra from nonlinear gyrokinetic simulations to understand the turbulence properties and the diagnostic effects relevant to RCDR measurements. Realistic turbulence spectra naturally yield a relationship $l_r(k_b) \sim C k_b^{-\alpha}$, illustrated in figures \ref{ccfx_lrky_escale_exb} and \ref{ccfx_lrky_iscale_exb}. In this section we discuss the consequences of modelling RCDR using turbulence spectra that are unable to capture the dependence of the radial correlation length on the binormal scale, which has become a common practice in the community \cite{gusakov_ppcf_2004, schirmer_ppcf_2007, gusakov_eps_2011, blanco_estrada_ppcf_2013, gusakov_ppcf_2014, gusakov_pop_2017, prisiazhniuk_ppcf_2018, pinzon_ppcf_2019, pinzon_nf_2019}.


The dependence of the radial correlation length on the binormal scale is encoded in the ensemble averaged turbulent wavenumber spectrum of the density fluctuations $\langle | \delta \hat{n}(k_n, k_b) |^2\rangle_T$. Importantly, the previous theoretical and numerical studies \cite{gusakov_ppcf_2014, blanco_estrada_ppcf_2013} (as well as many other works \cite{schirmer_ppcf_2007, gusakov_ppcf_2004, gusakov_pop_2017, prisiazhniuk_ppcf_2018, pinzon_ppcf_2019, pinzon_nf_2019}) used a separable turbulence wavenumber spectrum in the $k_n$ and $k_b$ components, written as a product of two independent functions $h_n$ and $h_b$ as $\langle |\delta \hat{n}(k_n, k_{b})|^2 \rangle_T = h_n(k_n)h_b(k_b)$. This results in $l_r(k_{b}) = l_r^\text{avg}$, which does not accurately represent magnetized plasma turbulence in the tokamak core. To understand the consequences of using a separable turbulence spectrum in $k_n$ and $k_b$, we use the characteristic expansion of the density fluctuation field $\delta n(x,y,\theta,t) = \sum_{k_n, k_b} \delta \hat{n}(k_n, k_b, \theta,t) \text{e}^{i k_n x + k_b y}$ as treated in a local flux-tube (equation (\ref{dn_fluxtube}), applicable to GYRO and GS2 in particular). We substitute this expression for $\delta n$ into the expression for the model scattered amplitude $A_s$ from equation (\ref{scatt_amp}), and use that to compute the synthetic $\text{CCF}^\text{syn}$ from equation (\ref{ccf_defs}), yielding equation (\ref{ccfsyn_app})

\begin{equation}
	\text{CCF}^\text{syn}(\Delta x) = \frac{\sum_{k_n, k_b} \text{e}^{-k_n^2W_n^2/2} \text{e}^{-(k_b-k_{b0})^2W_b^2/2} \langle |\delta \hat{n}(k_n, k_{b})|^2 \rangle_T \text{e}^{i k_n \Delta x} }{\sum_{k_n, k_b} \text{e}^{-k_n^2W_n^2/2} \text{e}^{-(k_b-k_{b0})^2W_b^2/2} \langle |\delta \hat{n}(k_n, k_{b})|^2 \rangle_T},
	\label{ccfsyn_app}
\end{equation}

which is valid for general $W_n$ and $W_b$. Next we set the experimentally relevant value of $W_b \gg 2\pi/k_{b0}$, we place ourselves in the best possible scenario where $W_n \ll l_r(k_{b0})$, and we neglect the effect of the varying $k_{b0}$ between the different DBS diagnostic channels. This yields

\begin{equation}
	\text{CCF}^\text{syn}(\Delta x) \rightarrow \frac{\sum_{k_n} \langle |\delta \hat{n}(k_n, k_{b0})|^2 \rangle_T \text{e}^{i k_n \Delta x} }{\sum_{k_n} \langle |\delta \hat{n}(k_n, k_{b0})|^2 \rangle_T} \equiv \text{CCF}^{k_{b0}} (\Delta x) \qquad \text{for $W_b \gg 2\pi/k_{b0}$ and $W_n \ll l_r(k_{b0})$} .
	\label{ccfsyn_app_ccfkb}
\end{equation}

Equation (\ref{ccfsyn_app_ccfkb}) analytically shows that the synthetic radial correlation function $\text{CCF}^\text{syn}$ will be a good representation of the scale-dependent $\text{CCF}^{k_b}$ under these assumptions. This was already observed in the previous sections, particularly in figures \ref{ccfx_lrky_wrscan_ky2} and  \ref{ccfx_lrky_wrscan_ky02}. Importantly, this correlation function $\text{CCF}^\text{syn}$ depends on the measured $k_{b}$ via the turbulence spectrum, which is not in general separable into its $k_n$ and $k_b$ coordinates (we have omitted the $\theta=0$ and time $t$ variables for clarity in equation (\ref{ccfsyn_app_ccfkb})). This expression contains the dependence of $l_r$ with $k_{b}$ encoded through the turbulence spectrum. Using a separable turbulence spectrum $\langle |\delta \hat{n}(k_n, k_{b})|^2 \rangle_T = h_n(k_n)h_b(k_b)$ (Gaussian, Lorentzian, top-hat, etc.) will yield a radial correlation length $l_r$ that becomes independent of $k_{b}$ (substitute $h_n(k_n)h_b(k_b)$ into equation (\ref{ccfsyn_app_ccfkb})). This means that a separable turbulence spectrum of the form $\langle |\delta \hat{n}(k_n, k_{b})|^2 \rangle_T = h_n(k_n)h_b(k_b)$ cannot contain the scale-by-scale variation of realistic turbulence spectra,  which can have important implications for interpreting and projecting RCDR measurements. The average correlation length $l_r^\text{avg}$ (computed from $\text{CCF}^\text{avg}$) and the scale dependent correlation length $l_r(k_{b})$ become identical for separable wavenumber spectra. 

A proposed analytical expression for non-separable, density fluctuation wavenumber power spectra inspired from the power law spectra characteristic of gyrokinetic simulations, and which contains the scale-by-scale dependence of the radial correlation length, is as follows

\begin{equation}
	\bigg\langle \Big| \frac{\delta \hat{n}}{n} (k_n, k_b) \Big|^2 \bigg\rangle_T = \frac{D}{1 + \big(\frac{|k_n|}{w_{k_n} }\big)^\gamma + \big(\frac{|k_b - k_{b*}|}{w_{k_b} } \big)^\beta }.
	\label{spec_analytic}
\end{equation}

In this expression, the spectral exponents $\gamma$ in $k_n$ and $\beta$ in $k_b$ are allowed to differ. $w_{k_n}$ and $w_{k_b}$ represent the spectral widths of the turbulence wavenumber spectrum, and are related to the average average turbulence radial and binormal correlation lengths by $w_{k_n} \propto 1/l_r$ and $w_{k_b} \propto 1/l_b$. The spectrum in $k_b$ is allowed to peak at a finite $k_b = k_{b*}$, consistent with the linear drive for typical micro-instabilities in the tokamak core peaking at finite values of $k_b$, and representing the injection driving scale of the turbulence (typically $k_{b*}\rho_s \approx 0.1 \textendash 0.6$ for ITG, $k_{b*}\rho_s \approx 2 \textendash 30$ for ETG, etc.). The spectrum from equation (\ref{spec_analytic}) is chosen to represent spectra in the absence of tilt angle $\theta_\text{tilt}$ ($\gamma_E=0$ and up-down symmetric flux surfaces at the outboard midplane $\theta=0$). This results in a symmetric spectrum in $k_n$, which is enforced by the absolute value $|k_n|$. Representing the effect of flow shear or non-symmetric spectra with a finite tilt angle could easily be obtained via a rotation of the $\mathbf{k}_\perp = (k_n, k_b)$ perpendicular wave-vector, which would result in a finite peak normal wavenumber $k_{n*}$. Equation (\ref{spec_analytic}) is normalized to $[\rho_s/a]$ in the case of GS2, but not in the case of GYRO.

\begin{figure}
	\begin{center}
		\includegraphics[height=7cm]{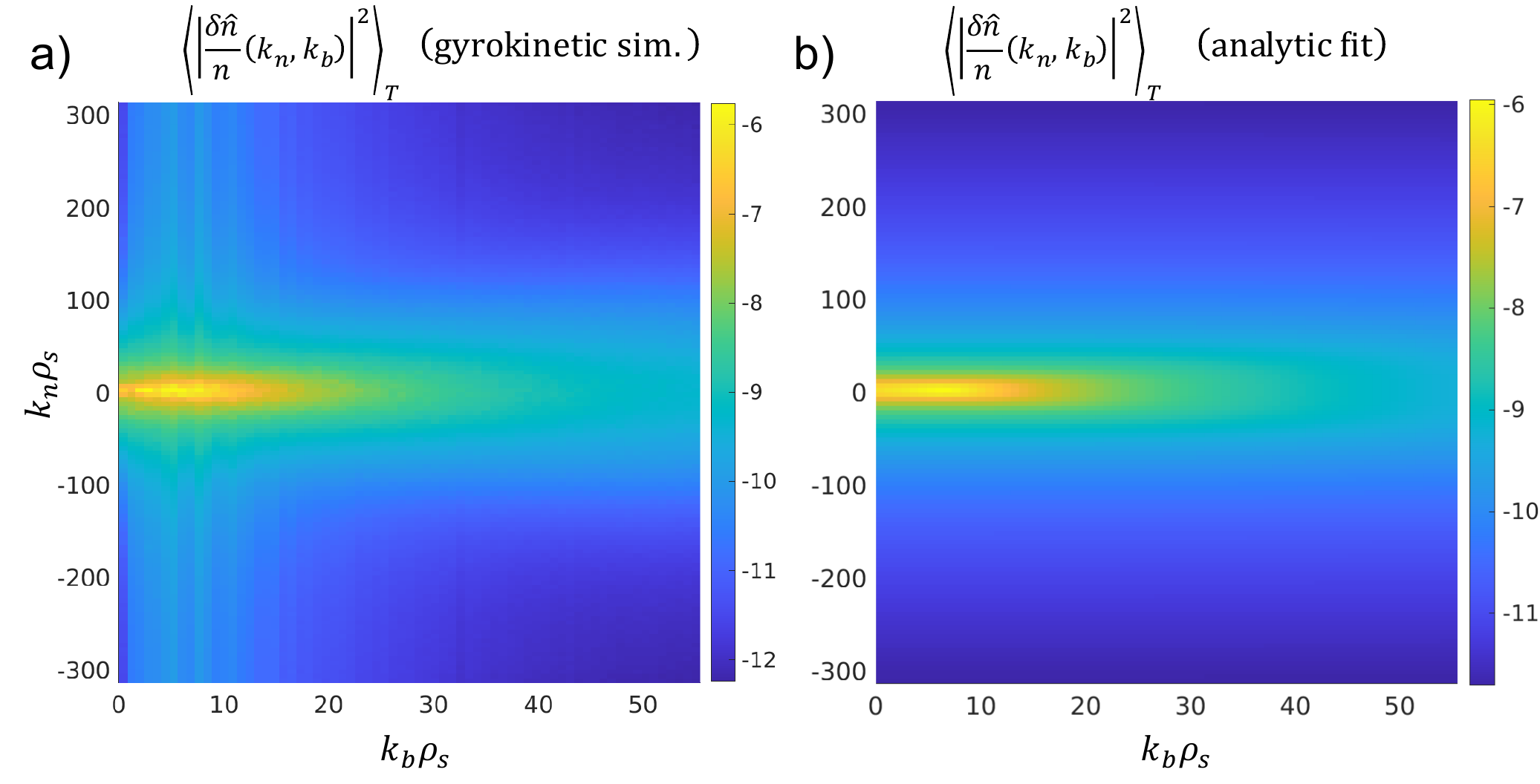}
	\end{center}
	\caption{\textbf{a)} Time-averaged density fluctuation power spectrum $\langle |\delta \hat{n}/n (k_n, k_b)|^2 \rangle_T$ from electron-scale nonlinear gyrokinetic simulation in section \ref{escale_gyro} ($\gamma_E=0$). \textbf{b)} Bidiminensional fit in $(k_n, k_b)$ to the simulated $\langle |\delta \hat{n}/n (k_n, k_b)|^2 \rangle_T$ using the analytical spectral spectral shape from equation (\ref{spec_analytic}).}   
	\label{dn2_knkb2d_gk_fit_comp}
\end{figure}

\begin{figure}
	\begin{center}
		\includegraphics[height=7.5cm]{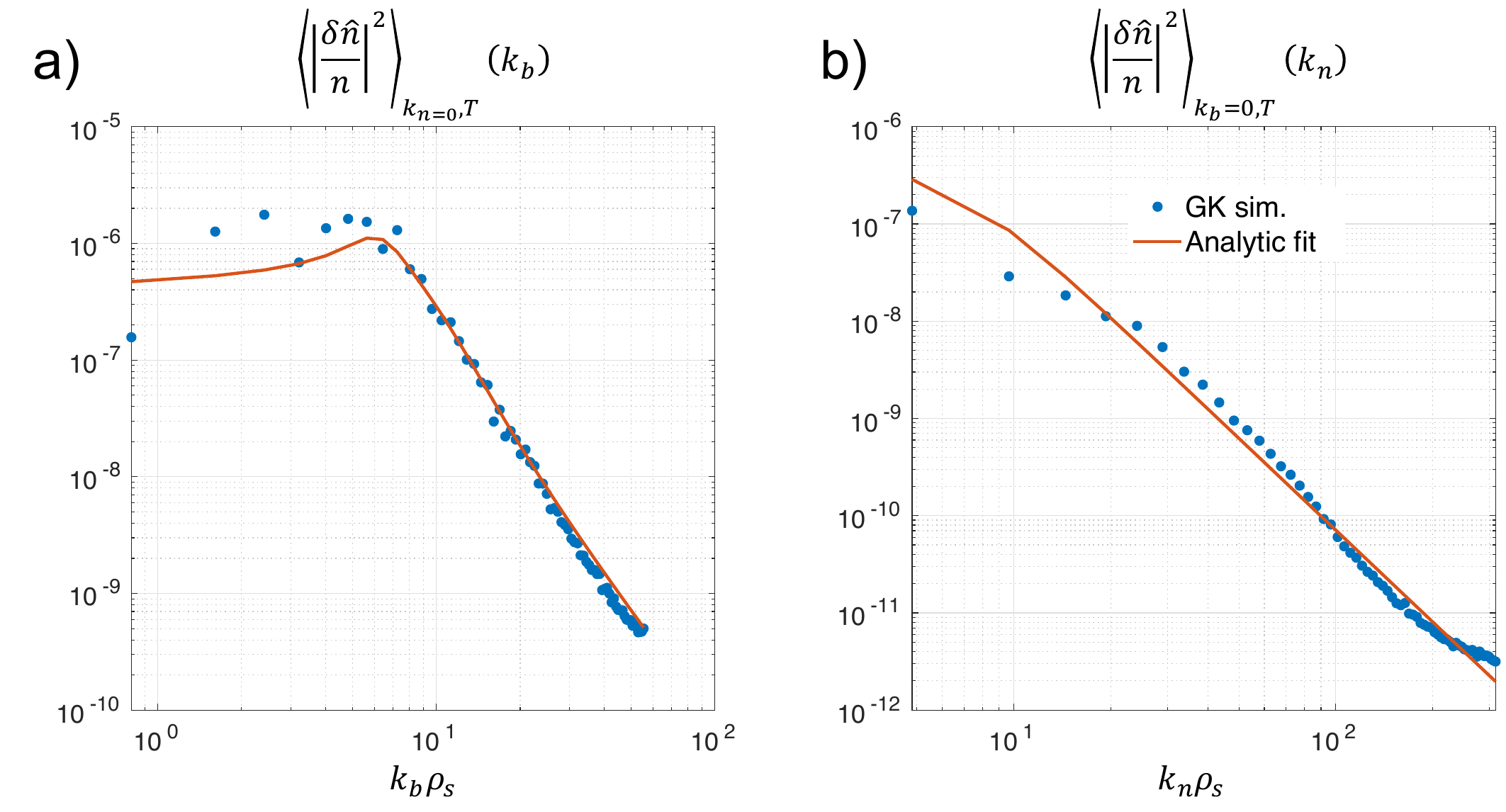}
	\end{center}
	\caption{\textbf{a)} Comparison between the time-averaged, gyrokinetic density fluctuation power spectrum (blue dots) and the analytic fit to the simulation spectrum using formula (\ref{spec_analytic}), plotted against binormal wavenumber $k_b\rho_s$. \textbf{b)} Gyrokinetic spectrum and analytic fit with respect to normal wavenumber $k_n\rho_s$. Both correspond to the electron-scale, ETG-driven condition in section \ref{escale_gyro}.}   
	\label{dn2_knkb_1d_gk_fit_comp}
\end{figure}

In order to illustrate the characteristics of the proposed analytic spectrum, the spectral shape from equation (\ref{spec_analytic}) is fit to the time-averaged spectrum computed from nonlinear gyrokinetic simulation corresponding to the strongly-driven ETG condition of section \ref{escale_gyro}. The 2D spectrum in $(k_n, k_b)$ from gyrokinetic simulation is compared with the analytical fit in figures \ref{dn2_knkb2d_gk_fit_comp}.\textbf{a)} and \ref{dn2_knkb2d_gk_fit_comp}.\textbf{b)}. The analytical fit is shown to qualitatively represent the dominant spectral features of the 2D spectrum in $(k_n, k_b)$. In order to quantitatively assess the fit, the simulated spectrum is compared to the analytical fit separately in the $k_n$ and $k_b$ directions in figures \ref{dn2_knkb_1d_gk_fit_comp}.$\textbf{a)}$ and $\textbf{b)}$. The spectral dependence with $k_b$ (for $k_n=0$) is shown to satisfactorily reproduce the main characteristics of the gyrokinetic spectrum, capturing a peak wavenumber $ k_{b}\rho_s \approx 5 \textendash 7$, and also the expected power law dependence of larger wavenumbers $k_b\rho_s \gtrsim 10$. The spectral dependence in $k_n$ ($k_b=0$) is monotonously decreasing, and the analytic expression is once more shown to recover the power law dependence in this dimension. The fitted analytical expression (\ref{spec_analytic}) for this specific strongly driven ETG simulation yields the following fit parameters: $\gamma \approx 2.88, \beta \approx 3.14, w_{k_n} \rho_s \approx 3.71, w_{k_b}\rho_s \approx 4.97, k_{b*}\rho_s \approx 5.60, D \approx 8.6 \ 10^{-7}$. 

To test how well the proposed spectral shape represents the dependence of the radial correlation length with $k_b$, we compute the scale-dependent correlation function $\text{CCF}^{k_b}$ directly from the spectrum (equations (\ref{ccfsyn_app_ccfkb}) and (\ref{ccf_kb_syn}) in appendix \ref{app4}), and then compute the scale-dependent correlation length $l_r(k_b)$ as in the rest of the manuscript. Luckily, the analytic expression of the spectrum allows an analytic expression of $l_r(k_b)$, which has the following functional dependence

\begin{equation}
	l_{r}(k_b) \sim \frac{\text{const.}}{w_{k_n} } \frac{1}{\Big[ 1 + \big(\frac{|k_b - k_{b*}|}{ w_{k_b} } \big)^\beta \Big]^{1/\gamma} }
	\label{lr_gk_analytic_comp}.
\end{equation}

\begin{figure}
	\begin{center}
		\includegraphics[height=7cm]{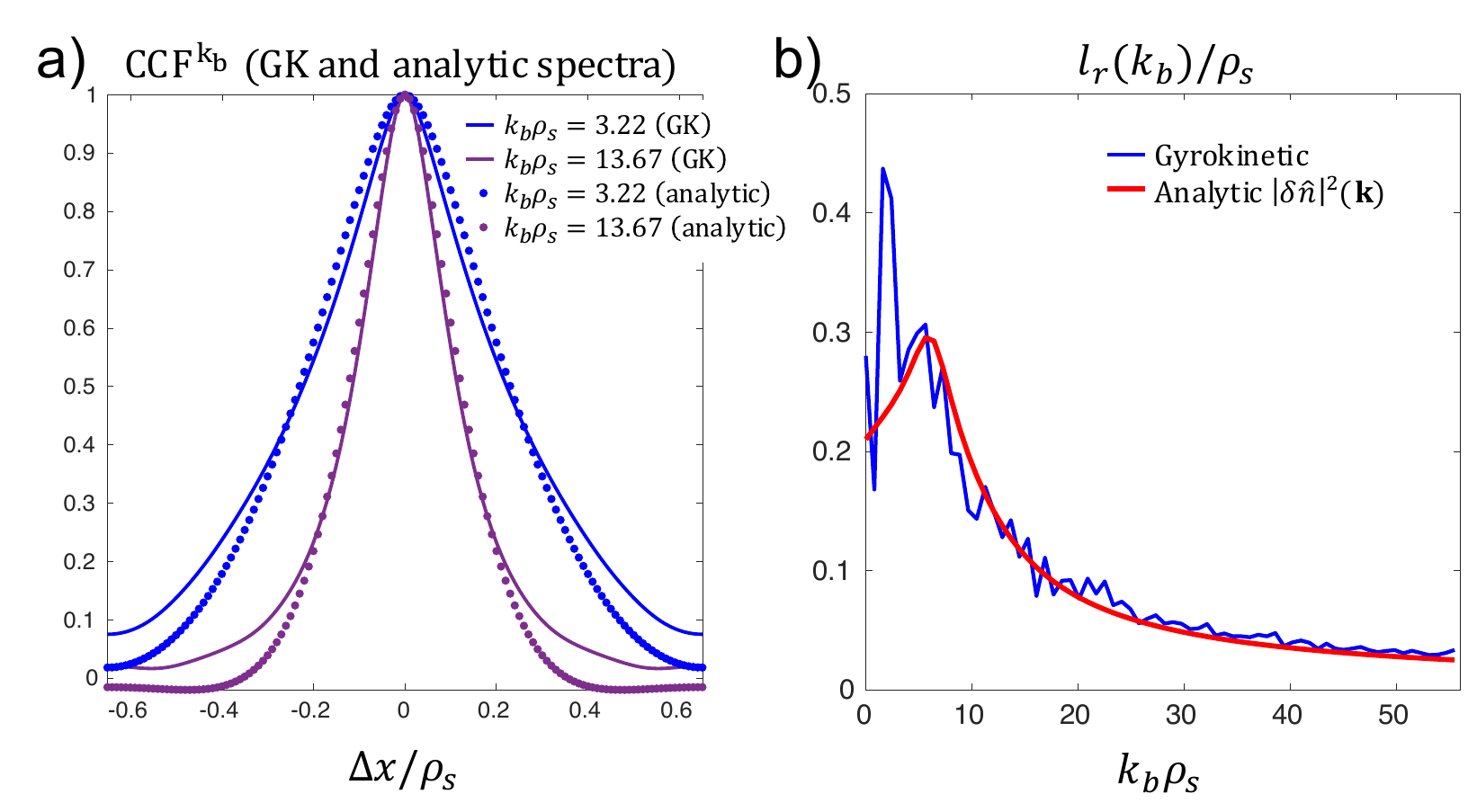}
	\end{center}
	\caption{\textbf{a)} Scale-dependent correlation function $\text{CCF}^{k_b}$ directly computed from gyrokinetic simulation (thick lines) from two wavenumbers $k_b\rho_s = 3.22, 13.67$, compared to the scale-dependent correlation function computed using the analytic fit to the wavenumber spectrum from equation (\ref{spec_analytic}) (colored dots). \textbf{b)} Scale-dependent correlation length for all wavenumbers $k_b$ in the simulation, using gyrokinetic spectrum (blue) and analytic fit from equation (\ref{lr_gk_analytic_comp}) (red).}   
	\label{ccfx_lrkb_gk_fit_comp}
\end{figure}

Similarly to the spectrum in equation (\ref{spec_analytic}), the analytic expression for $l_r(k_b)$ exhibits a peak correlation length at $k_b \approx k_{b*}\rho_s$. The correlation length is shown to be inversely proportional to the spectral width $w_{k_n}$, which follows from the Fourier transform relationship between $\langle |\delta \hat{n}(k_n, k_{b})|^2\rangle_T$ and $\text{CCF}^{k_{b}} (\Delta x)$ (equation (\ref{ccfsyn_app_ccfkb})). Equation (\ref{lr_gk_analytic_comp}) also exhibits a power law dependence for large $k_b$ as $l_r \sim k_b ^{-\beta/\gamma}$. Recalling that the turbulence eddies have a scale-invariant aspect-ratio when $l_r \sim k_b^{-1}$, this shows that the density fluctuation wavenumber power spectrum in equation (\ref{spec_analytic}) will result in scale-invariant eddy aspect-ratios when $\gamma = \beta$. In this specific condition, we find $\beta / \gamma \approx 1.1$, which is reasonably close to 1. The eddy aspect ratio here can be calculated as $l_r/l_b \propto w_{k_b}/w_{k_n} \approx 1.34$, which is reasonably close to the value $l_r/l_b \approx 1.54$ reported for this same condition in section \ref{escale_gyro}. In figure \ref{ccfx_lrkb_gk_fit_comp}.\textbf{a)} we compare the scale-dependent correlation function $\text{CCF}^{k_b}$ of the gyrokinetic (thick lines) and analytic spectra fit (colored dots), corresponding to two different wavenumbers $k_b\rho_s = 3.22, 13.67$. The scale-dependent correlation functions are shown to be satisfactorily reproduced by the analytic expression for the spectrum. The scale-dependent correlation length $l_r(k_b)$ computed for all $k_b$ is also shown to be well reproduced by the analytic spectrum in equation (\ref{spec_analytic}), as shown in figure \ref{ccfx_lrkb_gk_fit_comp}.\textbf{b)}. This discussion motivates a preferential use of realistic spectra, gyrokinetic or analytic expressions as in equation (\ref{spec_analytic}), above the use of separable spectra (Gaussian, Lorentzian, top-hat, etc.) in future modelling works of RCDR. For reference, similar analysis carried out for the ion-scale ITG-driven turbulence condition in section \ref{iscale_gs2} gives the following fit parameters: $\gamma \approx 3.90, \beta \approx 3.24, w_{ k_n} \rho_s \approx 0.17, w_{ k_b}\rho_s \approx 0.28, k_{b*}\rho_s \approx 0.24, D \approx 0.0395 $ (recall $\delta \hat{n}/n$ is normalized to $[\rho_s/a]$ in GS2). In this condition, we find the exponent $\beta/\gamma \approx 0.83$, while $l_r/l_b \approx w_{ k_b}/w_{ k_n} \approx 1.65$, which is consistent with the aspect ratio calculated from the 2D correlation function $l_r/l_b \approx 1.88$ in section \ref{iscale_gs2}. These parameters might be more relevant to predominantly ion-scale driven turbulent fluctuations (ITG/TEM) in the core of conventional tokamak scenarios, while the previous ETG case might be more relevant to predominantly ETG-driven and suppressed ion-scale turbulence, as in some plasma scenarios in the core of spherical tokamaks.


It is interesting to contrast the relationship $l_r \sim C \ k_b^{-\alpha}$ with other works, from experiments to modelling and theory. In this work, this is a mere consequence of the scale-by-scale dependence of the radial and binormal length scales of the turbulence, which is itself a consequence of the non-separable, power law character of the turbulent spectrum (eq. (\ref{spec_analytic})). The first RCDR radial correlation length measurements were performed by Schirmer \emph{et al}. for an ASDEX-U discharge in \cite{schirmer_ppcf_2007}. These pioneering measurements, coupled to full-wave simulations argued that the measured radial correlation length is independent of the measured $k_\perp$ (although only a restricted range of $k_\perp$ was analyzed, $k_\perp/k_i = 0.47, 0.69$). The same dependence is also observed in the non-linear response regime \cite{gusakov_ppcf_2005} as shown by full-wave simulations \cite{blanco_estrada_ppcf_2013} and experimental measurements \cite{fernandezmarina_nf_2014}, but was not the object of the present work. Analytical theory by Gusakov \emph{et al}. \cite{gusakov_eps_2011, gusakov_ppcf_2014} and full-wave simulations by Blanco \emph{et al}. \cite{blanco_estrada_ppcf_2013} suggested that the radial correlation length is a monotonically decreasing function of the antenna tilt angle when operating in the linear response regime. This is a similar observation to what we find. The incident angle of the launching microwave antenna with respect to the horizontal is directly related to the measured $k_b$ at the cutoff location ($k_b = 2 k_i \sin\theta_i$ in slab geometry, where $k_i$ is the microwave wavenumber and $\theta_i$ the vertical incident antenna angle). In \cite{gusakov_eps_2011, gusakov_ppcf_2014, blanco_estrada_ppcf_2013} it was argued that the poorly localized forward scattering contribution is suppressed once the antenna tilt angle surpasses a critical launch angle, which should result in correlation length measurements by RCDR similar to the true radial correlation length. Experiments by Fernández-Marina \emph{et al}. \cite{fernandezmarina_nf_2014} performed at the TJ-II stellarator seemed to confirm the theoretical and full-wave simulation results \cite{gusakov_eps_2011, gusakov_ppcf_2014, gusakov_pop_2017, blanco_estrada_ppcf_2013} in the linear response regime. Importantly, these works used a separable turbulence wavenumber spectrum in the $k_n$ and $k_b$ components (top-hat, Gaussian or Lorentzian in shape), which do not contain the scale-by-scale dependence characteristic of the turbulence, and result in $l_r(k_{b}) = l_r^\text{avg}$, which does not accurately represent magnetized plasma turbulence in the tokamak core. Therefore, one proposed explanation to the observed inverse proportionality of $l_r^\text{exp}$ with the incidence angle was the poorly localized forward scattering contribution of the microwave beam along the path, which can increase the measured radial correlation length at small antenna incidence angles, that is small $k_b$. It has been argued that this effect disappears for large enough antenna incidence angles (large $k_b$), therefore yielding an inverse proportionality relationship between $l_r$ and $k_b$. 

The proposed relationship $l_r(k_b) \sim k_b^{-\alpha}$ appears to be a universal, robust characteristic of the turbulence, as provided by nonlinear gyrokinetic simulation of ETG and ITG-driven turbulence in sections \ref{escale_gyro} and \ref{iscale_gs2}. This may explain, or contribute to the explanation of, the observed decrease of the measured radial correlation length with the $k_b$ selected by DBS \cite{fernandezmarina_nf_2014} (denoted $k_\perp$ in the literature). The fact that analytical work \cite{gusakov_eps_2011, gusakov_ppcf_2014} and full-wave simulations \cite{blanco_estrada_ppcf_2013} show that separable spectra (Gaussia, top-hat, Lorentzian, etc.) yield a measured $l_r$ dependent on $k_b$ suggests that a combination of two effects, the scale-by-scale variation of $l_r$ with $k_\perp$ and a diagnostic effect (forward scattering) may explain the experimental observations \cite{fernandezmarina_nf_2014}. For this reason, an analytic, realistic power law spectrum (equation (\ref{spec_analytic})) is proposed in this section, which could be used in future works to model the DBS response using full-wave or analytical calculations.


\section{Conclusions and discussion}
\label{conclusions}

In the present manuscript we have highlighted several important considerations to be taken into account when performing radial correlation measurements via the RCDR technique. The conclusions are reached by using a simplified model DBS for the scattering amplitude and deploying it on two specific tokamak core conditions: a highly unstable ETG regime in the outer core of an NSTX moderate-$\beta$ plasma, and a fully developed ITG turbulence regime in the mid core of a JET L-mode plasma. 

The two conditions are chosen as representative of electron and ion-scale turbulent fluctuations for potential RCDR measurements. The measured $k_\perp\rho_s$ from different DBS systems around the world \cite{gusakov_ppcf_2013, altukhov_ppcf_2016, altukhov_pop_2018a, altukhov_pop_2018b, fernandezmarina_nf_2014, happel_pop_2011, conway_ppcf_2004, schirmer_ppcf_2007, hillesheim_nf_2015a, hillesheim_nf_2015b, bourdelle_nf_2011, casati_prl_2009, meijere_ppcf_2014, windisch_rsi_2018, deboo_pop_2010, hillesheim_prl_2016, silva_ppcf_2018, silva_ppcf_2019, conway_irw_2011} (illustrated in figure \ref{tokamaks_kperprho_betae}) is shown to exhibit a close correlation with the local value of electron beta $\beta_e$. The correlation is shown to agree with a simple scaling $k_\perp\rho_s \sim ( m_i/m_e )^\frac{1}{2} \beta_e^\frac{1}{2}$, which can be explained by assuming O-mode polarization of the incoming microwaves and ignoring the important effect of the incident antenna tilt angle in the measured $k_\perp$. However, when scanning $\sim 4$ orders of magnitude in $\beta_e$, in very different plasma conditions (and different antenna incidence angles), the 'analytic' scaling seems to agree fairly well with the DBS measurements as shown in figure \ref{tokamaks_kperprho_betae}. This highlights that DBS can be sensitive to ion and electron scale density fluctuations: the former will likely take place in low-$\beta_e$ conditions (low performance plasmas, at the pedestal or in high-field devices), while the latter will preferentially take place for larger $\beta_e$ values (high performance plasmas, at the deep core or in spherical tokamaks). Perhaps the clearest example of this comes from the JET data points, where the low-$\beta_e$ cases correspond to edge measurements in low performance plasmas \cite{silva_ppcf_2018, silva_ppcf_2019}, while the high-$\beta_e$ cases correspond to deep core measurements in high $\beta_N$ discharges. This scaling is \emph{not} intended to be accurate for predicting the measured $k_\perp\rho_s$ in specific experiments, and does \emph{not} intend to be an equality between $k_\perp\rho_s$ and $\beta_e$, but merely seeks to illustrate how the measured $k_\perp\rho_s$ can be affected by large variations in $\beta_e$.  

This work has also highlighted that RCDR measurements are not sensitive to the average correlation length of the turbulence, but are sensitive to the scale-dependent correlation length $l_r(k_{b0})$ corresponding to the measured $k_{b0}$ (in conditions where $l_r(k_{b0}) \gg W_n$). Notably, this is a different quantity from the radial correlation length measured by other core fluctuation diagnostics such as BES \cite{fonck_prl_1993, mckee_nf_2001}, PCI \cite{coda_rsi_1995, coda_prl_2001, porkolab_ieee_2006}, CECE \cite{sattler_prl_1994, white_pop_2008} and conventional reflectometry \cite{nazikian_rsi_1995, nazikian_pop_2001}. This conclusion is reached by using a simple DBS model applied to realistic gyrokinetic simulations. We show that separable turbulence spectra (Gaussian, top hat, Lorentzian, etc., as in previous theoretical and modelling works \cite{schirmer_ppcf_2007, gusakov_ppcf_2004, blanco_estrada_ppcf_2013, gusakov_ppcf_2014, gusakov_pop_2017, prisiazhniuk_ppcf_2018, pinzon_ppcf_2019, pinzon_nf_2019}) do not contain the variation of $l_r(k_{b0})$ with $k_{b0}$. For those turbulence spectra, the average correlation length is equal to the scale-dependent correlation length by construction, $l_r^\text{avg} = l_r(k_{b0})$, which may be the reason why this has not been emphasized in the literature until now. This is an important realization for interpreting RCDR measurements and for potential future cross-diagnostic comparisons. An analytical expression for the density fluctuation turbulence wavenumber spectrum is proposed. The use of analytic, non-separable power law spectra to model magnetized plasma turbulence in tokamaks is not new (see for example \cite{staebler_prl_2013}), but it is new in the context of RCDR modelling. Fit to gyrokinetic simulation spectra, the proposed analytic spectrum is able to capture the dependence $l_r \propto k_b^{-\alpha} $ with $\alpha \approx 1$, exhibit a peak correlation length $l_r(k_b)$ for a finite $k_b$, and reasonably reproduce the eddy aspect ratio by the ratio of the spectral widths $l_r/l_b \approx w_{k_b}/w_{k_n} $. The functional shape for the spectrum (eq. (\ref{spec_analytic})) could be used in future studies to more accurately represent intrinsic turbulence characteristics, as well as to differentiate these from effects due to the diagnostic measurement (forward scattering).

The variation of the correlation length with the binormal wavenumber $l_r(k_{b}) \sim C k_b^{-\alpha}$ ($\sim C k_\perp^{-\alpha}$ in conventional notation, $\alpha \approx 1$) appears to be an intrinsic characteristic of the turbulence emanating from the non-separable, power law character of the turbulent spectrum (eq. (\ref{spec_analytic})). This is particularly true in the core of magnetic confinement fusion devices, as suggested by the ETG and ITG driven turbulence conditions analysed here. This relationship shows how the radial and binormal characteristic lengths of individual eddies vary scale by scale, and is independent of a particular Doppler reflectometer measurement. This work only discusses the scale-dependence of $l_r(k_b)$ with $k_b$ in the context of local, flux-tube turbulence simulations and in ITG/ETG turbulence driven regimes, and care should be taken to extrapolate these to other regimes. However, we deem it reasonable to think that turbulence naturally exhibits a scale-by-scale dependence of $l_r(k_{b})$ with $k_b$ in regimes where global-effects are important, as well as in conditions where other turbulence modes are predominantly driving turbulence. The particular spectral parameters and turbulence characteristics will likely be quantitatively different (eg. the spectral exponents, the spectral widths, the spectral peak, etc. from eq. (\ref{spec_analytic}). It appears reasonable to think that the non-separable, power law \emph{character} of the turbulent spectrum (and not its specific parameter values) carries over into the global and other turbulent regimes, however this issue will have to be addressed in detail in future works using global simulations and turbulence in the presence of other turbulent modes. The dependence of $l_r$ on $k_b$ found here agrees with previous experimental RCDR measurements carried out at the TJ-II stellarator \cite{fernandezmarina_nf_2014} when measuring in the linear response regime. Previous works pointed to a diagnostic effect, the poorly localized forward scattering contribution along the beam path \cite{gusakov_ppcf_2014, blanco_estrada_ppcf_2013} as an explanation. The forward scattering contribution to the DBS signal is absent in the present DBS model, and we cannot discriminate its importance compared to the variation of the turbulent characteristics with binormal wavenumber as an explanation to the observations in \cite{fernandezmarina_nf_2014}. The idea that the dependence of $l_r$ on $k_b$ may be physical was already mentioned in \cite{fernandezmarina_nf_2014}. Our results suggest that the variation of $l_r$ should at least be considered important in order to quantitatively interpret RCDR measurements.

The scale-dependent correlation length is only accessible when it satisfies $l_r(k_{b0}) \gg W_n$, that is for very small values of the radial resolution $W_n$. This applies to ion-scale RCDR measurements, and is consistent with the agreement between the experimental, simulated and synthetic correlation lengths obtained from low $\beta$, ion-scale measurements at the FT-2 tokamak \cite{gusakov_ppcf_2013, altukhov_ppcf_2016, krutkin_nf_2019}. For $W_n \gtrsim l_r(k_{b0})$, the measured $l_r^\text{exp}$ becomes dominated by the radial resolution $W_n$. This will likely happen for electron-scale measurements. These conclusions are reached using a local model for the scattered amplitude and a Gaussian filter shape (equation (\ref{uw_filters})), which can contain a combination of beam width and curvature affecting the radial resolution. This conclusion is independent of the specific shape of the Gaussian scattering filter, and applies to other non-Gaussian filters. Additional effects such as high amplitude turbulent fluctuations at the plasma edge are expected to produce microwave beam broadening, as is the case for electron cyclotron heating and current drive microwave beams \cite{kirov_ppcf_2002, tsironis_pop_2009, peysson_ppcf_2011, ram_pop_2013, sysoeva_nf_2015, ram_pop_2016, brookman_epj_2017, kohn_ppcf_2018, snicker_nf_2018, chellai_prl_2018, chellai_ppcf_2019, chellai_nf_2021}. Accurately taking these effects into account might be necessary for a quantitative interpretation of experiments, which will require more sophisticated beam-tracing, wave-kinetic or full-wave models.


Interestingly, in conditions when $l_r(k_{b0}) \gg W_n$ and the eddy tilt angle is negligible, this work has highlighted the possibility of measuring the eddy aspect ratio in the perpendicular dimensions to the background magnetic field $\bold{B}$. Since one expects $l_r \approx C k_b^{-\alpha}$ with $\alpha \approx 1$, the aspect ratio can be approximated by $l_r/l_b \approx C \arccos(1/\text{e})^{-\alpha} \approx l_rk_b/1.19$, suggesting that the eddy aspect ratio could directly be inferred from specific RCDR experiments by computing the radial correlation length by cross-correlation of the nearby DBS slave signals, and the measured $k_b$ from ray or beam tracing calculations. This estimate is shown to agree with the average eddy aspect ratio computed from the $2D$ correlation function $\text{CCF}^\text{avg}_{2D}(\Delta x, \Delta y)$ (details in appendix \ref{app5}). However, the eddy aspect ratio can only be accurately measured when there is negligible eddy tilt in the perpendicular plane to $\mathbf{B}$. As shown in sections \ref{escale_gyro} and \ref{iscale_gs2}, the background $E\times B$ shear can induce a strong eddy tilt, making it difficult to measure the aspect ratio by this method. Interestingly, both the ion and electron-scale conditions show that the intrinsic eddy aspect ratio along the tilted axes $l_r/l_b$ increases for finite tilt ($\gamma_E^\text{exp}$) with respect to the untilted case ($\gamma_E=0$), while the radial correlation length along the physical $x$-direction is reduced by $\gamma_E$ as expected. Although not shown here, additional effects such as off-midplane RCDR measurements (effect of magnetic shear $\hat{s}$) and up-down asymmetric geometries are expected to induce eddy tilt \cite{parra_pop_2011}, which should be taken into account when performing RCDR experiments. In conditions of finite eddy tilt, a combination of the estimate proposed here to characterize the eddy aspect ratio $(l_r/l_b \sim l_rk_b/1.19)$ and that proposed by Pinzón \emph{et al}. to measure the eddy tilt angle \cite{pinzon_ppcf_2019, pinzon_nf_2019} could be used in future studies to simultaneously characterize the eddy aspect ratio and eddy tilt angle.

In this manuscript we have strived to make the gyrokinetic simulation work easily interpretable to the experimental RCDR community by working in real-space coordinates $(x, y)$, representing real lengths $[m]$ in physical space. Obtaining quantitative comparisons between experimental turbulence measurements and direct numerical simulations is known to be a challenge for the validation of turbulent transport models \cite{terry_pop_2008, greenwald_pop_2010, holland_pop_2016}, and the gap often lies in non-trivial geometry considerations between experimental measurements (which often operate in conventional cylindrical coordinates) and gyrokinetic simulations (which operate in field-aligned geometry). The normal $x$ and binormal $y$ components can be easily mapped from the internal radial and binormal components from specific gyrokinetic codes (appendix \ref{app2}). Interestingly, inferring the eddy aspect ratio and the eddy tilt angle (although not extensively analyzed here) using the internal field-aligned coordinates in GYRO for the NSTX case yields error factors of $\times 5 \textendash 6$. This is due to the strongly shaped flux-surface geometry characteristic of spherical tokamaks, where parameters such as elongation and Shafranov's shift can strongly deform the field-aligned flux-tube at the outboard midplane. Although not as dramatically, inferring eddy tilt and aspect ratios for the JET L-mode case from the GS2 field-aligned coordinates can yield factors of $\approx 30 \textendash 60 \%$ error. This suggests that achieving quantitative comparisons of the eddy tilt and aspect ratio between experiments and simulation of conventional tokamaks requires carefully considering physical $(x,y)$ coordinates (or equivalent), and not the internally defined field-aligned coordinates characteristic of gyrokinetic codes. Additional details on this discussion are provided in appendix \ref{app5}. A quantitative interpretation of the measured radial correlation length, eddy aspect ratio and eddy tilt angle could be used to place strict constraints on turbulence and transport models. For example, it could prove useful to untangle the differing dependences that the turbulent heat fluxes and correlation length have exhibited with the isotope mass (isotope effect) in recent gyrokinetic simulations of ITER-extrapolated scenarios \cite{garcia_nf_2017}, which could have important consequences for the projections of future devices.

The highly simplified model DBS used in this work can be derived from a first principles, linear response DBS model recently developed for beam tracing \cite{valerian_ppcf_2021}. However, it contains important limitations that are known to affect the response of the backscattered signal, such as the problem of localization and scattering along the beam path contributing to poorly localized forward scattering \cite{gusakov_ppcf_2004, gusakov_ppcf_2014} (we assumed a localized signal response from the cutoff), the problem of wavenumber resolution \cite{lin_ppcf_2001, hirsch_ppcf_2004, conway_irw_2015, conway_irw_2019} (unimportant for our discussion as long as $W_b \gg 2\pi/k_{b0}$), the effect of the mismatch angle between the incident beam wave-vector and the binormal direction of scattering \cite{rhodes_rsi_2006, hillesheim_nf_2015b, valerian_ppcf_2021} (we assumed $\theta_\text{mis} = 0$), and multiple forward scattering yielding a nonlinear signal response to the density fluctuation amplitude (originating at high density fluctuation amplitudes \cite{gusakov_ppcf_2005, pinzon_ppcf_2017, krutkin_ppcf_2019, blanco_ppcf_2008, blanco_estrada_ppcf_2013, lechte_ieee_2009, stroth_nf_2015, happel_ppcf_2017, lechte_ppcf_2017, lechte_pst_2020}; we assumed a purely linear response, eq. (\ref{scatt_amp})), among other things. To understand these phenomena from a fundamental perspective, more sophisticated theoretical scattering models are necessary. For modelling purposes, full-wave simulations possibly yield the most accurate representation of these phenomena for direct comparisons with experimental measurements \cite{blanco_ppcf_2008, blanco_estrada_ppcf_2013, lechte_ieee_2009, stroth_nf_2015, happel_ppcf_2017, lechte_ppcf_2017, lechte_pst_2020}. For the purposes of transport model validation, if the question asked is whether gyrokinetic simulations can accurately represent all experimental turbulence phenomena possibly measured by Doppler backscattering (turbulent fluctuation wavenumber spectrum, radial correlation lengths, tilt angle, etc.), then arguably the most sophisticated models (full-wave) should be considered to try to \emph{match} experimental measurements. A particular concern when using full-physics models arises when agreement between experiments and simulations is not achieved. Is it to do with the gyrokinetic simulations, which might not accurately describe the turbulent fluctuations, or is it due to some important physical mechanism contained in a highly sophisticated model of Doppler backscattering (such as full-wave simulations)? In those circumstances, it is often difficult to reach a conclusion as to what the source of disagreement might be, and it might be instructive to break the problem into small pieces. In this manuscript we have opted to apply a bottom-up approach by considering a simple model DBS, which does \emph{not} contain all possible physical phenomena affecting the DBS signal, but which is simple enough to understand its direct implications on DBS measurements. Arguably, it is the simplest model for the response to a single $k_b$ originating from a finite radial region $W_n$. This has yielded insight into the scale-dependent correlation length that the DBS signal is sensitive to, as well as the effect of a finite $W_n$ on the measured correlation length. More sophisticated models will contain these, and also additional effects. In future work we will consider extending the present DBS model, in line with recent work \cite{gusakov_ppcf_2014, gusakov_pop_2017, gusakov_ppcf_2005, krutkin_ppcf_2019, valerian_ppcf_2021}, in order to account for some of the cited additional phenomena, and their implications to RCDR. These will also be useful for qualifying the predictions from the present model.

\section*{Acknowledgements}

The author would like to thank the whole NSTX team for providing the profile data used for the analysis of the plasma discharge presented here. Discussions with T. Estrada and D. Carralero have also been insightful for the preparation of this manuscript. This work has been supported by U.S. D.O.E. contract DE-AC02-09CH11466, and in part by the Engineering and Physical Sciences Research Council (EPSRC) [EP/R034737/1]. Computer simulations were carried out at the National Energy Research Scientific Computing Center, supported by the Office of Science of the U.S. D.O.E. under Contract No. DE-AC02-05CH11231, at the MIT-PSFC partition of the Engaging cluster at the MGHPCC facility (www.mghpcc.org), which was funded by D.O.E. grant number DE-FG02-91-ER54109, and the supercomputer ARCHER in the UK, which was funded by EPSRC [EP/R034737/1]. C. Holland was partially funded by U.S. D.O.E. grant number DE-SC0018287. This work has been carried out within the framework of the EUROfusion Consortium and has received funding from the Euratom research and training programme 2014-2018 and 2019-2020 under grant agreement No 633053. The views and opinions expressed herein do not necessarily reflect those of the European Commission.

\clearpage 
\appendix

\section{Definitions and normalizations in GYRO and GS2}
\label{app2}

In this appendix we include the definitions of the GYRO and GS2 internally defined wavenumber components, and their relation to the normal and binormal components $(k_n, k_b)$ employed throughout this manuscript. 

GYRO and GS2 solve the nonlinear gyrokinetic equation in a field-aligned coordinate system in which the magnetic field is written as $\bold{B} = \nabla \alpha \times \nabla \psi$  \cite{kruskal_kulsrud_physflu_1958}, where $\alpha$ is the field-line label and $\psi$ is the poloidal flux divided by $2\pi$. The density fluctuation field $\delta n$ is decomposed into a slowly varying part (parallel coordinate $\theta$ along the magnetic field $\mathbf{B}$) and a rapidly varying part $\text{e}^{iS}$ ($|S| \gg 1$), $\f(\bold{r}) = \sum_{\bold{k}_\perp} \hat{\f}_{\bold{k}_\perp}(\theta) \text{e}^{iS(\bold{r, k_\perp})}$, where $\bold{k_\perp} \equiv \nabla S$ is the perpendicular wave-vector to the magnetic field $\bold{B}$. In what follows we describe how this representation in GYRO and GS2 is related to the $(k_n, k_b)$ wavenumber components.

\subsection{GYRO}


In the case of GYRO, the eikonal function is written as $S = - n \alpha + {2\pi p r}/{L_r} $, where $r$ is the radial coordinate (half of the flux surface diameter for $\theta=0$), $n$ is the toroidal mode number, $p$ is the radial mode number and $L_r$ is the radial box size. The electron density fluctuation field $\delta n$ is expanded as $\delta n (\bold{r}, t) = \sum_{n, p} \delta n_{np}(\theta, t) \text{e}^{-in\alpha} \text{e}^{i 2\pi p r/L_r}$ \cite{gyro, gyro_guide}. In this work GYRO is run in the \emph{local} approximation in which the background profiles and gradients are taken as constant. Correspondingly, the function $\alpha$ is considered only as a function of $\theta$, that is $\alpha(r, \theta) \rightarrow \alpha(r_0, \theta)$, meaning that the decomposition $\sum_{n,p}$ is equivalent to a Fourier decomposition in $(n,p)$ ($r_0$ is taken as the center of the radial domain). The function $\alpha$ can be readily computed from knowledge of the flux-surface shape $\big( R(r,\theta), Z(r, \theta) \big)$. We use the Miller equilibrium parametrization \cite{miller_pop_1998}. 

The GYRO radial and poloidal wavenumbers are defined by $k_r = 2\pi p/L_r$ and $k_\theta = n q/r_0$ where $q$ is the value of the safety factor at $r_0$. These are normalized by the ion-sound gyro-radius $\rho_{s,\text{unit}} = \frac{\bar{c}_s}{e B_\text{unit}/m_ic}$, where $B_\text{unit}$ is the \emph{effective field strength} $B_\text{unit} = \frac{q}{r}\frac{\text{d}\psi}{\text{d}r}$ ($\psi $ is the poloidal flux divided by $2\pi$) and $\bar{c}_s = \sqrt{\bar{T}_e/m_i}$ the sound speed evaluated at the reference radius $r_0$. These definitions and normalization are internal to GYRO and differ from those implemented in GS2. 

\subsection{GS2}

GS2 uses equivalent coordinates to solve for the nonlinear gyrokinetic equation. The eikonal function in GS2 is written as $S = n(\alpha + q \theta_0) = \text{x} k_\text{x} + \text{y}k_\text{y}$, and the sum $\sum_\bold{k_\perp} = \sum_{k_\text{x}, k_\text{y}}$ (note the different notation between the GS2 $(\text{x, y})$ and the real-space $(x,y)$ employed throughout the manuscript). The field $\delta n$ is expanded as $\delta n(\bold{r}, t) = \sum_{k_\text{x}, k_\text{y}} \delta \hat{n} (k_\text{x}, k_\text{y}, \theta, t) \text{e}^{ik_\text{x} \text{x}} \text{e}^{i k_\text{y} \text{y}}$, consistent with the \emph{local}, flux-tube formulation of GS2. Here $k_\text{x}$ and $k_\text{y}$ are the spectral radial and 'poloidal' coordinates. The center of the flux-tube is assumed at $(\alpha_0, \psi_0)$. The variables $\text{x}, \text{y}, k_\text{x}$ and $k_\text{y}$ are defined by $k_\text{y} = \frac{n}{a} / \frac{d\psi_N}{d\rho}$, $k_\text{x} = k_\text{y} \hat{s}\theta_0$, $\text{x} = a\frac{q}{\rho}(\psi_N-\psi_{N0})$, and $\text{y} = a \frac{d\psi_N}{d\rho} (\alpha - \alpha_0)$, where $a$ is the half diameter at the last closed flux surface (LCFS), $\psi_N$ is the normalized poloidal flux $\psi_N = \psi/a^2B_\text{ref}$, $n$ is the toroidal mode number, $\rho = r/a$ is the normalized radial coordinate at the local $\psi = \psi_0$, $\hat{s}$ is the magnetic shear, $q$ is the safety factor, $B_\text{ref}$ the reference field in GS2 and $\theta_0$ the parameter relating $k_\text{x}$ and $k_\text{y}$ via $k_\text{x} = k_\text{y}\hat{s} \theta_0$. 


The standard GS2 normalization is that wavenumbers $k_\text{x}$ and $k_\text{y}$ are normalized by a reference gyro-radius $\rho_r$, where $\rho_r$ is defined by $\rho_r = v_{tr}/\Omega_r$, $v_{tr} = \sqrt{2T_r/m_r}$ and $\Omega_r = eB_\text{ref}/m_rc$ ($T_r$ is the reference temperature, $m_r$ the reference mass, $c$ the speed of light). In this manuscript we have chosen the deuterium ions as the reference species and $B_\text{ref}$ as the on-axis toroidal magnetic field strength. This normalization differs from that employed in GYRO by the $\sqrt{2}$ in $v_{tr}$, the reference field $B_\text{ref}$ vs. the GYRO $B_{\text{unit}}$, and the use of $T_i$ for normalizations instead of $T_e$. The quoted $\rho_s$ in the main text is mapped from $\rho_r$ by the use of the local toroidal field $B = |\mathbf{B_\varphi}|$, $T_e$ and the removal of the $\sqrt{2}$ factor. In the case of GS2, we also employ the Miller flux-surface parametrization for $\big( R(r,\theta), Z(r, \theta) \big)$.

\subsection{Normal and binormal components $k_n$ and $k_b$}

In order to reconcile the coordinate definitions in different gyrokinetic codes, throughout this manuscript we have opted to employ the normal and binormal wavenumber components $k_n$ and $k_b$ which are \emph{not} dependent on a specific gyrokinetic code and carry information about real-space length scales. They correspond to real inverse lengths in physical space. These components are routinely employed in experimental measurements to characterize the measured $\bold{k}$ (via ray tracing, beam tracing or full-wave simulations), and provide a natural common reference with which to compare experimental measurements and gyrokinetic simulation. 

The eikonal function $S$ used to expand the field $\delta n$ in a gyrokinetic simulation can be used to define the normal and binormal wavenumber components from those defined in a specific gyrokinetic code, via the relations $k_n = \bold{e_n} \cdot \bold{k}_\perp$, and $k_b = \bold{e_b} \cdot \bold{k}_\perp$ where the normal and binormal unit vectors are defined by $\bold{e_n} = \frac{\nabla r}{|\nabla r|}$, $\bold{e_b} = \bold{e_n} \times \bold{b} $ and $\bold{k}_\perp = \nabla S$ ($\mathbf{b} = \mathbf{B}/|\mathbf{B}|$). In this manuscript we restrict ourselves to fluctuations taken at the outboard midplane ($\theta=0$), which is the location where traditional drift-wave ballooning type instabilities tend to exhibit highest amplitude and where the vast majority of fluctuation measurements are performed. The components $k_n$ and $k_b$ take a simple form at the outboard midplane, and are related to the GYRO $(k_r, k_\theta)$ and GS2 $(k_\text{x}, k_\text{y})$ via

\begin{equation}
	\begin{aligned}
	& &k_n &= |\nabla r| k_r \qquad \text{(GYRO)} &&=  \frac{q}{\rho} \frac{d\psi_N}{d\rho} |\nabla r| k_\text{x} \qquad \text{(GS2)} \ , \\
	& &k_b &= -\bigg[ \frac{1}{R^2} + \Big( \frac{\partial \alpha}{\partial \theta} \Big)^2/(r^2\kappa^2) \bigg]^{1/2} \frac{r}{q}k_\theta  \qquad \text{(GYRO)} &&= a \frac{d\psi_N}{d\rho} \bigg[ \frac{1}{R^2} + \Big( \frac{\partial \alpha}{\partial \theta} \Big)^2/(\rho^2 a^2 \kappa^2) \bigg]^{1/2} k_\text{y} \qquad \text{(GS2)}, \\
	\end{aligned}
	\label{knkb}
\end{equation} 

where $\kappa$ is the flux-surface elongation as defined in the Miller geometry representation. With these definitions, we can write both the GYRO and GS2 eikonal function $S$ (at the outboard midplane) in a compact form as $S = x k_n + y k_b$, where 

\begin{equation}
	\begin{aligned}
	& x = (r-r_0)/|\nabla r| \qquad \text{(GYRO)} \qquad &= & \frac{\rho}{q \frac{d\psi_N}{d\rho} |\nabla r|} \text{x} \qquad \text{(GS2)} \ , \\
	& y = (\alpha -\alpha_0)/ \bigg[ \Big( \frac{\partial \alpha}{\partial \theta} \Big)^2 / (r^2 \kappa^2) + \frac{1}{R^2} \bigg]^{1/2} \qquad \text{(GYRO)} &= & \frac{\text{y}}{a \frac{d\psi_N}{d\rho}}/ {\bigg[ \Big( \frac{\partial \alpha}{\partial \theta} \Big)^2 / (\rho^2 a^2 \kappa^2) + \frac{1}{R^2} \bigg]^{1/2} } \qquad \text{(GS2)} .
	\end{aligned}
	\label{xpyp}
\end{equation}

Both GYRO and GS2 exhibit the same parametric, physical dependences on the local flux surface geometry (terms in $|\nabla r|, \partial \alpha/\partial \theta, \kappa, ...$), however the expressions differ in the specific definitions and normalizations from each code (use of variable $\alpha$ vs. $\text{y}$, $r$ vs. x, etc ...). The coordinates $(x, y)$ [m] are the normal and binormal coordinates employed throughout this manuscript, and correspond to real lengths in physical space. Note, for example, how the radial direction in physical space is related to the field-aligned radial direction by a factor $|\nabla r|$ which contains information about the compression of the flux-surface at the outboard midplane due to Shafranov's shift. The factor in square brackets $[.]^{1/2}$ in equations (\ref{knkb}) and (\ref{xpyp}) indicates that the binormal component at the outboard midplane is a combination of a toroidal component (term in $1/R$) and a vertical component along the $Z$ direction (term in $\frac{\partial \alpha}{\partial \theta}/(r\kappa)$ at the outboard midplane). The vertical component exhibits a dependence on $\kappa$, showing how elongation affects the vertical dimension via the stretching of the flux surface at the outboard midplane. 

Using these definitions, equation (\ref{scatt_amp}) is written using $\delta \hat{n}(\bold{k}_\perp, \theta, t) = \delta n_{np}(\theta, t)$ in the case of GYRO, while $\delta \hat{n}(\bold{k}_\perp, \theta, t) = \delta \hat{n}_{k_\text{x}, k_\text{y}}(\theta, t)$ in the case of GS2. The mappings from equations (\ref{knkb}) and (\ref{xpyp}) are employed throughout the manuscript to compute the wavenumber components $(k_n, k_b)$ and the real-space coordinates $(x,y)$ from the internal GYRO and GS2 definitions. As mentioned in the main text, at the outboard midplane of an up-down symmetric flux surface $k_n \propto k_r \ (\text{GYRO}) \propto k_\text{x} \ (\text{GS2})$ and $k_b \propto k_\theta \ (\text{GYRO}) \propto k_\text{y} \ (\text{GS2})$. This has allowed the expansion $\text{e}^{ik_n x + i k_b y}$ in equation (\ref{dn_fluxtube}) (a more complicated form would have been necessary for $\theta \neq 0$). For completeness, the general mapping between the GYRO and GS2 wavenumbers and $k_n$ and $k_b$ at arbitrary poloidal location $\theta$ is 

\begin{equation}
	\left\{
	\begin{aligned}
	& k_n = - n \frac{\nabla \alpha \cdot \nabla r}{| \nabla r |} + \frac{2\pi p}{L_r} |\nabla r| \ , \\
	& k_b = n \Big( \bold{b} \times \frac{\nabla r}{|\nabla r|} \Big) \cdot \nabla \alpha \ , \\
	\end{aligned}
	\right.
	\qquad (\text{GYRO})
	\label{k2d_mapping_outboardmp_app}
\end{equation} 

\begin{equation}
	\left\{
	\begin{aligned}
	& k_n = a \frac{d\psi_N}{d\rho} \frac{\nabla \alpha \cdot \nabla r}{| \nabla r |} k_\text{y} + \frac{q}{\rho} \frac{d\psi_N}{d\rho} |\nabla r| k_\text{x} \ , \\
	& k_b = - a \frac{d\psi_N}{d\rho} \Big( \bold{b} \times \frac{\nabla r}{|\nabla r|} \Big) \cdot \nabla \alpha \ k_\text{y} \ , \\
	\end{aligned}
	\right.
	\qquad (\text{GS2})
	\label{gs2_mapping}
\end{equation}

which reduces to equation (\ref{knkb}) for $\theta=0$. As a side note, equations (\ref{knkb}), (\ref{k2d_mapping_outboardmp_app}) and (\ref{gs2_mapping}) show that the GYRO $(k_r, k_\theta)$ components are related to the GS2 $(k_\text{x}, k_\text{y})$ by $k_r = \frac{q}{\rho} \frac{d\psi_N}{d\rho} k_\text{x}$ and $k_\theta = - \frac{q}{\rho} \frac{d\psi_N}{d\rho} k_\text{y}$, where the constant of proportionality is simply $B_\text{unit}/B_\text{ref}$. 

This discussion has focused on technical details without making any assessment of the impact on the real measurement. Appendix \ref{app5} discusses the important consequences of using the real space coordinates for interpreting experimental measurements and performing quantitative comparisons.

\section{Definitions of the radial correlation function}
\label{app4}

In this appendix we include the details to calculate the cross-correlation functions employed throughout this manuscript (equations (\ref{ccf_defs})). The average turbulence radial correlation function $\text{CCF}^{\text{avg}}$ is defined as 

\begin{equation}
	\begin{aligned}
		& \text{CCF}^{\text{avg}}(\Delta x) &&= \frac{\langle \delta n(x_0 + \Delta x, y_0) \delta n(x_0, y_0) \rangle_{x_0, y_0, T}}{ \langle |\delta n(x_0, y_0)|^2 \rangle_{x_0, y_0, T} } &= \frac{ \int_T{dt} \int{dx_0} \int{dy_0 \ \delta n(\bold{r_0} + \Delta x \bold{e_n} )} \delta n(\bold{r_0}) }{ \int_T{dt} \int{dx_0} \int{dy_0 \ |\delta n(\bold{r_0})}|^2 } \\ 
		& &&= \frac{ \sum_{k_n, k_b} \langle | \delta \hat{n}(k_n, k_b) |^2 \rangle_T \ \text{e}^{i k_n \Delta x} }{  \sum_{k_n, k_b} \langle | \delta \hat{n}(k_n, k_b) |^2 \rangle_T  } ,
		\end{aligned}
	\label{ccf_real_def}
\end{equation}

where $\bold{r_0} = (x_0, y_0, \theta=0)$, $\delta n(\bold{r_0})$ is expanded following equation (\ref{dn_fluxtube}), and the time variable $t$ was omitted for clarity. The integration $dx_0$ and $dy_0$ is equivalent in this model to an ensemble average, since essentially any $(x,y)$ location within a local flux-tube should give rise to the same turbulent fluctuations statistically. The brackets $\langle . \rangle_T$ denote a time average over the simulation time $T$. Expression (\ref{ccf_real_def}) also shows how the radial correlation function of the turbulence is the Fourier transform of the time averaged density fluctuation power spectrum. This remains valid for the subsequent correlation functions. Note the normalization ensures $\text{CCF}^{\text{avg}} = 1$ for $\Delta x = 0$. The correlation length from equation (\ref{ccf_real_def}) is represented in figures \ref{ccfx_lrky_escale_exb}, \ref{ccfx_lrky_wrscan_ky2}, \ref{ccfx_lrky_iscale_exb} and \ref{ccfx_lrky_wrscan_ky02} as a horizontal dashed line. 

A generalization of $\text{CCF}^{\text{avg}}$ is the 2D correlation function in $\Delta x$ and $\Delta y$, briefly introduced in the main body of the text to discuss 2D turbulence properties such as the eddy aspect ratio and eddy tilt. It is defined as 

\begin{equation}
	\begin{aligned}
		& \text{CCF}^{\text{avg}}_{2D}(\Delta x, \Delta y) &&= \frac{\langle \delta n(x_0+\Delta x, y_0+\Delta y) \delta n(x_0, y_0) \rangle_{x_0, y_0, T}}{ \langle |\delta n(x_0, y_0)|^2 \rangle_{x_0, y_0, T} } \\
		& &&= \frac{ \int_T{dt} \int{dx_0} \int{dy_0 \ \delta n(\bold{r_0} + \Delta x \bold{e_n} + \Delta y \bold{e_b})} \delta n(\bold{r_0}) }{ \int_T{dt} \int{dx_0} \int{dy_0 \ |\delta n(\bold{r_0})}|^2 } \\ 
		& &&= \frac{ \sum_{k_n, k_b} \langle | \delta \hat{n}(k_n, k_b) |^2 \rangle_T \ \text{e}^{i k_n \Delta x} \text{e}^{i k_b \Delta y} }{  \sum_{k_n, k_b} \langle | \delta \hat{n}(k_n, k_b) |^2 \rangle_T  } .
		\end{aligned}
	\label{ccf_2dreal_def}
\end{equation}

Equation (\ref{ccf_2dreal_def}) is represented in figures \ref{ccfxy2d_exbexb0_lrky_escale} and \ref{ccfxy2d_exbexb0_lrky_iscale} of the main body of the text. Note the sum over $k_b$ in equation \ref{ccf_2dreal_def}, which represents the sum over scales and motivates the denomination of 2D \emph{average} correlation function of the turbulence. By extension, the 2D scale-dependent correlation function for binormal wavenumber $k_b$, $\text{CCF}_{2D}^{k_b}$, is calculated by removing the $\sum_{k_b}$ and replacing the quantity under the summatory sign with the specific desired $k_b$. This leads to a cosinusoidal dependence of $\text{CCF}_{2D}^{k_b}$ in the $\Delta y$ direction, while only the sum over $k_n$ remains, giving 

\begin{equation}
	\begin{aligned}
		& \text{CCF}^{k_b}_{2D}(\Delta x, \Delta y) = \cos({k_b \Delta y}) \frac{ \sum_{k_n} \langle | \delta \hat{n}(k_n, k_b) |^2 \rangle_T \ \text{e}^{i k_n \Delta x} }{  \sum_{k_n} \langle | \delta \hat{n}(k_n, k_b) |^2 \rangle_T  } .
		\end{aligned}
	\label{ccf_2d-scaledep_def}
\end{equation}

This expression is used in sections \ref{escale_gyro1} and \ref{iscale_gs2_2} to compute the binormal correlation length $l_b$ as the $1/e$ value of $\text{CCF}_{2D}^{k_b}$ for $\Delta x = 0$, that is $\text{CCF}_{2D}^{k_b}(\Delta x = 0, l_b) = 1/e$, giving the quoted value $l_b = \arccos(1/e)/k_b$. The one-dimensional scale-dependent correlation function $\text{CCF}^{k_b}$ and the synthetic correlation functions $\text{CCF}^\text{syn}$ are defined as 

\begin{equation}
	\begin{aligned}
		& \text{CCF}^{k_b}(\Delta x) &&= \frac{\langle \delta \tilde{n}(x_0+\Delta x, k_b) \delta \tilde{n}(x_0, k_b)^* \rangle_{x_0, T}}{ \langle |\delta \hat{n}(x_0, k_b)|^2 \rangle_{x_0, T} } = \frac{ \sum_{k_n} \langle | \delta \hat{n}(k_n, k_b) |^2 \rangle_T \ \text{e}^{i k_n \Delta x} }{  \sum_{k_n} \langle | \delta \hat{n}(k_n, k_b) |^2 \rangle_T  } \ , \\
		&\text{CCF}^{\text{syn}}(\Delta x) &&= \frac{\langle A_s(\bold{r_0} + \Delta x \bold{e_n}, k_{b0} \bold{e_b}) A_s(\bold{r_0}, k_{b0} \bold{e_b})^* \rangle_{x_0, y_0, T}}{ \langle |A_s(\bold{r_0}, k_{b0} \bold{e_b})|^2 \rangle_{x_0, y_0, T} } \\
		& &&= \frac{ \sum_{k_n, k_b} \text{e}^{-k_n^2W_n^2/2} \text{e}^{-(k_b-k_{b0})^2W_b^2/2} \langle | \delta \hat{n}(k_n, k_b) |^2 \rangle_T \ \text{e}^{i k_n \Delta x} }{  \sum_{k_n, k_b} \text{e}^{-k_n^2W_n^2/2} \text{e}^{-(k_b-k_{b0})^2W_b^2/2} \langle | \delta \hat{n}(k_n, k_b) |^2 \rangle_T  } \ , \\
	\end{aligned}
	\label{ccf_kb_syn}
\end{equation}

which agree with the expressions used in equation (\ref{ccfsyn_app}). In equation (\ref{ccf_kb_syn}), $\delta \tilde{n}$ is Fourier decomposed in $y$ but not in $x$, while $\delta \hat{n}$ is Fourier decomposed in both $x$ and $y$. Note that the computation of $\text{CCF}^{k_b}$ only requires a sum over $k_n$, since by definition the $k_b$ component is selected, while the sum is made over $k_n$ and $k_b$ in $\text{CCF}^\text{avg}$ and $\text{CCF}^\text{syn}$. For the experimentally relevant values $W_b \gg 2\pi/k_{b0}$, we have

\begin{equation}
	\text{CCF}^{\text{syn}}(\Delta x) \rightarrow \frac{ \sum_{k_n} \text{e}^{-k_n^2W_n^2/2} \langle | \delta \hat{n}(k_n, k_{b0}) |^2 \rangle_T \ \text{e}^{i k_n \Delta x} }{  \sum_{k_n} \text{e}^{-k_n^2W_n^2/2}  \langle | \delta \hat{n}(k_n, k_{b0}) |^2 \rangle_T  } \qquad \text{for $W_b \gg 2\pi/k_{b0}$} .
	\label{ccf_kb_wn}
\end{equation}

Equation (\ref{ccf_kb_syn}) shows how the synthetic correlation function is in fact a convolution product between a filter $|W(\bold{k_\perp - k_{0}})|^2 = \text{e}^{-k_n^2W_n^2/2} \text{e}^{-(k_b-k_{b0})^2W_b^2/2}$ (equation (\ref{uw_filters}), reduced to $\text{e}^{-k_n^2W_n^2/2}$ in equation (\ref{ccf_kb_wn})) and the wavenumber spectrum of fluctuations. In real space, this means that the radial structure of $\text{CCF}^\text{syn}$ will be determined by the largest length scale between the Fourier transform of the filter $\propto \text{e}^{-x^2/2W_n^2}$, that is $W_n$, and the Fourier transform of the power spectrum, that is the scale-dependent correlation length for $k_b = k_{b0}$. This explains why for $W_n \ll l_r(k_{b0})$ we recover the scale-dependent correlation length (figures \ref{ccfx_lrky_wrscan_ky2} and \ref{ccfx_lrky_wrscan_ky02}), while for $W_n \gg l_r(k_{b0})$, the dominant scale will be $\approx W_n$.


\section{Physical radial correlation length, eddy aspect ratio and eddy tilt}
\label{app5}

In this appendix we discuss some geometric considerations to take into account when interpreting the physical radial correlation length, eddy aspect ratio and tilt angle in the perpendicular direction to $\mathbf{B}$ from local flux-tube gyrokinetic simulation outputs. We focus on the outboard midplane location $\theta=0$, and intend to provide intuition behind the mathematical mappings introduced in appendix \ref{app2} (equations (\ref{knkb}) and (\ref{xpyp})) \footnote{A general flux-tube cross section perpendicular to $\mathbf{B}$ would be twisted for finite finite shear $\hat{s}$ compared to the $\theta=0$ cross-section. This effect is not considered here.}.

\begin{figure}
	\begin{center}
		\includegraphics[height=12cm]{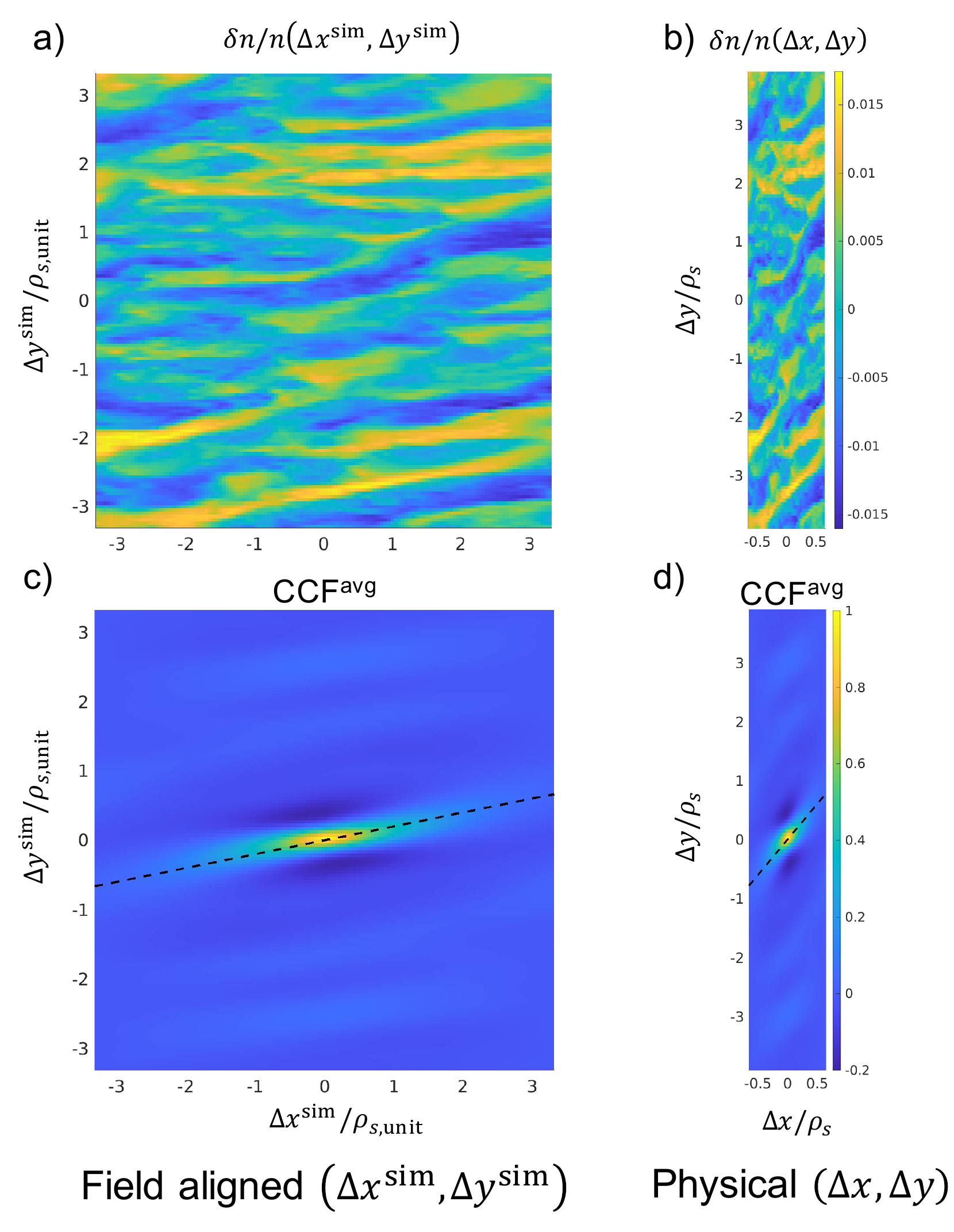}
	\end{center}
	\caption{Comparison of the electron-scale electron density fluctuation amplitude $\delta n$ and the average correlation function $\text{CCF}^\text{avg}$ as a function of the internal field-aligned coordinates (noted $\Delta x^\text{sim}$ and $\Delta y^\text{sim}$ here in \textbf{a)} and \textbf{c)}) corresponding to the electron-scale, ETG-driven turbulence condition in NSTX (section \ref{escale_gyro}). To be consistent, these are plotted as a function of $\rho_{s,\text{unit}}$, internally defined in GYRO \cite{gyro, gyro_guide}. The same quantities are plotted in real physical coordinates $(\Delta x, \Delta y)$, representing real lengths $[m]$, in \textbf{b)} and \textbf{d)}, at the outboard midplane. As in the rest of the manuscript, these are normalized to the local $\rho_s$. Note a square aspect ratio simulation domain in $(\Delta x, \Delta y)^\text{sim}$ is deformed in real physical space at the outboard midplane in highly shaped ST geometry.}   
	\label{ccf_dn_xnyb_xyfs_escale_comp}
\end{figure}

Figure \ref{ccf_dn_xnyb_xyfs_escale_comp} shows the electron-scale electron density fluctuation amplitude $\delta n$ and the average correlation function $\text{CCF}^\text{avg}$ as a function of the internal field-aligned coordinates (noted $\Delta x^\text{sim}$ and $\Delta y^\text{sim}$ here in \textbf{a)} and \textbf{c)}) corresponding to the electron-scale, ETG-driven turbulence condition in NSTX (section \ref{escale_gyro}). These are plotted as a function of $\rho_{s,\text{unit}}$, internally defined in GYRO \cite{gyro, gyro_guide}. The same quantities are plotted in the real physical coordinates $(\Delta x, \Delta y)$ employed throughout this manuscript, representing real lengths $[m]$, in \textbf{b)} and \textbf{d)}. These are normalized to the local $\rho_s$. A square aspect ratio simulation domain in $(\Delta x, \Delta y)^\text{sim}$ is compressed and stretched in real physical space owing to the highly shaped flux surface geometry characteristic of spherical tokamaks. The mapping from the simulation field-aligned $(\Delta x^\text{sim}, \Delta y^\text{sim})$ to the physical $(\Delta x, \Delta y)$ is made using equation (\ref{xpyp}). 

In the radial direction, the Shafranov's shift measured via the $|\nabla r|$ coefficient contributes to the compression of the flux surfaces in physical space. Additionally, the $\rho_{s,\text{unit}}$ value employed internally in GYRO can substantially differ from the local $\rho_s$ ($B_\text{unit}/B \approx 3.61$, table \ref{geo_coeffs_table}), implying a further size 'reduction' in $\Delta x / \rho_s$ with respect to $\Delta x^\text{sim} / \rho_{s,\text{unit}}$. Both factors contribute to a factor of $\approx \times 5 \textendash 6$ reduction in the physical radial domain with respect to the simulation radial domain. In the binormal direction, figure \ref{ccf_dn_xnyb_xyfs_escale_comp} does not show a substantial difference between $\Delta y^\text{sim} / \rho_{s,\text{unit}}$ and $\Delta y/\rho_s$ due to a combination of cancelling effects. The flux-surface elongation $\kappa$ contributes to stretching the real domain in the vertical direction, while the $1/R^2$ term represents the toroidal component of the binormal direction (along $\mathbf{e_b}$), further contributing to the stretching of the real domain in the binormal direction $\Delta y$ with respect to the simulation $\Delta y^\text{sim}$. These factors contributing to the stretching are counteracted by the differing $\rho_s$ definition, which 'reduces' the domain in physical space as previously described. Overall, the difference between the simulation $\Delta y^\text{sim}/\rho_{s,\text{unit}}$ and physical $\Delta y/\rho_s$ is not substantial.

The previous discussion is pertinent when interpreting physical radial correlation lengths, aspect ratio and tilt angle from local flux-tube gyrokinetic simulations. For comparisons to experiments, a factor of $\approx 3 \textendash 4$ error can be made in the radial correlation length if directly comparing to the field-aligned $(\Delta x^\text{sim}, \Delta y^\text{sim})$. Inferring the eddy tilt from the flux-tube $(\Delta x^\text{sim}, \Delta y^\text{sim})$ would result in $\theta_\text{tilt} \approx 11^o$, while the physical tilt angle is $\theta_\text{tilt} \approx 50^o$, a factor $\approx \times 5$ larger. The inferred eddy aspect ratio does also greatly differ between the field-aligned $(\Delta x^\text{sim}, \Delta y^\text{sim})$ and real space. In particular, the cases with $\gamma_E=0$ (zero tilt angle) exhibit an aspect ratio $l_r/l_b \approx 9$ in the field-line coordinates, while the physical aspect ratio is $l_r/l_b \approx 1.54$.

\begin{figure}
	\begin{center}
		\includegraphics[height=12cm]{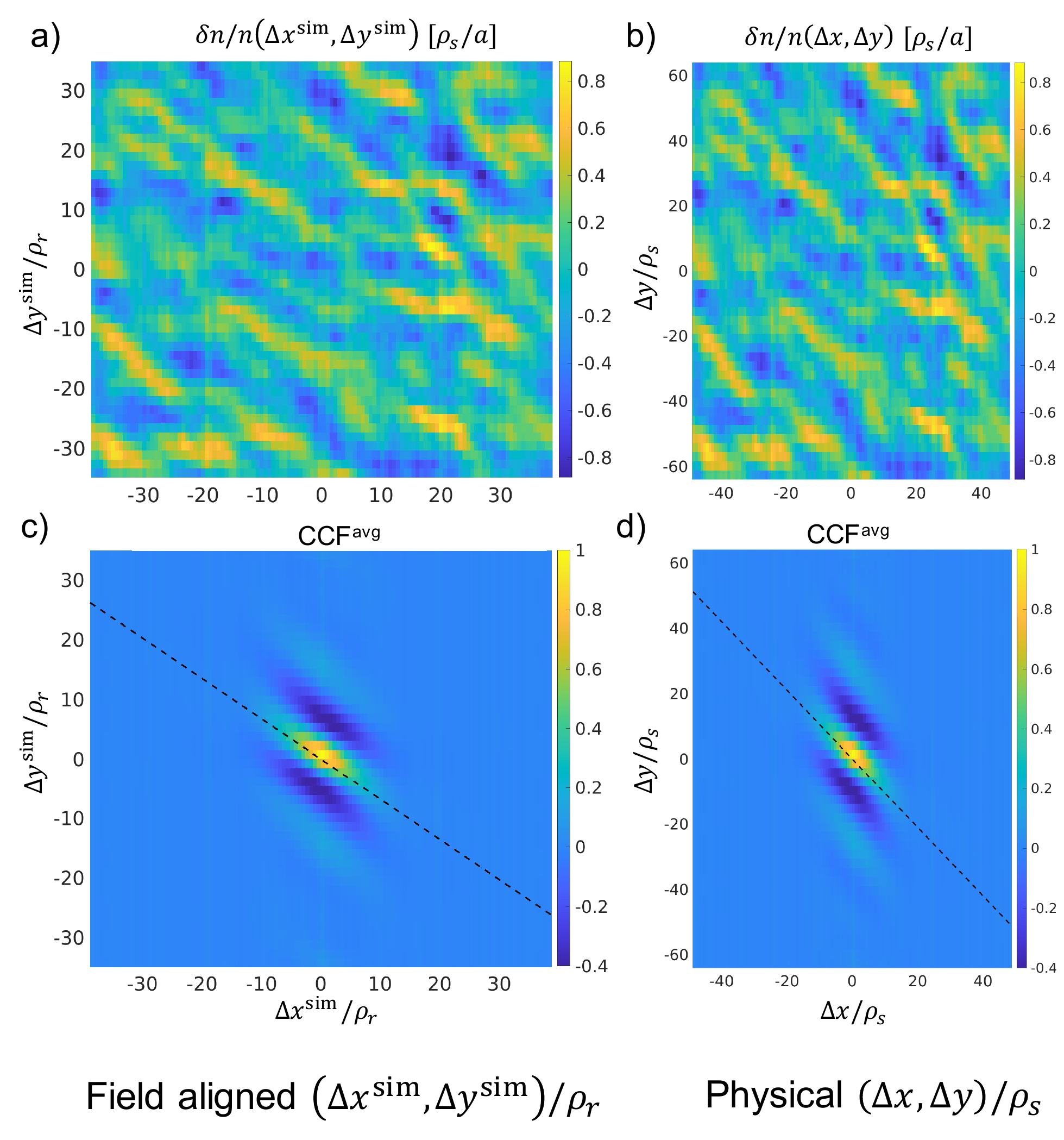}
	\end{center}
	\caption{Comparison of the electron density fluctuation amplitude $\delta n$ and the average correlation function $\text{CCF}^\text{avg}$ as a function of the internal field-aligned coordinates (noted $\Delta x^\text{sim}$ and $\Delta y^\text{sim}$ here in \textbf{a)} and \textbf{c)}, and normalized by the GS2 internally defined $\rho_r$) corresponding to the ion-scale, ITG-driven turbulence condition in JET (section \ref{iscale_gs2}). The same quantities are plotted in real physical coordinates $(\Delta x, \Delta y)$, representing real lengths $[m]$, in \textbf{b)} and \textbf{d)}, and normalized by the local $\rho_s$ at the outboard midplane. In this conventional tokamak core condition, the simulation domain in $(\Delta x, \Delta y)^\text{sim}$ is deformed when mapped to real space, but not as dramatically as the previous ST case.}   
	\label{ccf_dn_xnyb_xyfs_iscale_comp}
\end{figure}

The same care needs to be taken when considering conventional tokamak flux-surface geometry such as in the JET ion-scale conditions analyzed here, although one does not expect differences as large as in the NSTX case with higher shaping. Figure \ref{ccf_dn_xnyb_xyfs_iscale_comp} shows 2D plots of the density fluctuation amplitude $\delta n$ and the 2D average correlation function $\text{CCF}^\text{avg}$ corresponding to the JET ion-scale, ITG-driven condition. Visually the difference is less dramatic than in figure \ref{ccf_dn_xnyb_xyfs_escale_comp}. Quantitatively, the eddy tilt calculated in the simulation $(\Delta x, \Delta y)^\text{sim}$ would yield $\theta_\text{tilt} \approx -32.6^o$, while in physical space one finds $\theta_\text{tilt} \approx -42.9^o$. In these cases, the aspect ratio $l_r/l_b$ is only moderately modified by $\approx 10 \%$ when calculated in physical space vs. in field-aligned coordinates when $\gamma_E^\text{exp}$ is finite, but can change by factors of $\approx 60\%$ when $\gamma_E=0$ as we noted in the main text ($l_r/l_b \approx 2.99$ in the field-aligned frame while the physical $l_r/l_b \approx 1.88$). The geometric considerations discussed in this appendix are important to consider when comparing the output of local flux-tube gyrokinetic simulations to experimental measurements.


\begin{table}[]
\centering
{
\begin{tabular}{llllllll}
\toprule
$ $ & $|\nabla r| $ & $\frac{\partial \alpha}{\partial \theta}$ & $\kappa$ & ${R_0}/{a}, {R_\text{ref}}/{a}$ & $q$ & $r/a$ & $B$  \\ \hline
\textbf{NSTX: e- scale} &  $1.43$& $ 1.33$ & $ 2.11$ & $ 1.52$ & $ 3.79$ & $ 0.71$ & ${B_\text{unit}}/{B} \approx 3.61$  \\
\textbf{JET: ion scale} & $ 1.12$ & $1.02$ & $ 1.36 $ & $ 3.35$ & $ 1.43$ &  $ 0.51$ & $ B_\text{ref}/B \approx 1.09, \frac{d\psi_N}{d\rho} \approx 0.51 $ \\
\toprule
\end{tabular}
}
\caption{Local geometry parameters at the outboard midplane ($\theta=0$) necessary for mapping the gyrokinetic simulation perpendicular domain $(\Delta x^\text{sim}, \Delta y^\text{sim})$ to the real space $(\Delta x, \Delta y)$.}
\label{geo_coeffs_table}
\end{table}




\begin{thebibliography}{99}


\bibitem{schirmer_ppcf_2007} J. Schirmer, G. D. Conway, E. Holzhauer ,W. Suttrop, H. Zohm and the ASDEX Upgrade Team, Plasma Phys. Control. Fusion \textbf{49} 1019 (2007).

\bibitem{fernandezmarina_nf_2014} F. Fernández-Marina, T. Estrada and E. Blanco, Nucl. Fusion \textbf{54} 072001 (2014).

\bibitem{liewer_nf_1985} P. C. Liewer, Nucl. Fusion \textbf{25}, 5 (1985).

\bibitem{tynan_ppcf_2009} G. R. Tynan, A. Fujisawa and G. McKee, Plasma Phys. Control. Fusion \textbf{51}, 113001 (2009). 

\bibitem{horton_revmodphys_1999} W. Horton, Reviews of Modern Physics, \textbf{71}, 3 (1999). 

\bibitem{garbet_ppcf_2001} X. Garbet, Plasma Phys. Control. Fusion \textbf{43}, A251 (2001). 

\bibitem{candy_prl_2003} J. Candy and R. E. Waltz, Phys. Rev. Lett. \textbf{91}, 045001 (2003).

\bibitem{terry_pop_2008} P. W. Terry, M. Greenwald, J.-N. Leboeuf, G. R. McKee, D. R. Mikkelsen, W. M. Nevins, D. E. Newman, D. P. Stotler, Task Group on Verification and Validation, U.S. Burning Plasma Organization, and U.S. Transport Task Force, Phys. Plasmas \textbf{15}, 062503 (2008).

\bibitem{greenwald_pop_2010} M. Greenwald, Physics of Plasmas \textbf{17}, 058101 (2010).

\bibitem{holland_pop_2016} C. Holland, Phys. Plasmas \textbf{23}, 060901 (2016).

\bibitem{fonck_rsi_1990} R. J. Fonck, P. A. Duperrex, and S. F. Paul, Review of Scientific Instruments \textbf{61}, 3487 (1990).

\bibitem{fonck_prl_1993} R. J. Fonck, G. Cosby, R. D. Durst, S. F. Paul, N. Bretz, S. Scott, E. Synakowski, and G. Taylor, Phys. Rev. Lett. \textbf{70}, 3736 (1993).

\bibitem{mckee_nf_2001} G. R. McKee, C. C. Petty, R. E.Waltz, C. Fenzi, R.J. Fonck, J.E. Kinsey, T.C. Luce, K.H. Burrell, D.R. Baker, E. J. Doyle, X. Garbet, R. A. Moyer, C. L. Rettig, T. L. Rhodes , D. W. Ross, G. M. Staebler, R. Sydora, M. R. Wade, Nucl. Fusion \textbf{41}, 1235 (2001).

\bibitem{weisen_rsi_1990} H. Weisen, Review of Scientific Instruments \textbf{59}, 1544 (1988).

\bibitem{coda_rsi_1992} S. Coda, M. Porkolab, and T. N. Carlstrom, Review of Scientific Instruments \textbf{63}, 4974 (1992).

\bibitem{coda_rsi_1995} S. Coda, and M. Porkolab, Rev. Sci. Instrum. \textbf{66}, 454 (1995).

\bibitem{coda_prl_2001} S. Coda, M. Porkolab, and K. H. Burrell, Phys. Rev. Lett. \textbf{86}, 4835 (2001).

\bibitem{porkolab_ieee_2006} M. Porkolab, J. C. Rost, N. Basse, J. Dorris, E. Edlund, L. Lin, Y. Lin, S. Wukitch, IEEE Trans. Plasma Sci., \textbf{34}, pp. 229 (2006).

\bibitem{cima_pop_1995} G. Cima, R. V. Bravenec, A. J. Wootton, T. D. Rempel, R. F. Gandy, C. Watts, and M. Kwon, Phys. Plasmas \textbf{2}, 720 (1995).

\bibitem{sattler_prl_1994} S. Sattler, H. J. Hartfuss, and W7-AS Team, Phys. Rev. Lett. \textbf{72}, 653 (1994).

\bibitem{white_pop_2008} A. E. White, L. Schmitz, G. R. McKee, C. Holland, W. A. Peebles, T. A. Carter, M. W. Shafer, M. E. Austin, K. H. Burrell, J. Candy, J. C. DeBoo, E. J. Doyle, M. A. Makowski, R. Prater, T. L. Rhodes, G. M. Staebler, G. R. Tynan, R. E. Waltz, and G. Wang, Phys. Plasmas \textbf{15}, 056116 (2008).

\bibitem{cripwell_eps_1989} P. Cripwell, A. E. Costley, and A. E. Hubbard, in \emph{Proceedings of the 16th European Conference on Controlled Fusion and Plasma Physics}, Venice, 1989, edited by S. Segre, H. Knoepfel, and E. Sindoni (European Physical Society, Petit-Lancy, Switzerland, 1989), Vol. 12B, p. 75.

\bibitem{costley_rsi_1990} A. E. Costley, P. Cripwell, R. Prentice, and A. C. C. Sips, Review of Scientific Instruments \textbf{61}, 2823 (1990).

\bibitem{nazikian_rsi_1995} R. Nazikian, and E. Mazzucato, Rev. Sci. Instrum. \textbf{66}, 392 (1995).

\bibitem{nazikian_pop_2001} R. Nazikian, G. J. Kramer, and E. Valeo, Physics of Plasmas \textbf{8}, 1840 (2001).

\bibitem{rhodes_pop_2002} T. L. Rhodes, J.-N. Leboeuf, R. D. Sydora, R. J. Groebner, E. J. Doyle, G. R. McKee, W. A. Peebles, C. L. Rettig, L. Zeng, and G. Wang, Phys. Plasmas \textbf{9}, 2141 (2002).

\bibitem{hanson_rsi_1990} G. R. Hanson, J. B. Wilgen, E. Anabitarte, J. D. Bell, J. H. Harris, J. L. Dunlap, and C. E. Thomas, Review of Scientific Instruments \textbf{61}, 3049 (1990).

\bibitem{mazzucato_prl_1993} E. Mazzucato and R. Nazikinan, Phys. Rev. Lett. \textbf{71}, 1840 (1993).

\bibitem{rhodes_rsi_1992} T. L. Rhodes, W. A. Peebles, and E. J. Doyle, Review of Scientific Instruments \textbf{63}, 4661 (1992).

\bibitem{sanchez_branas_rsi_1993} J. Sanchez, B. Brañas, E. de la Luna, T. Estrada, Review of Scientific Instruments \textbf{64}, 487 (1993).

\bibitem{conway_ppcf_1997} G. D. Conway, Plasma Phys. Control. Fusion \textbf{39}, 407 (1997).

\bibitem{hutchinson_ppcf_1992} I. H. Hutchinson, Plasma Phys. Control. Fusion \textbf{34} 1225 (1992).

\bibitem{gusakov_ppcf_2004_nlrcr} E. Z. Gusakov and A. Yu Popov, Plasma Phys. Control. Fusion \textbf{46}, 1393 (2004). 

\bibitem{gusakov_ppcf_2002_linrcr} E. Z. Gusakov and B. O. Yakovlev, Plasma Phys. Control. Fusion \textbf{44}, 2525 (2002). 

\bibitem{holzhauer_ppcf_1998} E. Holzhauer, M. Hirsch, T. Grossmann, B. Brañas and F. Serra, Plasma Phys. Control. Fusion \textbf{40} 1869 (1998).

\bibitem{hirsch_rsi_2001} M. Hirsch, E. Holzhauer, J. Baldzuhn, and B. Kurzan, Review of Scientific Instruments \textbf{72}, 324 (2001).

\bibitem{hennequin_nf_2006} P. Hennequin, C. Honoré, A. Truc, A. Quéméneur, C. Fenzi-Bonizec, C. Bourdelle, X. Garbet, G.T. Hoang and the Tore Supra team, Nucl. Fusion \textbf{46}, S771 (2006).

\bibitem{hillesheim_nf_2015b} J. C. Hillesheim, N. A. Crocker, W. A. Peebles, H. Meyer, A. Meakins, A. R. Field, D. Dunai, M. Carr, N. Hawkes and the MAST Team, Nucl. Fusion \textbf{55}, 073024 (2015).

\bibitem{hirsch_ppcf_2004} M. Hirsch and E. Holzhauer, Plasma Phys. Control. Fusion \textbf{46} 593 (2004).

\bibitem{hennequin_rsi_2004} P. Hennequin, C. Honoré, A. Truc, A. Quéméneur, N. Lemoine, J.-M. Chareau, and R. Sabot, Review of Scientific Instruments \textbf{75}, 3881 (2004).

\bibitem{hillesheim_prl_2016} J. C. Hillesheim, E. Delabie, H. Meyer, C. F. Maggi, L. Meneses, E. Poli, and JET Contributors, Phys. Rev. Lett. \textbf{116}, 065002 (2016).

\bibitem{gusakov_ppcf_2014} E. Gusakov, M. Irzak and A. Popov, Plasma Phys. Control. Fusion \textbf{56}, 025009 (2014).

\bibitem{gusakov_pop_2017} E. Z. Gusakov, M. A. Irzak, A. Yu. Popov, S. A. Khitrov, and N. V. Teplova, Phys. Plasmas \textbf{24}, 022119 (2017).

\bibitem{gusakov_ppcf_2013} E. Gusakov, A. B. Altukhov, V. V. Bulanin, A. D. Gurchenko, J. A. Heikkinen, S. J. Janhunen, S. Leerink, L. A. Esipov, M. Yu Kantor, T. P. Kiviniemi, T. Korpilo, D. V. Kouprienko, S. I. Lashkul, A. V. Petrov and N. V. Teplova, Plasma Phys. Control. Fusion \textbf{55}, 124034 (2013).

\bibitem{altukhov_ppcf_2016} A. B. Altukhov, A. D. Gurchenko, E. Z. Gusakov, L. A. Esipov, M. A. Irzak, M. Yu Kantor, D. V. Kouprienko, S. I. Lashkul, S. Leerink, P. Niskala, A. Yu Stepanov and N. V. Teplova, Plasma Phys. Control. Fusion \textbf{58}, 105004 (2016).

\bibitem{altukhov_pop_2018a} A. B. Altukhov, A. D. Gurchenko, E. Z. Gusakov, M. A. Irzak, P. Niskala, L. A. Esipov, T. P. Kiviniemi and S. Leerink, Phys. Plasmas \textbf{25}, 082305 (2018).

\bibitem{altukhov_pop_2018b} A. B. Altukhov, A. D. Gurchenko, E. Z. Gusakov, M. A. Irzak, P. Niskala, L. A. Esipov, T. P. Kiviniemi, O. L. Krutkin, and S. Leerink, Phys. Plasmas \textbf{25}, 112503 (2018).

\bibitem{prisiazhniuk_ppcf_2018} D. Prisiazhniuk, G. D. Conway, A. Krämer-Flecken, U. Stroth and the ASDEX Upgrade Team, Plasma Phys. Control. Fusion \textbf{60}, 075003 (2018).

\bibitem{krutkin_nf_2019} O. L. Krutkin, A. B. Altukhov , A. D. Gurchenko, E. Z. Gusakov, M. A. Irzak, L. A. Esipov, A. V. Sidorov, L. Chôné, T. P. Kiviniemi, S. Leerink, P. Niskala, C. Lechte, S. Heuraux and G. Zadvitskiy, Nucl. Fusion \textbf{59} 096017 (2019).

\bibitem{happel_pop_2011} T. Happel, T. Estrada, E. Blanco, C. Hidalgo, G. D. Conway, U. Stroth, and TJ-II Team, Phys. Plasmas \textbf{18}, 102302 (2011).

\bibitem{conway_ppcf_2004} G. D. Conway, J. Schirmer, S. Klenge, W. Suttrop, E. Holzhauer and the ASDEX Upgrade Team, Plasma Phys. Control. Fusion \textbf{46}, 951 (2004).

\bibitem{hillesheim_nf_2015a} J. C. Hillesheim, F. I. Parra, M. Barnes, N. A. Crocker, H. Meyer, W. A. Peebles, R. Scannell, A. Thornton and the MAST Team, Nucl. Fusion \textbf{55}, 032003 (2015).

\bibitem{bourdelle_nf_2011} C. Bourdelle, T. Gerbauda, L. Vermare1, A. Casatib, T. Aniel, J.F. Artaud, V. Basiuk, J. Bucalossi, F. Clairet, Y. Corre, P. Devynck, G. Falchetto, C. Fenzi, X. Garbet, R. Guirlet, O. Gurcan, S. Heuraux, P. Hennequin, G.T. Hoang, F. Imbeaux, L. Manenc, P. Monier-Garbet, P. Moreau, R. Sabot, J.-L. Ségui, A. Sirinelli, D. Villegas and the Tore Supra Team, Nucl. Fusion \textbf{51}, 063037 (2011).

\bibitem{meijere_ppcf_2014} C. A. de Meijere, S. Coda, Z. Huang, L. Vermare, T. Vernay, V. Vuille, S. Brunner, J. Dominski, P. Hennequin, A. Kramer-Flecken, G. Merlo, L. Porte and L. Villard, Plasma Phys. Control. Fusion \textbf{56}, 072001 (2014).

\bibitem{casati_prl_2009} A. Casati, T. Gerbaud, P. Hennequin, C. Bourdelle, J. Candy, F. Clairet, X. Garbet, V. Grandgirard, O. D. Gurcan, S. Heuraux, G. T. Hoang, C. Honoré, F. Imbeaux, R. Sabot, Y. Sarazin, L. Vermare, and R. E.Waltz, Phys. Rev. Lett. \textbf{102}, 165005 (2009).

\bibitem{windisch_rsi_2018} T. Windisch, S. Wolf, G. M. Weir, S. A. Bozhenkov, H. Damm, G. Fuchert, O. Grulke, M. Hirsch, W. Kasparek, T. Klinger, C. Lechte, E. Pasch, B. Plaum, E. A. Scott, and W7-X Team, Review of Scientific Instruments \textbf{89}, 10H115 (2018).

\bibitem{deboo_pop_2010} J. C. DeBoo, C. Holland, T. L. Rhodes, L. Schmitz, G. Wang, A. E. White, M. E. Austin, E. J. Doyle, J. C. Hillesheim, W. A. Peebles, C. C. Petty, Z. Yan, and L. Zeng, Phys. Plasmas \textbf{17}, 056105 (2010).

\bibitem{silva_ppcf_2018} C. Silva, J. C. Hillesheim, L. Gil, C. Hidalgo, L. Meneses, F. Rimini and JET Contributors, Plasma Phys. Control. Fusion \textbf{60}, 085006 (2018).

\bibitem{silva_ppcf_2019} C. Silva, J. C. Hillesheim, L. Gil, C. Hidalgo, C. F. Maggi, L. Meneses, E. R. Solano and JET Contributors, Plasma Phys. Control. Fusion \textbf{61}, 075007 (2019).

\bibitem{conway_irw_2011} G. D. Conway, A. Stegmeir, E. Poli and E. Strumberger, Proc. 10th Int. Reflectometry Workshop, Padova (2011).

\bibitem{gusakov_ppcf_2004} E. Z. Gusakov and A. V. Surkov, Plasma Phys. Control. Fusion \textbf{46}, 1143 (2004).

\bibitem{gusakov_eps_2011} E. Z. Gusakov and A. Yu Popov, \emph{Proc. 38th EPS Conference on Plasma Physics}, Strasbourg, 27 June–1 July, P4.056 (2011).

\bibitem{blanco_estrada_ppcf_2013} E. Blanco and T. Estrada, Plasma Phys. Control. Fusion \textbf{55}, 125006 (2013).


\bibitem{ruizruiz_pop_2015} J. Ruiz Ruiz, Y. Ren, W. Guttenfelder, A. E. White, S. M. Kaye, B. P. Leblanc, E. Mazzucato, K. C. Lee, C. W. Domier, D. R. Smith, and H. Yuh, Phys. Plasmas \textbf{22}, 122501 (2015).

\bibitem{ruizruiz_ppcf_2019} J. Ruiz Ruiz, W. Guttenfelder, A. E. White, N. T. Howard, J. Candy, Y. Ren, D. R. Smith, N. F. Loureiro, C. Holland and C. W. Domier, Plasma Phys. Control. Fusion \textbf{61}, 115015 (2019). 

\bibitem{ruizruiz_ppcf_2020} J. Ruiz Ruiz, W. Guttenfelder, A. E. White, N. T. Howard, J. Candy, Y. Ren, D. R. Smith, and C. Holland, Plasma Phys. Control. Fusion \textbf{62}, 075001 (2020). 


\bibitem{ren_nf_2020} Y. Ren, W. X. Wang, W. Guttenfelder, S. M. Kaye, J. Ruiz Ruiz, S. Ethier, R. Bell, B. P. Leblanc, E. Mazzucato, D. R. Smith, C. W. Domier and H. Yuh, Nucl. Fusion \textbf{60}, 026005 (2020).

\bibitem{gyro} J. Candy and R. E. Waltz, J. Comput. Phys. \textbf{186}, (2003) 545.

\bibitem{kotschenreuther_cpc_1995} M. Kotschenreuther \emph{et al}, Comput. Phys. Commun. \textbf{88}, 128 (1995).

\bibitem{ruizruiz_pop_2020} J. Ruiz Ruiz, W. Guttenfelder, A. E. White, N. T. Howard, J. Candy, Y. Ren, D. R. Smith, and C. Holland, Phys. Plasmas \textbf{27}, 122505 (2020). 

\bibitem{gyro_guide} J. Candy and E. Belli, \emph{GYRO Technical Guide}, General Atomics, P.O. Box 85608, San Diego, CA 92186-5608, USA.

\bibitem{christen_jpp_2021} N. Christen, M. Barnes, and F. Parra, J. Plasma Phys. \textbf{87}(2), (2021). 

\bibitem{sugama_pop_1998} H. Sugama and W. Horton, Phys. Plasmas \textbf{5}, 2560 (1998).

\bibitem{valerian_ppcf_2021} Hall-Chen \emph{et al}., (in preparation). 

\bibitem{holland_pop_2009} C. Holland, A. E. White, G. R. McKee, M. W. Shafer, J. Candy, R. E. Waltz, L. Schmitz, and G. R. Tynan, Phys. Plasmas \textbf{16}, 052301 (2009).

\bibitem{holland_nf_2012} C. Holland, J. C. DeBoo, T. L. Rhodes, L. Schmitz, J. C. Hillesheim, G. Wang, A. E. White, M. E. Austin, E. J. Doyle,W. A. Peebles, C. C. Petty, L. Zeng and J. Candy, Nucl. Fusion \textbf{52}, 063028 (2012).

\bibitem{hillesheim_rsi_2012} J. C. Hillesheim, C. Holland, L. Schmitz, S. Kubota, T. L. Rhodes, and T. A. Carter, Rev. Sci. Instrum. \textbf{83}, 10E331 (2012).

\bibitem{rhodes_rsi_2006} T. L. Rhodes, W. A. Peebles, X. Nguyen, M. A. VanZeeland, J. S. deGrassie, E. J. Doyle, G. Wang, and L. Zeng, Rev. Sci. Instrum. \textbf{77}, 10E922 (2006)

\bibitem{belli_ppcf_2008} E. A. Belli and J. Candy, Plasma Phys. Control. Fusion \textbf{50}, 095010 (2008).

\bibitem{transp} R. J. Hawryluk, \emph{Physics of Plasma Close to Thermonuclear Conditions}, (Pergamon, New York, 1981).

\bibitem{guttenfelder_nf_2013} W. Guttenfelder, J. L. Peterson, J. Candy, S. M. Kaye, Y. Ren, R. E. Bell, G. W. Hammett, B. P. LeBlanc, D. R. Mikkelsen, W. M. Nevins and H. Yuh, Nucl. Fusion \textbf{53}, 093022 (2013).

\bibitem{miller_pop_1998} R. L. Miller. M. S. Chu, J. M. Greene, Y. R. Lin-Liu, and R. E. Waltz, Phys. Plasmas \textbf{5}, 973 (1998).

\bibitem{drake_prl_1988} J. F. Drake, P. N. Guzdar, and A. B. Hassam, Phys. Rev. Lett \textbf{61}, 2206 (1988).

\bibitem{dorland_prl_2000} W. Dorland, F. Jenko, M. Kotschenreuther, and B. N. Rogers, Phys. Rev. Lett. \textbf{85}, 5579 (2000).

\bibitem{jenko_pop_2000} F. Jenko, W. Dorland, M. Kotschenreuther, and B. N. Rogers, Phys. Plasmas \textbf{7}, 1904 (2000).

\bibitem{roach_ppcf_2005} C. M. Roach, D. J. Applegate, J. W. Connor, S. C. Cowley, W. D. Dorland, R. J. Hastie, N. Joiner, S. Saarelma, A. A. Schekochihin, R. J. Akers, C. Brickley, A. R. Field, M. Valovic and the MAST Team, Plasma Phys. Control. Fusion \textbf{47}, B323 (2005).

\bibitem{roach_ppcf_2009} C. M. Roach, I. G. Abel, R. J. Akers,W. Arter, M. Barnes, Y. Camenen, F. J. Casson, G. Colyer, J. W. Connor, S. C. Cowley, D. Dickinson, W. Dorland, A. R. Field, W. Guttenfelder, G. W. Hammett, R. J. Hastie, E. Highcock, N. F. Loureiro, A. G. Peeters, M. Reshko, S. Saarelma, A. A. Schekochihin2, M. Valovic and H.R. Wilson, Plasma Phys. Control. Fusion \textbf{51}, 124020 (2009).

\bibitem{guttenfelder_pop_2011} W. Guttenfelder and J. Candy, Phys. Plasmas \textbf{18}, 022506 (2011).

\bibitem{nevins_pop_2006} W. M. Nevins, J. Candy, S. Cowley, T. Dannert, A. Dimits, W. Dorland, C. Estrada-Mila, G. W. Hammett, F. Jenko, M. J. Pueschel, and D. E. Shumaker, Phys. Plasmas \textbf{13}, 122306 (2006).

\bibitem{jenko_pop_2002_w7as} F. Jenko, and A. Kendl, Phys. Plasmas \textbf{9}, 4103 (2002).

\bibitem{biglari_physflub_1990} H. Biglari, P. H. Diamond, and P. W. Terry, Phys. Fluids B: Plasma Physics \textbf{2}, 1 (1990).

\bibitem{burrell_pop_1997} K. H. Burrell, Phys. Plasmas \textbf{4}, 1499 (1997).

\bibitem{parra_pop_2011} Felix I. Parra, Michael Barnes, and Arthur G. Peeters, Phys. Plasmas \textbf{18}, 062501 (2011).

\bibitem{shafer_pop_2012} M. W. Shafer, R. J. Fonck, G. R. McKee, C. Holland, A. E. White, and D. J. Schlossberg, Phys. Plasmas \textbf{19}, 032504 (2012).

\bibitem{fox_ppcf_2017_symbreak} M. F. J. Fox, F. van Wyk, A. R. Field, Y-c Ghim, F. I. Parra, A. A. Schekochihin and the MAST Team, Plasma Phys. Control. Fusion \textbf{59}, 034002 (2017).

\bibitem{fox_ppcf_2017_2dbes} M. F. J. Fox, A. R. Field, F. van Wyk, Y-c Ghim, A. A. Schekochihin and the MAST Team, Plasma Phys. Control. Fusion \textbf{59}, 044008 (2017).

\bibitem{pinzon_ppcf_2019} J. R. Pinzón, T. Estrada, T. Happel, P. Hennequin, E. Blanco, U. Stroth, the ASDEX Upgrade and TJ-II Teams, Plasma Phys. Control. Fusion \textbf{61}, 105009 (2019).

\bibitem{pinzon_nf_2019} J. R. Pinzón, T. Happel, P. Hennequin, C. Angioni, T. Estrada, A. Lebschy, U. Stroth, and the ASDEX Upgrade Team, Nucl. Fusion \textbf{59}, 074002 (2019).

\bibitem{maj_pop_2009} O. Maj, G. V. Pereverzev, and E. Poli, Phys. Plasmas \textbf{16}, 062105 (2009).

\bibitem{maj_ppcf_2010} O. Maj, A. A. Balakin, and E. Poli, Plasma Phys. Control. Fusion \textbf{52}, 085006 (2010).

\bibitem{conway_irw_2015} G. D. Conway, C. Lechte, A. Fochi, and the ASDEX Upgrade Team, Proc. 12th Intl. Reflectometry Workshop - IRW12, Julich (2015).

\bibitem{conway_irw_2019} G. D. Conway, C. Lechte, E. Poli, O. Maj and the ASDEX Upgrade Team, Proc. 14th Intl. Reflectometry Workshop - IRW14, Lausanne, (2019).

\bibitem{dimits_pre_1993} A. M. Dimits, Phys. Rev. E \textbf{48}, 4070 (1993).

\bibitem{beer_pop_1995} M. Beer \emph{et al}., Phys. Plasmas \textbf{2} (1995).

\bibitem{mckee_rsi_2003} G. R. McKee, C. Fenzi, R. J. Fonck, and M. Jakubowski, Rev. Sci. Instrum. \textbf{74}, 2014 (2003).

\bibitem{ghim_prl_2013} Y.-c. Ghim, A. A. Schekochihin, A. R. Field, I. G. Abel, M. Barnes, G. Colyer, S. C. Cowley, F. I. Parra, D. Dunai, S. Zoletnik, and the MAST Team, Phys. Rev. Lett. \textbf{110}, 145002 (2013).

\bibitem{field_ppcf_2014} A. R. Field, D. Dunai, Y.-c. Ghim, P. Hill, B. McMillan, C. M. Roach, S. Saarelma, A. A. Schekochihin, S. Zoletnik and the MAST Team, Plasma Phys. Control. Fusion \textbf{56}, 025012 (2014).

\bibitem{gusakov_ppcf_2005} E. Z. Gusakov, A. V. Surkov and A. Yu Popov, Plasma Phys. Control. Fusion \textbf{47}, 959 (2005).

\bibitem{staebler_prl_2013} G. M. Staebler, R. E. Waltz, J. Candy, and J. E. Kinsey, Phys. Rev. Lett. \textbf{110}, 055003 (2013).

\bibitem{kirov_ppcf_2002} K. K. Kirov, F. Leuterer, G. V. Pereverzev, F. Ryter, W. Suttrop and ASDEX Upgrade team, Plasma Phys. Control. Fusion \textbf{44}, 2583 (2002).

\bibitem{tsironis_pop_2009} C. Tsironis, A. G. Peeters, H. Isliker, D. Strintzi, I. Chatziantonaki, and L. Vlahos2, Phys. Plasmas \textbf{16}, 112510 (2009).

\bibitem{peysson_ppcf_2011} Y. Peysson, J. Decker, L. Morini and S. Coda, Plasma Phys. Control. Fusion \textbf{53}, 124028 (2011).

\bibitem{ram_pop_2013} A. K. Ram, K. Hizanidis, and Y. Kominis, Phys. Plasmas \textbf{20}, 056110 (2013).

\bibitem{sysoeva_nf_2015} E.V. Sysoeva, F. da Silva, E. Z. Gusakov, S. Heuraux and A. Yu. Popov, Nucl. Fusion \textbf{55}, 033016 (2015).

\bibitem{ram_pop_2016} A. K. Ram and K. Hizanidis, Phys. Plasmas \textbf{23}, 022504 (2016).

\bibitem{brookman_epj_2017} M. W. Brookman, M. E. Austin, K. W. Gentle, C. C. Petty, D. E. Ernst, Y. Peysson, J. Decker, and K. Barada, EPJ Web Conf. \textbf{147}, 03001 (2017).

\bibitem{kohn_ppcf_2018} A. K\"{o}hn, L. Guidi, E. Holzhauer, O. Maj, E. Poli, A. Snicker and H. Weber, Plasma Phys. Control. Fusion \textbf{60}, 075006 (2018). 

\bibitem{snicker_nf_2018} A. Snicker, E. Poli, O. Maj, L. Guidi, A. Köhn, H. Weber, G. Conway, M. Henderson and G. Saibene, Nucl. Fusion \textbf{58}, 016002 (2018). 

\bibitem{chellai_prl_2018} O. Chella\"{i}, S. Alberti, M. Baquero-Ruiz, I. Furno, T. Goodman, F. Manke, G. Plyushchev, L. Guidi, A. Koehn, O. Maj, E. Poli, K. Hizanidis, L. Figini, and D. Ricci, Phys. Rev. Lett. \textbf{120}, 105001 (2018).

\bibitem{chellai_ppcf_2019} O. Chella\"{i}, S. Alberti, M. Baquero-Ruiz, I. Furno, T. Goodman, B. Labit, O. Maj, P. Ricci, F. Riva, L. Guidi, E. Poli and the TCV team, Plasma Phys. Control. Fusion \textbf{61}, 014001 (2019).

\bibitem{chellai_nf_2021} O. Chella\"{i}, S. Alberti, I. Furno, T. Goodman, O. Maj, G. Merlo, E. Poli, P. Ricci, F. Riva, H. Weber, and the TCV team, Nucl. Fusion in press (2021), \text{https://doi.org/10.1088/1741-4326/abf43f}.

\bibitem{lin_ppcf_2001} Y. Lin, R. Nazikian, J. H. Irby and E. S. Marmar, Plasma Phys. Control. Fusion \textbf{43}, L1 (2001).

\bibitem{garcia_nf_2017} J. Garcia, T. Görler, F. Jenko and G. Giruzzi, Nucl. Fusion \textbf{57}, 014007 (2017). 

\bibitem{pinzon_ppcf_2017} J. R. Pinzón, T. Happel, E. Blanco, G. D. Conway, T. Estrada and U. Stroth, Plasma Phys. Control. Fusion \textbf{59}, 035005 (2017).

\bibitem{krutkin_ppcf_2019} O. L. Krutkin, E. Z. Gusakov, S. Heuraux and C. Lechte, Plasma Phys. Control. Fusion \textbf{61} 045010 (2019).

\bibitem{blanco_ppcf_2008} E. Blanco and T. Estrada, Plasma Phys. Control. Fusion \textbf{50}, 095011 (2008). 

\bibitem{lechte_ieee_2009} C. Lechte, IEEE Trans. Plasma Sci., \textbf{37}, pp. 1099, (2009).

\bibitem{stroth_nf_2015} U. Stroth, A. Ba$\tilde{n}$\'{o}n Navarro, G. D. Conway, T. G\"{o}rler, T. Happel, P. Hennequin, C. Lechte, P. Manz, P. Simon, A. Biancalani, E. Blanco, C. Bottereau, F. Clairet, S. Coda, T. Eibert, T. Estrada, A. Fasoli, L. Guimarais, \"{O}. G\"{u}rcan, Z. Huang, F. Jenko, W. Kasparek, C. Koenen, A. Kr\"{a}mer-Flecken, M. E. Manso, A. Medvedeva, D. Molina, V. Nikolaeva, B. Plaum, L. Porte, D. Prisiazhniuk, T. Ribeiro, B. D. Scott, U. Siart, A. Storelli, L. Vermare, S. Wolf and the ASDEX Upgrade team, Nucl. Fusion \textbf{55}, 083027 (2015).

\bibitem{happel_ppcf_2017} T. Happel, T G\"{o}rler, P. Hennequin, C. Lechte, M. Bernert, G. D. Conway, S. J. Freethy, C. Honor\'{e}, J. R. Pinz\'{o}n, U. Stroth and The ASDEX Upgrade Team, Plasma Phys. Control. Fusion \textbf{59}, 054009 (2017).

\bibitem{lechte_ppcf_2017} C. Lechte, G. D. Conway, T G\"{o}rler, C. Tr\"{o}ster-Schmid and the ASDEX Upgrade Team, Plasma Phys. Control. Fusion \textbf{59}, 075006 (2017).

\bibitem{lechte_pst_2020} C. Lechte, G. D. Conway, T. G\"{o}rler, T. Happel and the ASDEX Upgrade Team, Plasma Sci. Technol. \textbf{22}, 064006 (2020).

\bibitem{kruskal_kulsrud_physflu_1958} M. D. Kruskal, and R. M. Kulsrud, Phys. Fluids \textbf{1}, 265 (1958).











\end{thebibliography}
\end{document}